\documentclass[11pt]{article}   
\usepackage{amsmath}\usepackage{hyperref}\usepackage{wasysym}
\usepackage[font=footnotesize]{caption}\usepackage[T1]{fontenc}\usepackage{latexsym}
\usepackage[english]{babel}\usepackage{cite}\usepackage{enumerate}
\usepackage{bbold}\usepackage[shortlabels]{enumitem}\usepackage[utf8]{inputenc}
\usepackage{dsfont}\usepackage{diagbox}\usepackage{amsfonts}\usepackage{mathtools}      
\usepackage{amssymb}\usepackage{euscript}\usepackage{braket}\usepackage{amssymb}
\usepackage{starfont}\usepackage{color,soul}\usepackage{braket}\usepackage{tensor}        
\usepackage{amsthm}\usepackage{graphicx}\usepackage{slashed}\usepackage{leftidx}
\usepackage{subfigure}\usepackage{bbm}\usepackage{empheq}\usepackage{color}
\usepackage{amsfonts}\usepackage{amssymb}\usepackage{graphicx}
\usepackage{amssymb,epsfig,subfigure}\usepackage{amssymb}\usepackage{comment}

\usepackage{skak}
\usepackage{makecell}
\usepackage[T1]{fontenc}\usepackage{latexsym}\usepackage{cancel}
\usepackage{titlesec}\usepackage[numbers,sort&compress]{natbib}  
\usepackage{float}\usepackage[dvipsnames]{xcolor}
\usepackage[header,title,page,titletoc]{appendix}  
\usepackage[numbers,sort&compress]{natbib}  
\usepackage[paper=a4paper,margin=1in]{geometry}  
\usepackage{wasysym} 
\allowdisplaybreaks[4]   
\numberwithin{equation}{section}
\makeatletter
\renewcommand\section{\@startsection {section}{1}{\z@}%
                                 {-3.5ex \@plus -1ex \@minus -.2ex}%nn
                                   {2.3ex \@plus.2ex}%
                                   {\normalfont\large\bfseries}}
\renewcommand\subsection{\@startsection{subsection}{2}{\z@}%
                                   {-3.25ex\@plus -1ex \@minus -.2ex}%
                                     {1.5ex \@plus .2ex}%
                                     {\normalfont\bfseries}}
\renewcommand\subsubsection{\@startsection{subsubsection}{3}{\z@}%
                                   {-3.25ex\@plus -1ex \@minus -.2ex}%
                                     {1.5ex \@plus .2ex}%
                                     {\normalfont\itshape}}
\makeatother
\def\pplogo{\vbox{\kern-\headheight\kern -29pt
\halign{##&##\hfil\cr&{\ppnumber}\cr\rule{0pt}{2.5ex}&\ppdate\cr}}}
\makeatletter
\def\ps@firstpage{\ps@empty \def\@oddhead{\hss\pplogo}%
  \let\@evenhead\@oddhead 
}
\def\maketitle{\par
 \begingroup
 \def\thefootnote{\fnsymbol{footnote}}
 \def\@makefnmark{\hbox{$^{\@thefnmark}$\hss}}
 \if@twocolumn
 \twocolumn[\@maketitle]
 \else \newpage
 \global\@topnum\z@ \@maketitle \fi\thispagestyle{firstpage}\@thanks
 \endgroup
 \setcounter{footnote}{0}
 \let\maketitle\relax
 \let\@maketitle\relax
 \gdef\@thanks{}\gdef\@author{}\gdef\@title{}\let\thanks\relax}
\makeatother            
\definecolor{rossocorsa}{rgb}{0.83, 0.0, 0.0}
\definecolor{navyblue}{rgb}{0.0, 0.0, 0.5}
\hypersetup{
    colorlinks,
    citecolor=rossocorsa,
    filecolor=navyblue,
    linkcolor=navyblue,
    urlcolor=navyblue
} 
\newcommand\bea{\begin{eqnarray}}\newcommand\eea{\end{eqnarray}}
\newcommand{\be}{\begin{equation}}\newcommand{\ee}{\end{equation}}
\newcommand{\ba}{\begin{align}}\newcommand{\ea}{\end{align}}
\newcommand{\e}{\varepsilon}\newcommand{\s}{\sigma}\newcommand{\p}{\times}

\begin{document} 

\begin{titlepage}

\begin{center}

\phantom{ }
\vspace{1cm}
{\bf \Large{Selection rules for RG flows of minimal models}}
\vskip 1cm
Valentin Benedetti${}^{a}$, Horacio Casini${}^{b}$, Javier M. Mag\'an${}^{c}$
\vskip 0.05in
${}^{a}$\small{ \textit{The Abdus Salam International Centre for Theoretical Physics}}
\vskip -.4cm
\small{\textit{Strada Costiera 11, Trieste 34151, Italy}}
\vskip 0.05in
${}^{b,c}$\small{ \textit{Instituto Balseiro, Centro At\'omico Bariloche}}
\vskip -.4cm
\small{\textit{ 8400-S.C. de Bariloche, R\'io Negro, Argentina}}

\begin{abstract}

Minimal $d=2$ CFTs are usually classified through modular invariant partition functions. There is a finer classification of ``non complete'' models when $S$-duality is not imposed. We approach this classification by starting with the local chiral algebra and adding primaries sequentially. At each step, we only impose locality (T-duality) and closure of the operator algebra. For each chiral algebra, this produces a tree-like graph. Each tree node corresponds to a local $d=2$ CFT, with an intrinsic Jones index measuring the size of Haag duality violation. This index can be computed with the partition function and is related to the total quantum dimension of the category of superselection sectors of the node, and to the relative size between the node and a modular invariant completion. In this way, we find in a very explicit manner a  classification of local minimal ($c<1$) $d=2$ CFTs.  When appropriate, this matches Kawahigashi-Longo's previous results. We use this finer classification to constrain RG flows. For a relevant perturbation, the flow can be restricted to the subalgebra associated with it, typically corresponding to a non-modular invariant node in the tree. The structure of the graph above such node needs to be preserved by the RG flow. In particular, the superselection sector category for the node must be preserved. This gives selection rules that recover in a unified fashion several known facts while unraveling new ones.

\end{abstract}

\end{center}

\small{\vspace{7.0 cm}\noindent ${}^{a}$valentin.benedetti@gmail.com\\
${}^{b}$horaciocasini@gmail.com\\
${}^{c}$javier.magan.physics@gmail.com}

\end{titlepage}

\setcounter{tocdepth}{2}

{\parskip = .4\baselineskip \tableofcontents}
\newpage

%%%%%%%%%%%%%%%%%%%%%%%%%%%%%%%%%%%%%%%%%%%%%%%%%%%%%%%%%%%%%%%%%
%%%%%%%%%%%%%%%%%%%%%%%%%%%%%%%%%%%%%%%%%%%%%%%%%%%%%%%%%%%%%%%%%
%%%%%%%%%%%%%%%%%%%%%%%%%%%%%%%%%%%%%%%%%%%%%%%%%%%%%%%%%%%%%%%%%
%%%%%%%%%%%%%%%%%%%%%%%%%%%%%%%%%%%%%%%%%%%%

\section{Introduction}\label{Sec1}

Locality is one of the main pillars of relativistic Quantum Field Theory (QFT). But, it is important to recall that there are two sensible notions of locality. The first expresses the fact that operators supported in a certain region are ultimately generated by those supported in smaller regions. This is called additivity, and the algebra $\mathcal{A}(R)$ of all operators that can be locally generated is called the additive algebra. The second expresses the fact that observables commute at spatial distances. This corresponds to the causality of the theory.  It is then natural to ask whether both notions agree. Formally, this question concerns the analysis of Haag duality for arbitrary regions $R$ \cite{Casini:2020rgj,Review}, namely the analysis of the (non)-saturation of the following inclusion of algebras
\be 
{\cal A}(R)\subseteq ({\cal A}(R'))'\equiv \hat{\cal A}(R) \,. \label{ff}
\ee
Here $R'$ is the set of points spatially separated from $R$, the causal complement of $R$, and ${\cal A}'$ is the commutant of ${\cal A}$, i.e. the algebra of operators commuting with all operators in ${\cal A}$. The left-hand side is the natural algebra of operators ${\cal A}(R)$ for region $R$ from the perspective of additivity, while the right-hand side is the natural algebra of operators $\hat{\cal A}(R)$ for region $R$ from the perspective of causality. Intuitively, the left-hand side is the ``invariant part'' of the right-hand side under the action of a generalized symmetry \cite{Casini:2020rgj,Review}.

Quite insightfully, the difference between the two meanings of locality, formalized in the previous equation, allows for a unified characterization of phases in relativistic QFT. This includes e.g. confinement phases and more standard symmetry-breaking scenarios \cite{Casini:2019kex,Casini:2020rgj}. Different QFT's and phases are characterized by the topology of the region $R$ where such inclusion is not saturated, and by the scaling of the expectation values of the operators differentiating the two algebras, i.e. those producing a violation of Haag duality. These Haag Duality Violating (HDV) operators are the generalized order parameters in the sense of the Landau paradigm. These subtleties regarding the notion of locality also suggest an intrinsic definition of completeness, namely that the previous inclusion is saturated for all possible regions \cite{Review}.

The purpose of this article is to deepen on these basic QFT features in the familiar context of $d=2$ CFTs, where we can match the algebraic language with the standard, thoroughly studied, approach to $d=2$ CFTs \cite{francesco2012conformal}. More precisely, our purpose is twofold. On one hand, we will see how these features motivate a finer classification of local $d=2$ CFTs, which we achieve in the case of minimal models. On the other hand, we will describe how this finer classification allows us to find selection rules constraining certain classes of RG flows in those theories.

Let's expand on these two directions. The standard classification of $d=2$ minimal CFTs is known as the ADE classification \cite{ADE}. This classification describes all modular invariant partition functions arising by combining left/right irreducible representations of the Virasoro (or chiral) algebra \cite{francesco2012conformal}. But by enforcing modular invariance one misses a large class of local  $d=2$ CFTs. This has been recently discussed in detail in Ref. \cite{Benedetti:2024dku}. Briefly, while $T$ invariance is necessary for locality, $S$ invariance is not mandatory for a relativistic $d=2$ CFT. Equivalently, while modular invariance for $c<1$ entails locality, associativity and closure of the operator algebra, there are more theories satisfying the latter and not the former. More precisely, Ref. \cite{Benedetti:2024dku} shows that $S$ invariance is a form of completeness of the theory that has a precise meaning as Haag duality for arbitrary multi-interval regions, or equivalently as the absence of superselection sectors.\footnote{The relation between $S$ invariance and the absence of superselection sectors was first conjectured by Rehren \cite{R6}. This conjecture was later proved by  Y. Kawahigashi and R. Longo, and independently by M. M\"uger, but the proof remained unpublished, see \cite{muger2010superselection,Benedetti:2024dku}.} Then, the obvious question concerns the classification of $d=2$ CFTs, where we only impose locality, namely $T$ invariance, and closure of the operator algebra. 

The first objective of this article is to obtain such finer classification of local unitary minimal models, using standard tools in $d=2$ CFTs \cite{francesco2012conformal}. The idea is simple and constructive. We start with the smallest chiral algebra. These algebras typically show strong violations of Haag duality, as we explicitly compute. This means the theory is incomplete and that extensions might be constructed. We explicitly construct such extensions by adding local operators one at a time and in all possible ways, respecting locality and closure of the resulting Operator Product Expansion (OPE) algebra. For each chiral algebra, we obtain a tree-like graph. Each node in the tree represents a local d = 2 CFT, and it is partially characterized by the size of Haag duality violation in two intervals. The
bottom node is the chiral algebra, while extremal nodes correspond to modular invariant
theories.  This way we find multiple local $d=2$ CFTs in between chiral algebras and the standard modular invariant theories. These are determined by the spectrum of local primaries and associated category of superselection sectors.

 Another perspective on this classification is the following. Complete, i.e. modular invariant, models have been classified by the ADE classification \cite{ADE}. As shown for example in \cite{Benedetti:2024dku}, all possible minimal models are local submodels of the complete ones. Then, given the fusion rules for each complete model, the classification we are seeking is reduced to the computational problem of finding all closed OPE subalgebras of the complete ones.

The classification of possible superselection categories for $c<1$ CFTs with chiral parity symmetry was accomplished previously by Kawahigashi and Longo  \cite{kawahigashi2001multi,Kawahigashi:2002px,Kawahigashi:2003gi}. Their more abstract approach uses the algebraic description of QFT, where the possible $Q$-systems of minimal chiral algebras were laid down. This uses the Doplicher-Haag-Roberts (DHR) approach to superselection sectors \cite{Doplicher:1971wk,Doplicher:1973at,Doplicher:1990pn} and the machinery of $\alpha$-induction introduced in \cite{Longo:1994xe,Bockenhauer:1998ca,Bockenhauer:1998in,Bockenhauer:1998ef}, a method for producing DHR endomorphisms of extended CFTs from DHR endomorphisms of the smaller ones. Our results match Kawahigashi and Longo (KL) classification when appropriate. But our result gives, in addition, the explicit field content of each model on top of its symmetry, i.e. the precise spectrum of local primaries. It also keeps track of the inclusion relations in the tree of submodels, and we find instances where a given category symmetry appears more than once as different submodels. We also find some submodels that are not parity symmetric and are not contained in KL classification table. Except for some examples for the lowest central charges we will not attempt a general classification of fermionic models. 

We will then use this finer classification and understanding of unitary minimal models to describe aspects of RG flows between them. The analysis of RG flows between minimal models has a long history. A seminal starting contribution was that of Zamolodchikov \cite{Zamolodchikov1987}. In such reference, the flow between subsequent minimal models (i.e. from labels $m+1$ to $m$ as will be described below), in the limit where $m$ is large and the flow is perturbative, was studied. Such flow is triggered by the least relevant perturbation in the UV. Later, the flow from tricritical $m=4$ to Ising $m=3$ was also analyzed using Bethe ansatz techniques \cite{ZamolodchikovBethe1,ZamolodchikovBethe2}. The analysis of this flow was deepened by considering all possible perturbations of tricritical ising \cite{CardyTricritical} and further justified using supersymmetry \cite{Susy1,Susy2,Susy3,Susy4}.  A new twist was introduced by Gaiotto \cite{GaiottoDomain}, who conjecture the mapping between UV and IR via domain walls and conformal boundary conditions. This approach already implicitly uses the concept of generalized symmetry, through the notion of topological defect. The analysis of these flows was extended to higher orders in \cite{PoghossianNexto}. Related to these developments, extensions of these flows to non-diagonal minimal models were described in \cite{NonDiagonalRavanini,NonDiagonalKlassen}, and to gapped phases, where one sees the emergence of topological field theories \cite{Reshetikhin:1989qg,Smirnov:1991uw,OBrien:2017wmx,Aasen:2016dop,Aasen:2020jwb,ShaoMinimal,CordovaMinimal}. We also note that a quantum information perspective was put forward in \cite{CalabreseTricritical} using Renyi entropies. Further from the context of the present paper, there are also some studies for non unitary models \cite{Nakayama:2024msv,Katsevich:2024jgq}. 

Given this context, the second objective of this article is to describe a new unifying perspective on these RG flows using the previous finer classification of $d=2$ CFTs. We first notice that, whenever a flow is triggered by a certain perturbation in the UV,  the flow might as well be thought as happening in the combined algebra generated by the stress tensor plus the perturbation itself. Interestingly, this combined algebra may not form a modular invariant theory, e.g. this is the case for the Zamolodchikov RG flows \cite{Zamolodchikov1987}. Equivalently, we can think the CFT where the RG flow happens is one of the local incomplete $d=2$ CFTs classified above. In turn, the fact that an RG flow happens in such subtheory implies that all the possible completions, i.e. all the tree paths that start in the given incomplete node and end in one modular invariant completion, must remain intact along the flow. In particular, all possible Jones indices above the subtheory must match between different scales. As could be anticipated, this can be seen as the preservation of generalized symmetries along the flow, where by generalized symmetry we here mean the relative category of superselection sectors unbroken by the perturbation. We will describe how this perspective explains known selection rules for RG flows while unraveling many more. Indeed, it provides a unified view of diagonal and non-diagonal flows, and one can proceed further and identify the charges of all operators under these generalized symmetries, providing non-pertubative selection rules for operators along the flow.

\noindent \textbf{Organization of the article}

 To begin, in section \ref{MinimalModels}, we classify all possible local minimal models, show their mutual relationships, and compute their indices. This classification will be performed constructively and we will describe first in full detail the cases of $m=3,4,5$. These simplest examples pave the way to understanding the generalizations. In section \ref{mrgflows} we use this finer classification to provide a new perspective on selection rules and RG flows between minimal models. We analyze both the diagonal and the non-diagonal cases. In section \ref{discussionrg} we will end with some open discussion.  We refer to Appendix \ref{HaagDuality} for a brief review of aspects of the Jones index associated with the inclusion of algebras, and its relation with Haag duality in QFT and the structure of superselection sectors. In particular, we explain how such indices relate to the quantum dimensions of the associated models and the intrinsic categories behind them. Appendix \ref{su2} provides relevant formulas for the $su(2)_k$ and $D_{2n}^{\textrm{even}}$ categories, which are the ones that appear in the context of minimal models.

\section{Haag duality and the classification of Minimal Models }\label{MinimalModels}

We now classify local relativistic unitary $d=2$ CFTs for $c<1$. The modular invariant part is known as the ADE classification \cite{ADE}. The full classification we seek to obtain is related to the classification of $Q$-systems for $c<1$ CFTs, accomplished in \cite{kawahigashi2001multi,Kawahigashi:2002px,Kawahigashi:2003gi}. We will comment on the relation in due time. Our approach will use standard tools, namely the OPE. We will see how the violation of Haag duality characterizes this finer classification of local minimal models.

We begin in section \ref{m1} with a brief review of minimal models. Then, in section \ref{m2}, we explain how to compute the global index of a given minimal model, see also the appendix \ref{HaagDuality} for the general discussion. In section \ref{sec23} we discuss the general strategy and rules for constructing local $d=2$ CFTs. In Section \ref{m345}, we present warm-up calculations describing all allowed models for $m=3,4,5$, where the modular invariant Ising, Tricritial Ising, and Three-State Potts models reside, respectively, and where we include the classification of fermioninc models as well. Then, we discuss the general case $m\geq 5$ in \ref{m6} for spinless models, i.e. submodels of diagonal modular invariants, namely the $A$ series. In section \ref{m7} we discuss models with integer spin which are submodels of non-diagonal modular invariants of the $D$ series. Finally in section \ref{mE} we discuss the submodels of the lowest non-diagonal $E$ series.

\subsection{Minimal models, partition functions and the ADE classification \label{m1} } 

We start by recalling some known facts about minimal models. This serves to set up conventions. The family of unitary minimal models is characterized by the central charge
\be 
c=1-\frac{6}{m(m+1)}\,. \label{cm}
\ee
The integer parameter $m\geq 3$ unequivocally determines the possible chiral parts of the spectrum of the theory. Every member of the family has a spectrum composed of a finite number of representations of the Virasoro algebra. These representations (chiral superselection sectors) are labeled by two integers $(r,s)$ obeying $1\leq r \leq m-1$ and $1\leq s \leq m$. This is the Kac table. These parameters determine the lowest energy of the representation, i.e. the conformal dimension of the primary, to be
\be 
h_{r,s}= \frac{[(m+1)r-m s]^2-1}{4 m(m+1)}\,. \label{hrs}
\ee
The representations $(r,s)$ and $(m-r,m+1-s)$  are the same, in particular, the equality of dimensions can be seen from (\ref{hrs}). This implies that a model defined by $m$ has up to $m(m-1)/2$ distinct chiral building blocks.

We will consider $d=2$ models with the two chiralities having the same central charge $c=\bar{c}$. The torus partition function of a given  minimal model defined by the parameter $m$ can be expressed as a function of the modular parameter $\tau$ as 
\be
Z(\tau)=\sum_{r,s}\sum_{r',s'} M_{r,s;\, r',s'}\,\chi_{r,s}(\tau)\overline{\chi}_{r',s'}\,(\overline{\tau})\,, \label{Zmin}
\ee
where $\chi_{r,s}$ represent the Virasoro characters associated with representations of conformal weight (\ref{hrs}) described by $(r,s)$ in a model of central charge (\ref{cm}). The sum over representations should be understood  without duplications in the Kac table. These characters are purely chiral partition functions in the given sector and can be written as a function of $q=e^{2\pi i \tau}$ in the form
\be
\chi_{r,s}(q)= \frac{q^{-\frac{1}{24}}}{\left[\prod_{{k}=1}^{\infty}(1-q^{k})\right]}\,\sum_{n=-\infty}^\infty\left(q^{\frac{(2m(m+1)n+mr+(m+1)s)^2}{4m(m-1)}} -q^{\frac{(2m(m+1)n+mr-(m+1)s)^2}{4m(m-1)}} \right)\,.
\label{char}
\ee
The coupling matrix $M$ appearing in the partition function is a square matrix of positive and integer entries of size $m(m-1)/2$. These integers determine which field representations, are formed by the two chiral components, and how many times, appear in the model.  

There is an action of the modular group $SL(2,\mathbb{R})/\mathbb{Z}_2$ on the characters. The modular group acts as Moebius transformations over $\tau$. Such group is generated by T-transformations acting as $\tau\to\tau+1$ and S-transformations acting as $\tau\to -1/\tau$. The characters provide a representation space for the modular group, where the modular transformations act linearly. More precisely
\be 
\chi_{r,s}(\tau+1)= \sum_{r',s'} T_{r,s;\, r',s'} \chi_{r,s}(\tau)\,,\quad \chi_{r,s}(-1/\tau)= \sum_{r',s'} S_{r,s;\, r',s'} \chi_{r,s}(\tau)\,.
\ee
where the $T$ and $S$ are unitary matrices given by
\bea
&T_{r,s;\, r',s'}& = \delta_{rr'}\delta_{ss'}e^{2\pi i(h_{r,s}-c/24)} \,, \label{tmat} \\
&S_{r,s;\, r',s'}& =  \sqrt{\frac{8}{m(m+1)}}(-1)^{(1+sr'+rs')}\sin{\Big(\frac{\pi (m+1)}{m} r r'\Big)}\sin{\Big(\frac{\pi m}{(m+1)} s s'\Big)}\,. \label{smat}
\eea
Therefore, the action of the modular group over the partition function (\ref{Zmin}) reads
\be
T[Z]\equiv Z(\tau+1)=\sum_{r,s}\sum_{r',s'} (T.M.T^\dagger)_{r,s;\, r',s'}\,\chi_{r,s}(\tau)\overline{\chi}_{r',s'}\,(\overline{\tau})\,, \label{Ztd}
\ee
\be
S[Z]\equiv Z(-1/\tau)=\sum_{r,s}\sum_{r',s'} (S.M.S^\dagger)_{r,s;\, r',s'}\,\chi_{r,s}(\tau)\overline{\chi}_{r',s'}\,(\overline{\tau})\,. \label{Zsd}
\ee
Modular invariant partition functions are then classified by coupling matrices commuting with the modular transformations. This is known as the ADE classification of minimal models \cite{ADE}. The easiest examples arise by including one copy of each spin-zero field. Equivalently, we take $M_{r,s;\, r',s'}= \delta_{rr'}\delta_{ss'}$, which obviously commutes with the unitary modular matrices. This is known as the $(A,A)$ series of modular invariant minimal models. Their partition functions are
\be
Z_{AA}(\tau)=\frac{1}{2}\sum_{r,s}|\chi_{r,s}(\tau)|^2\,. \label{ZAA}
\ee
From this point, there is a standard procedure to obtain non-diagonal modular invariants. This uses the fact that all models of the $(A,A)$ series have a $\mathbb{Z}_2$ symmetry. This becomes transparent by choosing a unique description for the fields obeying that $r+s$ is even. Then the $\mathbb{Z}_2$ symmetry transforms the fields as $\phi_{(r,s)}\to (-1)^{r+1} \phi_{(r,s)}$.  In this context, we can perform an orbifold of this $\mathbb{Z}_2$ symmetry and add the twisted sectors to recover modular invariance. The final partition function is
\be 
Z_{\tau}=\frac{1}{2}\Big( Z + Z_{-} + T[Z_{-}] + T[\,S[Z_{-}]\,] \Big)\,,
\label{morb}
\ee
where $Z_-$ is the twisted partition function of the form 
\be
Z_{-} (\tau)=\sum_{r,s}\,(-1)^{(r+1)}\chi_{r,s}(\tau)\overline{\chi}_{r,s}\,(\overline{\tau})\,. \label{Z-}
\ee
 The model described by $Z_{\tau}$ is always modular invariant.  For odd $m$ this is known as the $(A,D)$ series, and for even $m$ as the $(D, A)$ series. For $m=3,4$ the process yields the same diagonal model belonging to the $(A,A)$ series. This is because for $m< 5$  the $(A,D)$ or $(D,A)$ series coincides with the $(A,A)$. Finally we have the exceptional $E$ series. We will describe the spectrum of the lowest of these series later.

\subsection{Jones index from the partition function \label{m2}}

In $d=2$, the violation of Haag duality can only happen in regions with non-trivial $\pi_0$, i.e. disconnected regions. For a two interval region $R=R_1\cup R_2$ we have the following inclusion of algebras
\be 
{\cal A}(R_1\cup R_2)\subseteq  \hat{{\cal A}}(R_1\cup R_2)\;.
\ee
The violation of duality can be originated by vertex anti-vertex operators located in $R_1$ and $R_2$ respectively, but not belonging to the additive algebra. The Jones index $[ \mathcal{M}:\mathcal{N} ]$ for a general inclusion of algebras $\mathcal{N}\subset \mathcal{M}$ measures the relative size of $\mathcal{N}$ in $\mathcal{M}$,  see \cite{Jones1983,L11,Longo:1994xe} for original definitions and Appendix \ref{HaagDuality} for a brief account. The Jones index $[ \hat{{\cal A}}(R_1\cup R_2):{\cal A}(R_1\cup R_2) ]$ of the present inclusion is called the global index $\mu$ of the model \cite{kawahigashi2001multi}, and it is independent of the particular two intervals. It is a measure of the amount of Haag duality violation.  In the limit in which the two intervals touch each other, it can be computed with an entropic order parameter. This is a relative entropy, in the maximal algebra of the two intervals,  between the vacuum  state and a state  in which we have set to zero the expectation values of non local operators. By definition this entropic order parameter is upper bounded by the Jones index. The relation between the entropic order parameter for the Jones index of touching intervals was established by different means and in different scenarios in \cite{Casini:2020rgj,Casini:2019kex,longo2018relative,Xu:2018uxc,hollands2020variational,Pedro,magan2021proof}, and the sum of entropic order and disorder parameters is controlled by the Jones index in all geometric scenarios \cite{Casini:2019kex,Magan:2020ake,hollands2020variational}.

There is a simpler, more practical approach put forward recently in Ref. \cite{Benedetti:2024dku}. This is based on the computation of the second Renyi mutual information, whose value can be mapped to the computation of a standard partition function. The end result is the following explicit formula for the global index  
\be \label{mucft}
\mu^{-1/2}=\lim _{l\to 0}\frac{Z(il)}{Z(i/l)}=\lim _{l\to \infty}\frac{Z(i/l)}{Z(il)}\,,\quad \tau=il\,.
\ee
Notice this transparently shows how the violation of Haag duality is related to the violation of $S$-duality in the model \cite{Benedetti:2024dku}.

Let us compute (\ref{mucft}) explicitly here. Using the formulas from (\ref{m1}) to write both partition functions in terms of $l$, we get
\be
\frac{Z(i/l)}{Z(il)} =\frac{\sum_{r,s}\sum_{r',s'} (S.M.S^\dagger)_{r,s;\, r',s'} \,\chi_{r,s}(il)\overline{\chi}_{r',s'}\,(i\overline{l})}{\sum_{\rho,\s}\sum_{\rho',\s'} M_{\rho,\s;\, \rho',\s'}\,\chi_{\rho,\s}(il)\overline{\chi}_{\rho',\s'}\,(i\overline{l})}\,.
\ee
We are interested in the limit $l \to \infty $. Equivalently, this is $q\rightarrow 0$ for $q=e^{-2\pi l}$. In such limit,  the leading contributions of  the Virasoro characters come from
\be 
\chi_{r,s}(q)= q^{h_{r,s}}q^{-\frac{c}{24}}\big(a^{r,s}_0 + a^{r,s}_1 q  + a_2^{r,s} q^2 + O(q^3)\big)\,, \quad q\ll 1 \,.
\ee
Therefore, we get that the leading contributions come from the character of the stress tensor as
\be
\mu^{-1/2}= \frac{(SMS^\dagger)_{1,1;\, 1,1}\,\chi_{1,1}(q)\overline{\chi}_{1,1}\,(q)}{M_{1,1;\, 1,1}\,\chi_{1,1}(q)\overline{\chi}_{1,1}\,(q)}= \frac{\sum_{r,s}\sum_{r',s'} d_{r,s} M_{r,s;\, r',s'} d_{r',s'}}{\sum_{r,s} d^2_{r,s} }\nonumber\,,
\ee

where $d_{r,s}$ describe the quantum dimensions of the representations $(r,s)$. These explicitly read
\be 
d_{r,s} = \frac{S_{1,1;\, r,s}}{S_{1,1;\, 1,1}} = (-1)^{(r+s)}\frac{\sin{(\frac{\pi (m+1)r}{m} )}\sin{(\frac{\pi m s}{(m+1)})}}{\sin{(\frac{\pi (m+1)}{m})}\sin{(\frac{\pi m}{(m+1)})}}\,.
\ee
We have also used, from the unitarity of $S_{r,s;\, r',s'}$, that
\be 
\sum_{r,s} d^2_{r,s} =  \frac{1}{S^2_{1,1;\, 1,1}}\,. \label{dims}
\ee
Summarizing, the global Jones index $\mu$ of a model defined with a partition function of the form (\ref{Zmin}) is
\be 
\mu = \left( \frac{\sum_{r,s} d^2_{r,s} }{\sum_{r',s'}\sum_{r'',s''} d_{r',s'} M_{r',s';\, r'',s''} d_{r'',s''}}\right)^2\,.
\label{esteindice}
\ee
As described in \cite{Benedetti:2024dku}, this solves the problem of finding  Haag duality violations directly from the bootstrap data. Notice that for modular invariant theories we obtain $\mu=1$, as it should from the direct relation between modular invariance and completeness expressed in (\ref{mucft}).

Given a model ${\cal A}$ and a submodel ${\cal B}\subset {\cal A}$ we can also compute the Jones index of the inclusion of algebras of single intervals $R$, $\lambda_{AB}=[{\cal A}(R):{\cal B}(R)]$. This is also  independent of the particular interval $R$. In Ref. \cite{kawahigashi2001multi} it was proven that
\be
\mu_B=\mu_A \, \lambda_{AB}^2\,.
\ee
See Appendix \ref{HaagDuality} for an intrinsic account of $\lambda_{AB}$, in particular how it is computed  independently of the global indices $\mu$. Then the previous relation becomes a crosscheck for our computations.

\subsection{Rules for constructing local submodels}\label{sec23}

We now describe the rules for constructing local submodels of the complete ones described previously. This is a minimal set of requirements, in particular, we do not impose $S$ invariance. We can express these rules in terms of the coupling matrix $M$ appearing in the partition function of the model. These rules are given by
\begin{itemize}
    \item \textbf{Inclusion of the stress tensor net:} A first fundamental requirement for $M$ is $M_{1,1;\, 1,1}=1$. This ensures the identity net generated by the stress tensor is in the model with multiplicity one. 

    \item \textbf{Locality:} We should require invariance under T transformations. This is because, at the local level, T invariance is the requirement of causality of the fields.\footnote{ Given a primary field $\phi$, with scaling dimensions $(h,\bar{h})$, the Euclidean correlation function is $ z^{-2 h}\, \bar{z}^{-2 \bar{h}}$. This must be real analytic everywhere except at the coincidence point. It must also be permutation invariant. This implies $
h=\bar{h}+k\,,\quad k \in \mathbb{Z}$, that coincides with the requirement of $T$-invariance for the parity symmetric case $c=\bar{c}$ (for simplicity we are restricting to this case).}  This can be upgraded to include fermions as usual. More precisely, each of the representations $(r,s)$ only describes the chiral part of a given field, and it is generically not self-local. To construct a local field we need to combine chiral and anti-chiral parts properly. Considering that the difference in conformal dimension represents the helicity of the field, Lorentz symmetry and locality requires that $2(h_{(r,s)}-\overline{h}_{(r',s')})\in \mathbb{Z}$. Then, the diagonal elements of $M$ are spin $0$ fields, while non-diagonal elements represent fields of spin greater or equal to $1$ (half integers for fermionic models).

    \item \textbf{Closure of the operator algebra:} We should include sets of fields that define a closed algebra under the OPEs.  If we have a model with certain field $\phi$, we necessarily have all the fields that appear in the OPE with itself, and so forth with the generated local fields. In this step we are using the OPE of fields that belong to a complete model and that have been computed in the literature. All possible models are submodels of a complete one \cite{Benedetti:2024dku}. 
    
\end{itemize}

When considering incomplete models a subtle distinction has to be made. On one hand, we have the superselection sector structure of the model. This can be mathematically represented by certain endomorphisms (DHR endomorphisms) that can be composed and decomposed into irreducible ones. These define a category. In particular, fusion rules can be defined for DHR endomorphisms. This structure is completely intrinsic to the model and it contains the information of all possible completions. On the other hand, given a particular completion of the model, we have certain classes of charged operators that belong to the complete model and are closed under the action of the submodel. These ``charged'' operators of the complete model with respect to the incomplete one close an OPE algebra, but do not, strictly speaking, form a fusion algebra. For more details see appendix \ref{HaagDuality}.

In particular, for the Virasoro algebra, the chiral fields in the Kac table can be associated with DHR sectors. These sectors satisfy standard fusion rules
\be 
(r,s)\p (r',s') = \sum_{(r'',s'')} \mathcal{N}^{(r'',s'')}_{(r,s)\,; (r',s')}(r'',s'')\,,\,\,\,
\ee
for $\mathcal{N}^{(r'',s'')}_{(r,s)\,; (r',s')}$ given by Verlinde's formula 
\be 
\mathcal{N}^{(r'',s'')}_{(r,s)\,; (r',s')}= \sum_{(\rho,\sigma)}\frac{S_{r,s\,; \rho,\sigma}S_{r',s'\,; \rho,\sigma}S_{r'',s''\,; \rho,\sigma}}{S_{1,1\,; \rho,\sigma}}\;.
\ee
The S-Matrix is the chiral one defined in (\ref{smat}). This implies
\be 
(r,s)\p (r',s') = \sum^{r_{\text{max}}}_{\substack{r''=1-|r-r'|\\\text{Mod 2}}}\,\,\,\,\sum^{s_{\text{max}}}_{\substack{s''=1-|s-s'|\\ \text{Mod 2}}} (r'',s'')\,,
\label{fusion} \ee 
where $\text{Mod 2}$ denotes that both sums run by increments of two, and the values of $r_{\text{max}}$ and $s_{\text{max}}$ are given by 
\be 
r_{\text{max}}={\text{min}}\big[r+r'-1,\,2m-1-r-r'\big]\,,\,\,\,s_{\text{max}}={\text{min}}\big[s+s'-1,\,2m+1-s-s'\big]\,.
\ee

But these do not give directly the OPE algebra rules we need for computing models generated by primary fields with two chiral components. These OPE algebras have been studied in the literature. For diagonal completions it is the case one can use the chiral fusion rules, now thought as controlling the OPE of the local diagonal fields, i.e. whether an operator appears or not in the OPE. For non-diagonal fields, one needs to be more careful. As explained by S. Ribault \cite{Ribault:2016sla} the OPE associated with non-diagonal fields of spin $>0$ in many cases can also be determined from (\ref{fusion}) assuming the conservation of diagonality. We will be more specific about this point in the following sections.

\subsection{The \texorpdfstring{$m=3,4,5$}{Lg} examples \label{m345}}  

Before the general classification, we start with the simplest examples. This will clarify the main ideas. Also, the models discussed in this section are the ones with more relevant applications. The case $m=2$ is the trivial CFT. The first non-trivial case is $m=3$.

\subsubsection{The case \texorpdfstring{$m=3$}{Lg} and the Ising Model \label{m3}}

 The unitary minimal model with $m=3$ has central charge $c=1/2$. This model has three allowed chiral representations for the Virasoro algebra. These are usually denoted as $1$, $\e$, and $\s$. Their   Kac labels $(r,s)$, conformal dimension $h$, and quantum dimensions $d$ are depicted in Fig. \ref{kac3}:
\begin{figure}[H]
\begin{minipage}[l]{0.45\textwidth}
    \centering
    \vspace{1cm}
    \includegraphics[width=1 \textwidth]{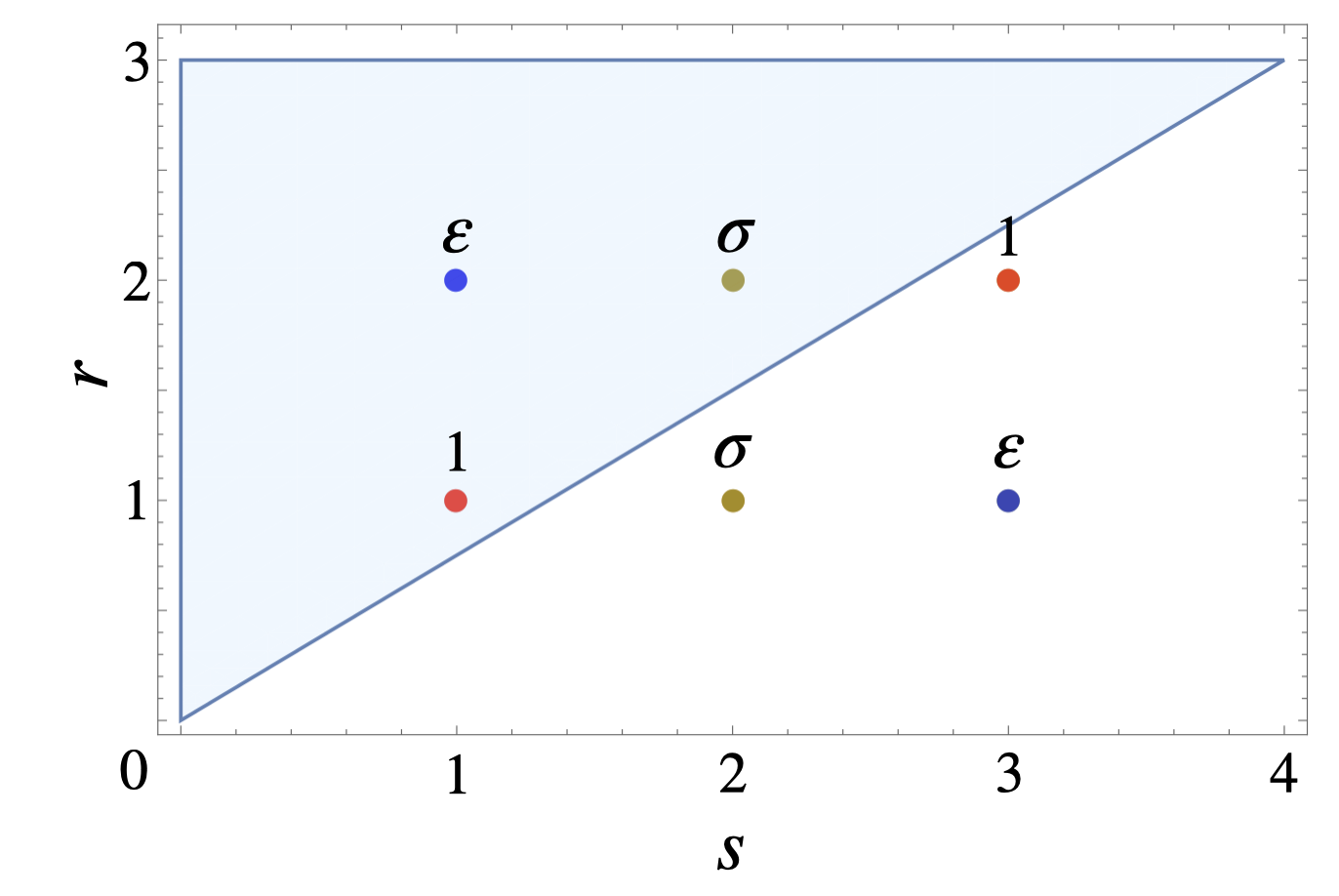}
\end{minipage}
\begin{minipage}[r]{0.48\textwidth}
\begin{table}[H]
    \centering
\begin{tabular}{cccc}
\cline{1-4}
\multicolumn{1}{|c|}{}              & \multicolumn{1}{c|}{$(r,s)$}         & \multicolumn{1}{c|}{$h$}   & \multicolumn{1}{c|}{$d$}               \\ \cline{1-4}
\multicolumn{1}{|c|}{$1$}           & \multicolumn{1}{c|}{$(1,1)\,\text{or}\,(2,3)$} & \multicolumn{1}{c|}{$0$}     & \multicolumn{1}{c|}{$1$}               \\ \cline{1-4}
\multicolumn{1}{|c|}{$\varepsilon$} & \multicolumn{1}{c|}{$(1,3)\,\text{or}\,(2,1)$} & \multicolumn{1}{c|}{$1/2$}  & \multicolumn{1}{c|}{$1$}      \\ \cline{1-4}
\multicolumn{1}{|c|}{$\sigma$}      & \multicolumn{1}{c|}{$(1,2)\,\text{or}\,(2,2)$} & \multicolumn{1}{c|}{$1/16$}  & \multicolumn{1}{c|}{$\sqrt{2}$}  \\ \cline{1-4}
\end{tabular}
\end{table}
\end{minipage}
\caption{Allowed representations of the $m=3$  minimal model, classified by their Kac label $(r,s)$ in the Kac table (left), and table with the corresponding conformal dimensions $h$ and quantum dimensions $d$ (right).}\label{kac3}
\end{figure}

\begin{figure}[H]
    \centering
    \includegraphics[width=0.45 \textwidth]{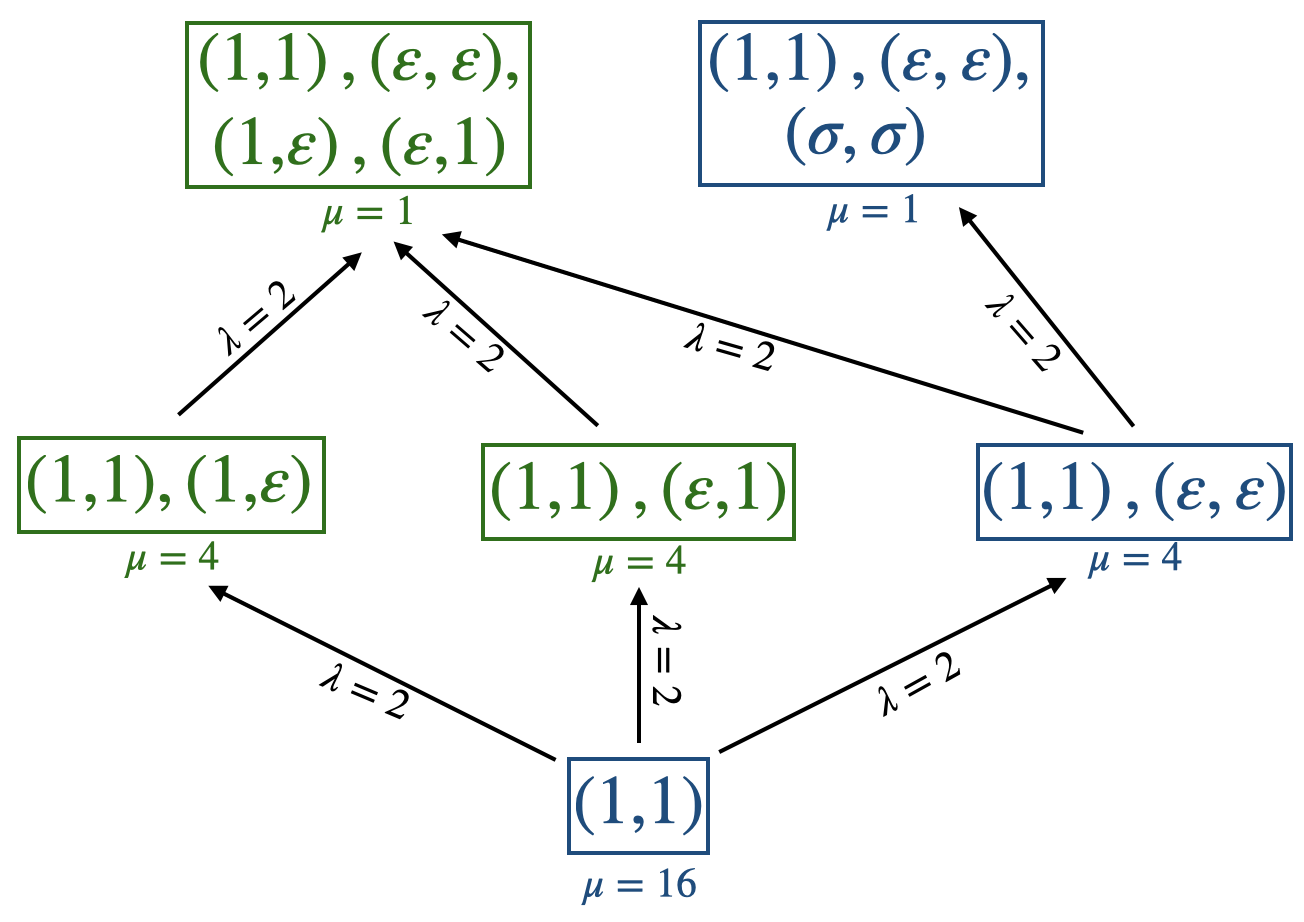}
    \caption{Classification of $d=2$ CFTs for $m=3$. Each node has a global Jones index $\mu$ and we also write the relative Jones index between all immediate inclusions $\lambda$. The blue boxes represent algebras composed purely of spin zero primary fields (excepting the stress tensor) and the green ones include spin $1/2$ fermion fields.}
    \label{43}
\end{figure}

The chiral sectors $1$, $\e$, and $\s$ obey the fusion rules that follow from (\ref{fusion}) as\footnote{A useful crosscheck is the conservation of quantum dimensions in the product. Namely, for fields labeled as $\varphi_i$ obeying $\varphi_i \times \varphi_j= \sum_k \mathcal{N}^k_{ij} \varphi_k$ we should have $d_i d_j= \sum_k \mathcal{N}^k_{ij} d_k$. For $m=3$ this is trivial but we have found it useful for general $m$.}
\be 
\e \p \e = 1\, ,\qquad \e \p \s =\s \, ,\qquad \s \p \s =1+\e\,. \label{fusion3}
\ee
Local (bosonic or fermionic) fields are constructed from these chiral sectors by combining in appropriate ways chiral and anti-chiral parts. In particular, the difference in the conformal dimension obeys that $2h\in \mathbb{Z}$. The possible spectrum of fields is then
\begin{itemize}
    \item Spin $0$: $(1,1)\,,\,(\e,\e) \,,\,\text{or}\,\,(\s,\s)$,
    \item Spin $1/2$: $(1,\e)\,\text{or}\,\,(\e,1)$.
\end{itemize}
To form a local $d=2$ model we must choose a set of local fields that close an algebra. We then need to consider the extension of the chiral fusion rules (\ref{fusion3}) to ones pertaining to the local fields. These follow from the OPE. At the practical level, these are uniquely defined by (\ref{fusion}) and the conservation of diagonality of the products. We can write them as
\begin{align} 
&(\e,\e) \times (\e,\e) = (1,1)\,,\,\,\, &(\e,\e) \times (1,\e) = (\e,1)\,, \nonumber\\ &(\e,\e) \times (\e,1) = (1,\e)\,,\,\,\, &(\e,\e) \times (\s,\s) = (\s,\s)\,, \nonumber\\
&(1,\e) \times (1,\e) = (1,1)\,,\,\,\,  &(1,\e) \times (\e,1) = (\e,\e)\,, \label{IsingFus} \nonumber
\end{align}
In Fig. \ref{43} we show all local models constructed with the previous fields, together with the corresponding global indices $\mu$, as computed using the formula (\ref{esteindice}), and the order of inclusions and corresponding relative Jones indices. The figure has the form of a tree in which each node  is a $d=2$ CFT, which is either local or Fermi-local, i.e. including fermion fields.
At the bottom of the tree, we only have the field $(1,1)$. This is the theory of the  stress tensor alone. This node has the maximal violation of duality, as seen by its global index. At the top of the tree we find the complete modular invariant models. In this case there are two possibilities, $\{ (1,1),(\e,\e),(\s,\s)\}$, famously known as the Ising model, and  $\{ (1,1),(\e,\e),(1,\e),(\e,1)\}$ which is a complete model including fermion fields, the free real fermion field with two chiral components. All nodes verify (graded) T-duality.

We remark that, as a crosscheck, all the Jones indices we obtain are in the allowed range \cite{Jones1983} (see appendix \ref{HaagDuality}).  Indeed, if a candidate model exhibits an index out of the allowed range, this is a sign of a problem. For example

\begin{itemize}
    \item The candidate submodel given by $\{(1,1),(\s,\s)\}$ has $\mu=16/9$, which is not inside Jones classification. This model fails because it is not closed under OPE.
    
    \item Another possibility is $\{(1,1),(1,\s)\}$. However, $(1,\s)$ is not a self-local field and we get a non-allowed value $\mu=4(1-\sqrt{2})^2$.
    
    \item We could try to overcomplete a modular invariant model as $\{(1,1),(\e,\e),(\s,\s),(1,\e),(\e,1)\}$. This model is closed under OPE and composed of local fields. However, it exhibits a value lower than $1$, in this case, $\mu=4/9$. This is because  the fields are not mutually local,  i.e. they do not commute with each other at spatial distance.
\end{itemize}

\subsubsection{The case \texorpdfstring{$m=4$}{Lg} and the Tricritical Ising Model \label{m4}}  

The next case is $m=4$. It has central charge $c=7/10$, and it can be treated similarly to the previous one. The allowed chiral representations are usually denoted as $1$, $\e$, $\e'$ ,$\e''$, $\s$ and $\s'$. In this case, the corresponding Kac labels $(r,s)$, conformal dimension $h$, and quantum dimensions $d$ are depicted in Fig \ref{kac4}:
\begin{figure}[H]
\begin{minipage}[l]{0.45\textwidth}
    \centering
    \vspace{1cm}
    \includegraphics[width=1 \textwidth]{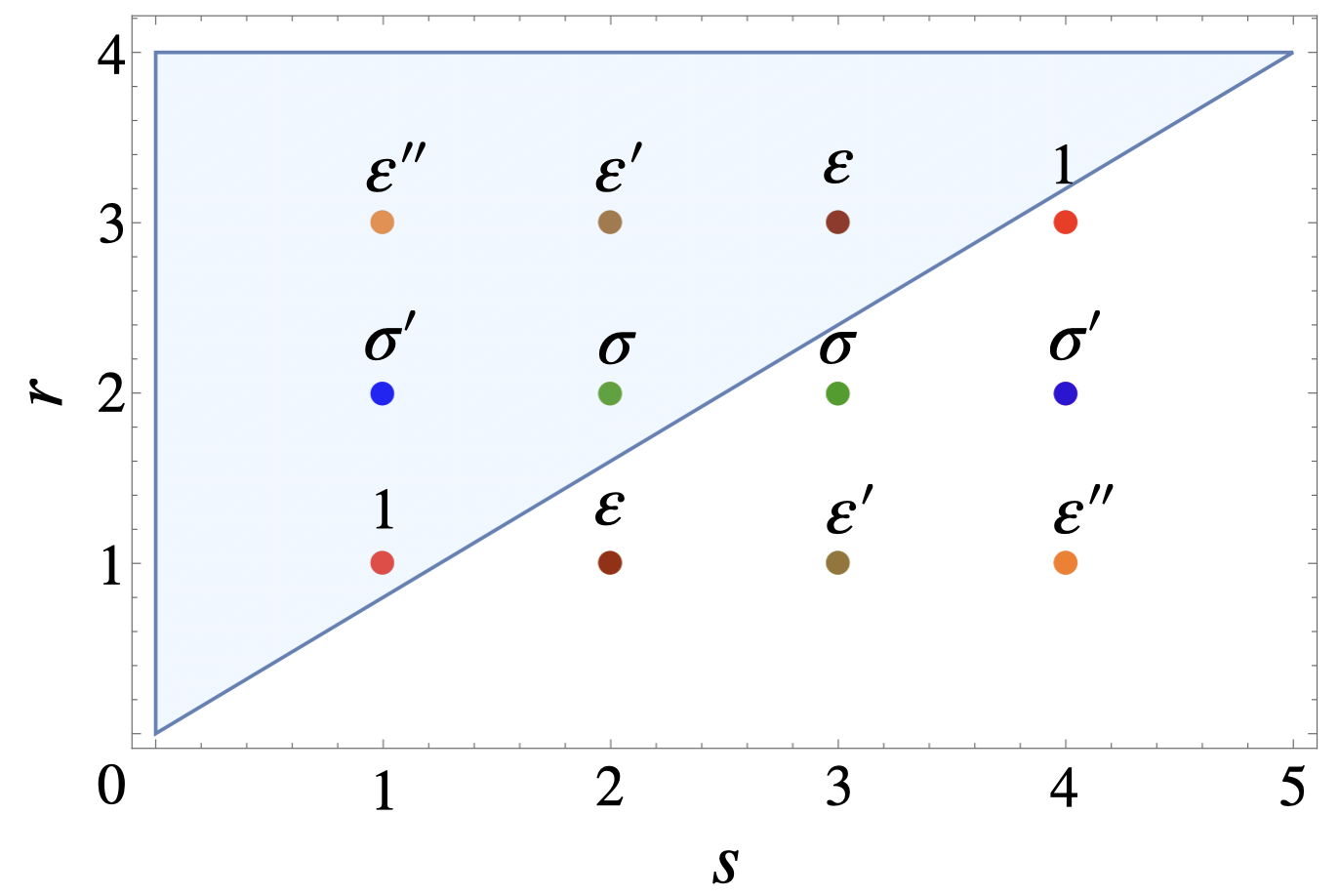}
\end{minipage}
\begin{minipage}[r]{0.48\textwidth}
\begin{table}[H]
    \centering
\begin{tabular}{cccc}
\cline{1-4}
\multicolumn{1}{|c|}{}              & \multicolumn{1}{c|}{$(r,s)$}         & \multicolumn{1}{c|}{$h$}   & \multicolumn{1}{c|}{$d$}               \\ \cline{1-4}
\multicolumn{1}{|c|}{$1$}           & \multicolumn{1}{c|}{$(1,1)\,\text{or}\,(3,4)$} & \multicolumn{1}{c|}{$0$}     & \multicolumn{1}{c|}{$1$}               \\ \cline{1-4}
\multicolumn{1}{|c|}{$\varepsilon$} & \multicolumn{1}{c|}{$(1,2)\,\text{or}\,(3,3)$} & \multicolumn{1}{c|}{$1/10$}  & \multicolumn{1}{c|}{($1+\sqrt{5})/2$}      \\ \cline{1-4}
\multicolumn{1}{|c|}{$\varepsilon'$}      & \multicolumn{1}{c|}{$(1,3)\,\text{or}\,(3,2)$} & \multicolumn{1}{c|}{$3/5$}  & \multicolumn{1}{c|}{$(1+\sqrt{5})/2$}    \\ \cline{1-4}
\multicolumn{1}{|c|}{$\varepsilon''$}      & \multicolumn{1}{c|}{$(1,4)\,\text{or}\,(3,1)$} & \multicolumn{1}{c|}{$3/2$}  & \multicolumn{1}{c|}{$1$}    \\ \cline{1-4}
\multicolumn{1}{|c|}{$\sigma$}      & \multicolumn{1}{c|}{$(2,2)\,\text{or}\,(2,3)$} & \multicolumn{1}{c|}{$3/80$}  & \multicolumn{1}{c|}{$\sqrt{3+\sqrt{5}}$}    \\ \cline{1-4}
\multicolumn{1}{|c|}{$\sigma'$}      & \multicolumn{1}{c|}{$(2,1)\,\text{or}\,(2,4)$} & \multicolumn{1}{c|}{$7/16$} & \multicolumn{1}{c|}{$\sqrt{2}$}    \\ \cline{1-4}
\end{tabular}
\end{table}
\end{minipage}    
\caption{Allowed representations of the $m=4$  minimal model, classified by their Kac label $(r,s)$ in the Kac table (left), and table with the correspondig conformal dimensions $h$ and quantum dimensions $d$ (right).}\label{kac4}
\end{figure}

In this case, the chiral fusion rules (\ref{fusion}) take the following form
\begin{align}
& \e \p \e=1+\e'\,,&\quad & \e'\p \e' = 1+\e'\,,& \quad & \e'' \p \s = \s\,,\nonumber \\ \nonumber
& \e \p \e'=\e+\e'' \,,&\quad & \e'\p \e''=\e \,,& \quad & \e''\p \s'= \s'\,,\\ \nonumber
& \e \p \e''= \e'\,,&\quad & \e' \p \s=\s+\s' \,,& \quad & \s\p \s= 1+\e+\e'+\e''\,,\\ \nonumber
& \e \p \s= \s+\s'\,,&\quad & \e'\p \s'=\s \,,&\quad & \s \p \s'=\e+\e'\,, \\ \nonumber
& \e \p \s'=\s \,,&\quad &\e''\p \e''= 1 \,,& \quad & \s'\p \s'=1+\e''\,.
\end{align}
The set of self-local fields arising by combinations of chiral and anti-chiral representations are
\begin{itemize}
    \item Spin $0$: $(1,1)\,,\,(\e,\e)\,,\,(\e',\e')\,,\,(\e'',\e'')\,,\,(\s,\s) \,,\,\text{or}\,\,(\s',\s')$,
    \item Spin $1/2$: $(\e,\e')\,\text{or}\,\,(\e',\e)$,
    \item Spin $3/2$: $(1,\e'')\,\text{or}\,\,(\e'',1)$.
\end{itemize}
As before, the fusion rules again extend to the OPE of these fields unequivocally, when we take into account the notion of diagonality mentioned above. 

\begin{figure}[H]
    \centering
    \includegraphics[width=0.85\textwidth]{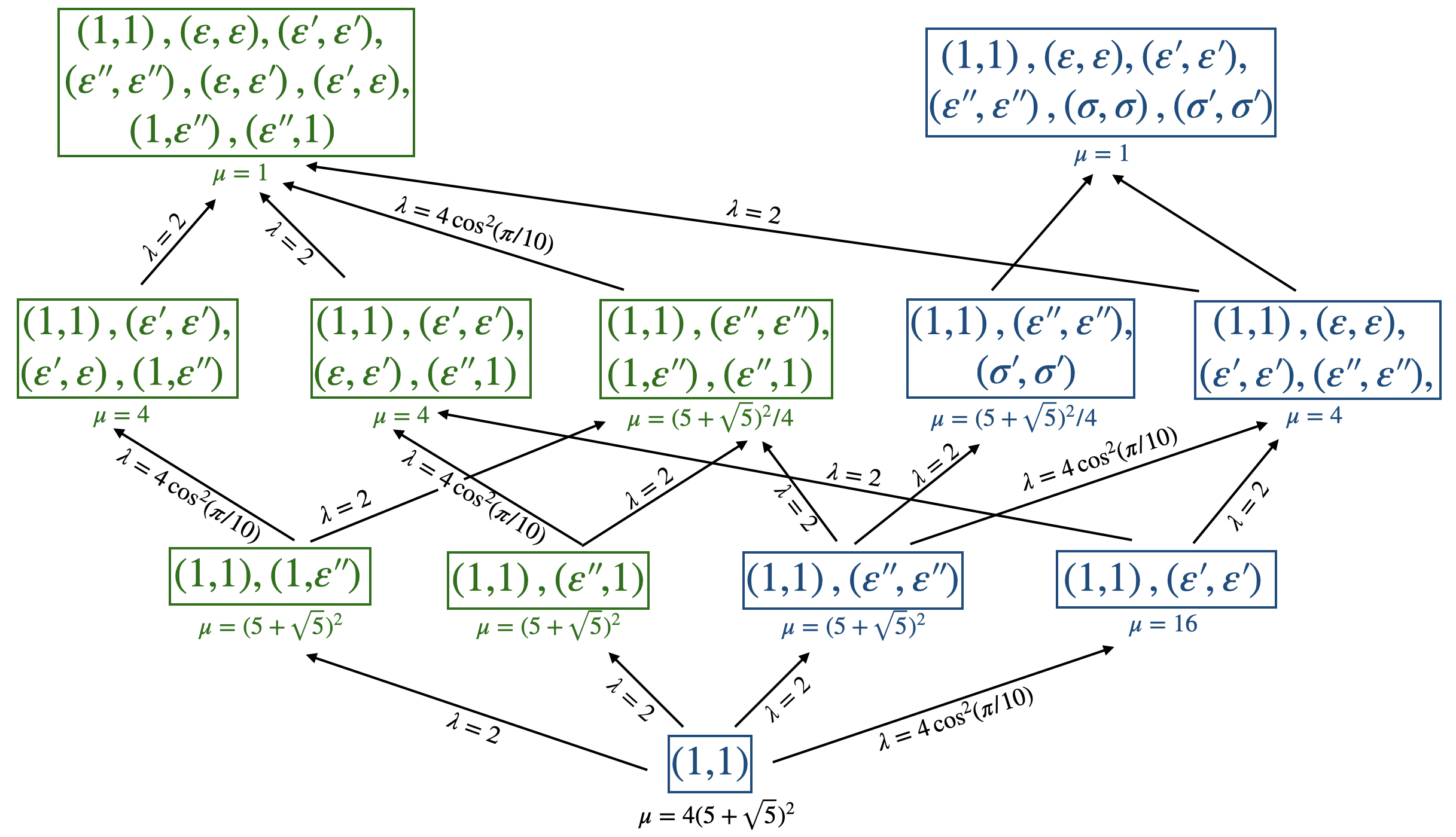}
    \label{54}
      \caption{Classification of $d=2$ CFTs for $m=4$. Each node has a global Jones index $\mu$ and we also write the relative Jones index between all immediate inclusions $\lambda$. The blue boxes represent algebras composed purely of spin zero primary fields and the green ones include spin $1/2$ and/or $3/2$ fermion fields.}\label{m4fig}
\end{figure}

Using this information we can classify the zoo of $m=4$ minimal models. This is depicted in Figure \ref{m4fig}. Again this takes the form of a tree. Each node corresponds to a $d=2$ CFT and has an intrinsic global index. As in the case of $m=3$, we also have here two complete modular invariant models. We have the diagonal one formed by spin zero fields. This is known as the tricritical Ising model. Also, we have a fermionic completion.

\subsubsection{The case \texorpdfstring{$m=5$}{Lg} and the Three-State Potts Model \label{m5}}

The last particular case we study in this section is $m=5$. This has central charge $c=4/5$. This case is interesting because it is the first value of the central charge that allows for a bosonic non-diagonal modular invariant model, the Three-State Potts Model. As before, we first list the allowed chiral representations and their conformal dimensions $h$, and quantum dimensions $d$ in Figure \ref{kac5}:
\begin{figure}[H]
\begin{minipage}[l]{0.45\textwidth}
    \centering
    \vspace{1cm}
    \includegraphics[width=1 \textwidth]{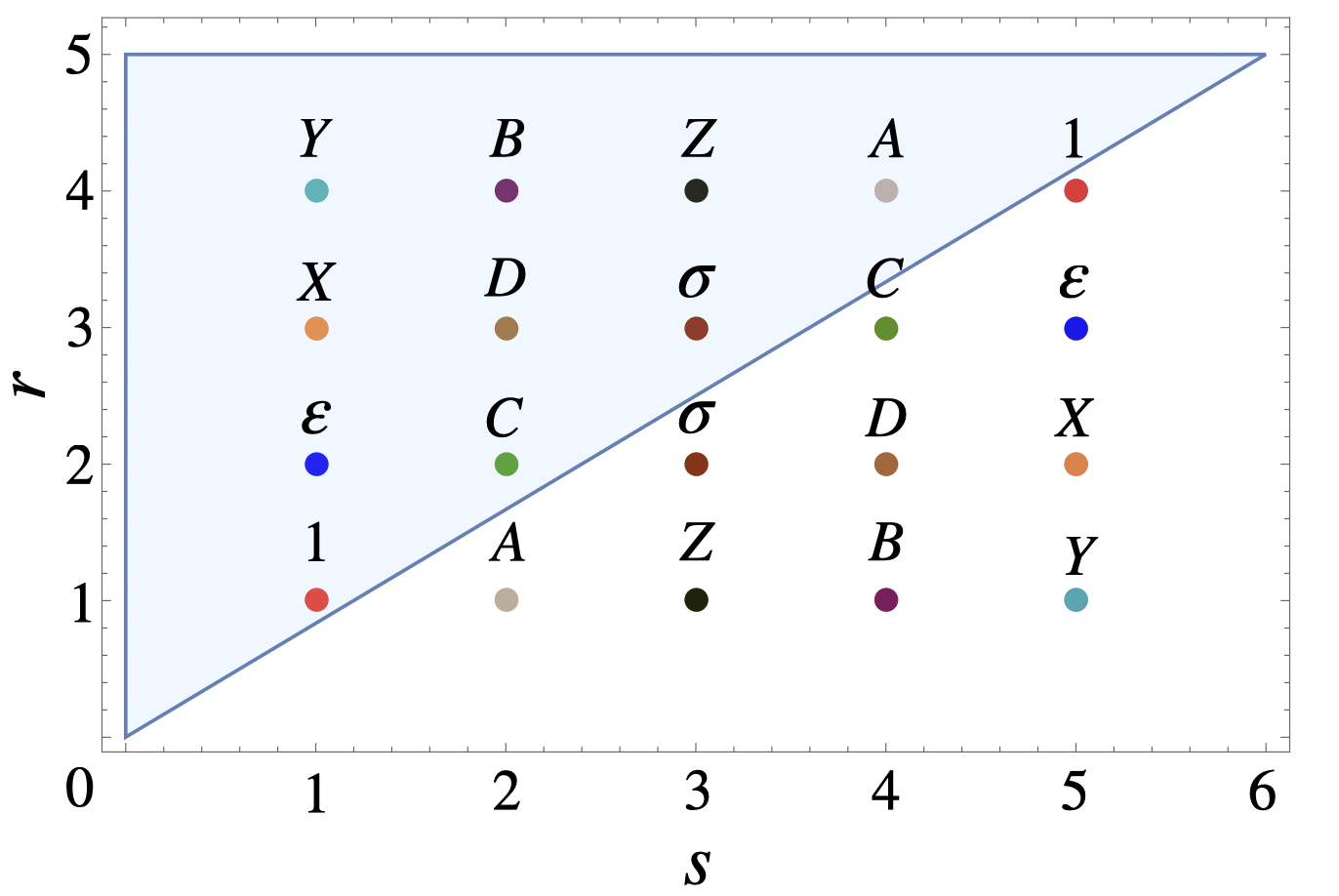}
\end{minipage}
\begin{minipage}[r]{0.48\textwidth}
\begin{table}[H]
    \centering
\begin{tabular}{cccc}
\cline{1-4}
\multicolumn{1}{|c|}{}              & \multicolumn{1}{c|}{$(r,s)$}         & \multicolumn{1}{c|}{$h$}   & \multicolumn{1}{c|}{$d$}               \\ \cline{1-4}
\multicolumn{1}{|c|}{$1$}           & \multicolumn{1}{c|}{$(1,1)\,\text{or}\,(4,5)$} & \multicolumn{1}{c|}{$0$}     & \multicolumn{1}{c|}{$1$}               \\ \cline{1-4}
\multicolumn{1}{|c|}{$\varepsilon$} & \multicolumn{1}{c|}{$(2,1)\,\text{or}\,(3,5)$} & \multicolumn{1}{c|}{$2/5$}  & \multicolumn{1}{c|}{($1+\sqrt{5})/2$}      \\ \cline{1-4}
\multicolumn{1}{|c|}{$\sigma$}      & \multicolumn{1}{c|}{$(2,3)\,\text{or}\,(3,3)$} & \multicolumn{1}{c|}{$1/5$}  & \multicolumn{1}{c|}{$1+\sqrt{5}$}    \\ \cline{1-4}
\multicolumn{1}{|c|}{$X$}      & \multicolumn{1}{c|}{$(3,1)\,\text{or}\,(2,5)$} & \multicolumn{1}{c|}{$7/5$}  & \multicolumn{1}{c|}{$(1+\sqrt{5})/2$}    \\ \cline{1-4}
\multicolumn{1}{|c|}{$Y$}      & \multicolumn{1}{c|}{$(1,5)\,\text{or}\,(4,1)$} & \multicolumn{1}{c|}{$3$}  & \multicolumn{1}{c|}{$1$}    \\ \cline{1-4}
\multicolumn{1}{|c|}{$Z$}      & \multicolumn{1}{c|}{$(1,3)\,\text{or}\,(4,3)$} & \multicolumn{1}{c|}{$2/3$} & \multicolumn{1}{c|}{$2$}     \\ \cline{1-4}
\multicolumn{1}{|c|}{$A$}      & \multicolumn{1}{c|}{$(1,2)\,\text{or}\,(4,4)$} & \multicolumn{1}{c|}{$1/8$} & \multicolumn{1}{c|}{$\sqrt{3}$}     \\ \cline{1-4}
\multicolumn{1}{|c|}{$B$}      & \multicolumn{1}{c|}{$(1,4)\,\text{or}\,(4,2)$} & \multicolumn{1}{c|}{$13/8$}& \multicolumn{1}{c|}{$\sqrt{3}$}  \\ \cline{1-4}
\multicolumn{1}{|c|}{$C$}      & \multicolumn{1}{c|}{$(2,2)\,\text{or}\,(3,4)$} & \multicolumn{1}{c|}{$1/40$}  & \multicolumn{1}{c|}{$\sqrt{(3/2)(3+\sqrt{5})}$}    \\ \cline{1-4}
\multicolumn{1}{|c|}{$D$}      & \multicolumn{1}{c|}{$(3,2)\,\text{or}\,(2,4)$} & \multicolumn{1}{c|}{$21/40$}   & \multicolumn{1}{c|}{$\sqrt{(3/2)(3+\sqrt{5})}$}   \\ \cline{1-4}
\end{tabular}
\end{table}
\end{minipage}
\caption{Allowed representations of the $m=5$  minimal model, classified by their Kac label $(r,s)$ in the Kac table (left), and table with the corresponding conformal dimensions $h$ and quantum dimensions $d$ (right).}\label{kac5}
\end{figure}

\begin{figure}[H]
    \centering
    \includegraphics[width=0.9\textwidth]{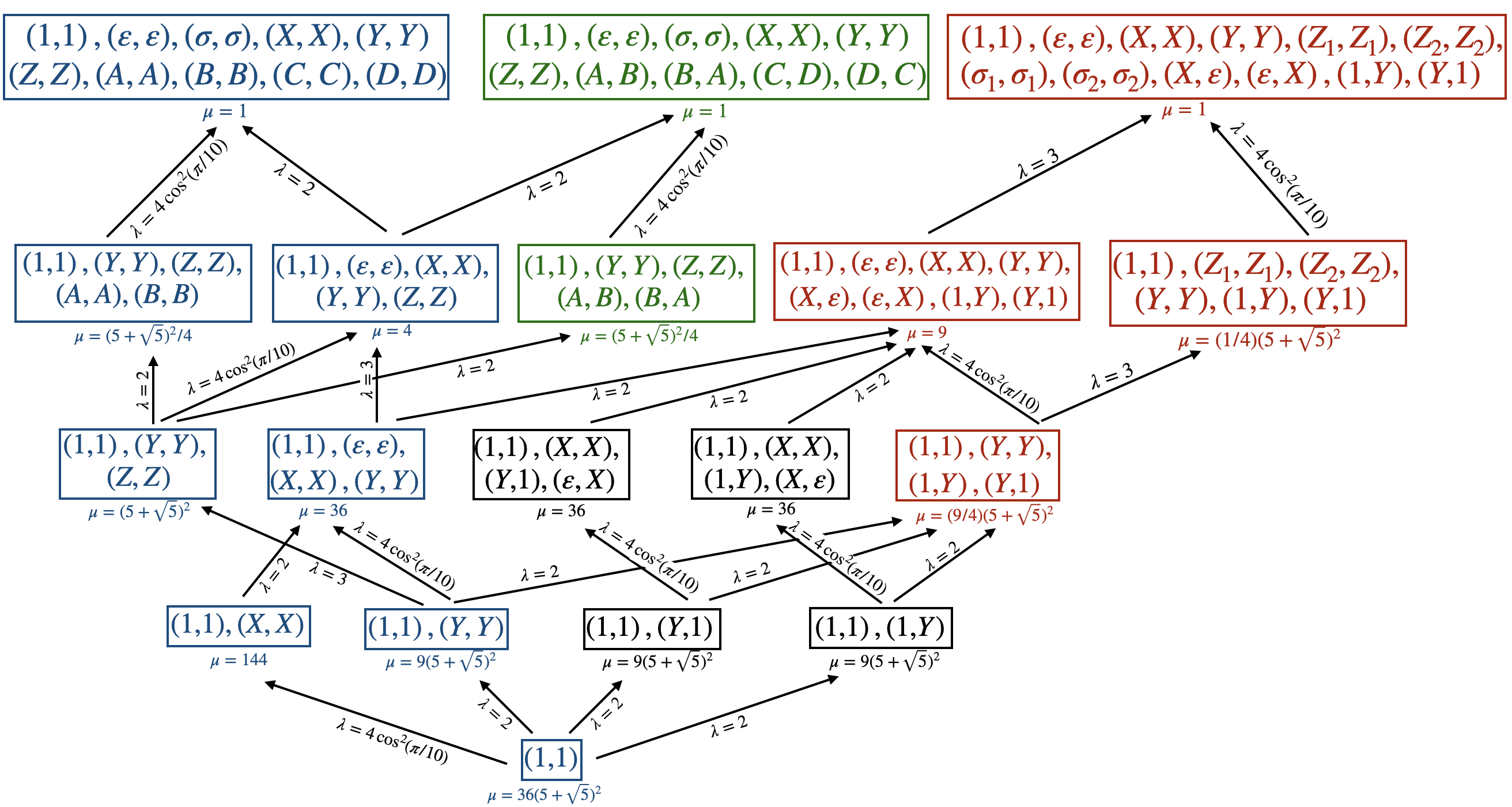}
    \label{65}
    \caption{Classification of $d=2$ CFTs for $m=5$. Each node has a global Jones index $\mu$ and we also write the relative Jones index between all immediate inclusions $\lambda$. The blue boxes represent algebras composed purely of spin zero fields. The green ones include fermionic fields while the red and black ones include bosonic fields of spin $\geq 1$ with or without parity symmetry respectively. }\label{m5fig}
\end{figure}

\noindent The chiral fusion rules obtained from (\ref{fusion}) can be explicitly written as
\begin{align}
& \e \p \e= 1+ X\,,&\quad & A\p A =1+Z \,,& \quad & B\p Y =A\,, \nonumber \\ \nonumber
& \e \p \s= \s+ Z\,,&\quad & A\p B =Y+Z \,,& \quad & B\p Z= A+B\,,\\ \nonumber
& \e \p X= \e+ Y\,,&\quad & A\p C=\e+\s \,,& \quad & C\p C=1+\s+X+Z\,,\\ \nonumber
& \e \p Y= X\,,&\quad & A\p D=\s+ X \,,&\quad & C\p D=\e+\s+Y+Z\,,\\ \nonumber
& \e \p Z= \s\,,&\quad &A\p \e= C \,,& \quad & C\p \e=A+D\,,\\ \nonumber
& \s \p \s= 1+\e+\s+X+Y+Z\,,&\quad & A\p \s =C+D\,,& \quad & C\p \s= A+B+C+D \,,\\\nonumber 
& \s \p X= \s+Z\,,&\quad & A\p X=D\,,& \quad & C\p X = B+C\,,\\ \nonumber
& \s \p Y= \s\,,&\quad & A\p Y=B\,,& \quad & C\p Y= D\,,\\ \nonumber
& \s \p Z= \e+\s+X\,,&\quad & A\p Z= A+B\,,& \quad & C\p Z= C+D\,,\\ \nonumber
& X \p X= 1+X\,,&\quad & B\p B= 1+ Z\,,& \quad & D\p D =1 +\s +X + Z\,,\\ \nonumber
& X \p Y= \e\,,&\quad & B\p C=\s + X \,,&\quad & D\p \e= B+C\,,\\ \nonumber
& X \p Z= \s\,,&\quad & B\p D=\e+\s\,,& \quad &  D\p \s = A+B+C+D\,, \\ \nonumber
& Y \p Y= 1\,,&\quad &  B\p \e=D \,,&\quad &  D\p  X= A+ D\,,\\ \nonumber
& Y \p Z= Z\,,&\quad &  B \p \s=C+D\,,& \quad & D\p  Y =C\,, \\ \nonumber
& Z \p Z= 1+ Y+Z\,, & \quad &  B\p X =C \,,&\quad & D\p Z= C+ D \,.
\end{align}
The possible local fields (with semi-integer spin) are 
\begin{itemize}
    \item Spin $0$: $(1,1)\,,\,(\e,\e)\,,\,(\s,\s)\,,\,(X,X)\,,\,(Y,Y)\,,\,(Z,Z)\,,\,(A,A)\,,\,(B,B)\,,\,(C,C) \,,\,\text{or}\,\,(D,D)$,
    \item Spin $1/2$: $(C,D)\,\text{or}\,\,(D,C)$,
    \item Spin $1$: $(\e,X)\,\text{or}\,\,(X,\e)$,
    \item Spin $3/2$: $(A,B)\,\text{or}\,\,(B,A)$,
    \item Spin $3$: $(1,Y)\,\text{or}\,\,(Y,1)$.
\end{itemize}
In this case, we again have two complete models, the one corresponding to the bosonic $(A,A)$ series and a fermionic one. However, if we take the complete model of the $(A,A)$ series, make a $\mathbb{Z}_2$ orbifold, and add the twisted sectors, we recover a modular invariant bosonic non-diagonal model.\footnote{The same orbifolding process can be performed for $m=3,4$. However, in those cases, the resulting model is the same as the one we started with. This is because the first models of the $(A,D)$ and $(D,A)$ series coincide with models of the $(A,A)$ series.} This is a complete model of the $(A,D)$ series which is commonly known as the Three-State Potts Model. The partition function recovered from (\ref{morb}) is 
\be 
Z_{\tau}=\sum_{r=1,2}|\chi_{r,1}+\chi_{r,5}|^2+2|\chi_{r,3}|^2 = \sum_{r=1,2}|C_{r,1}|^2+2|C_{r,3}|^2 \,,\label{Zpotts}
\ee
where we have introduced the notation for the conformal blocks as
\be 
C_{r,1}=\chi_{r,1}+\chi_{r,5}\,,\quad C_{r,3}=\chi_{r,3}\,, \quad r=1,2\,.
\label{cb}
\ee
The modular transformations acting on these conformal blocks are closed on themselves. This can be checked from the S matrix (\ref{smat}) and their definition (\ref{cb}). Also, note that a factor of two appears before the $(r,3)$ representations. This implies there are two copies of the corresponding fields: $Z$ is separated in $Z_1$ and $Z_2$ and $\s$ in $\s_1$ and $\s_2$. This generates a $\mathbb{Z}_3$ symmetry under which these fields are charged as
\be 
Z_1\to e^{\frac{2\pi i }{3}} Z_1\,,\quad Z_2\to e^{\frac{-2\pi i }{3}} Z_2\,, \quad \s_1\to e^{\frac{2\pi i }{3}}\s_1\,,\quad \s_2\to e^{\frac{-2\pi i }{3}}\s_2\,.
\ee
The duplication of these fields also generates an ambiguity in the fusion rules involving $\s$ and $Z$. But, following \cite{francesco2012conformal}, we can use the $\mathbb{Z}_3$ symmetry to recover the chiral OPEs as
\begin{align}
& \e \p \s_1= \s_1+Z_1\,,&\quad & \s_2 \p X =\s_2+Z_2 \,,& \quad & X\p Z_1 =\s_1\,, \nonumber \\ \nonumber
& \e \p \s_2= \s_2+Z_2\,,&\quad & \s_1 \p Y =\s_1 \,,& \quad & X\p Z_2 =\s_2\,, \\ \nonumber
& \e \p Z_1= \s_1\,,&\quad & \s_2 \p Y =\s_2 \,,& \quad & Y\p Z_1 =Z_1\,, \\ \nonumber
& \e \p Z_2= \s_2\,,&\quad &\s_1 \p Z_1=\s_2  \,,&\quad & Y\p Z_2 =Z_2\,,\\ \nonumber
& \s_1 \p \s_2= 1+\e+X+Y\,,&\quad &\s_1 \p Z_2=\e+X\,,& \quad & Z_1 \p Z_1 = Z_2\,,\\ \nonumber
& \s_1 \p \s_1= \s_2+Z_2\,,&\quad & \s_2\p Z_1= \e+X \,,& \quad & Z_2 \p Z_2 = Z_1\,,\\\nonumber 
& \s_2 \p \s_2= \s_1+Z_1\,,&\quad & \s_2\p Z_2 =\s_1\,,& \quad & Z_1 \p Z_2 = 1+Y\,,\\ \nonumber
& \s_1 \p X= \s_1+Z_1\,.&\quad & \,\,\,\,\, & \quad & \,\,\,\,\,
\end{align}
Note we have not included the fusion rules of $A,B,C$ and $D$ in the latter list. This is because these fields do not appear in the complete model of the $(A,D)$ series (nor in any of the corresponding submodels). This is because they were charged under the original $\mathbb{Z}_2$ symmetry that we orbifolded to go from the $(A,A)$ to the $(A,D)$ series. Now using the chiral building blocks above we can construct the following local fields to analyze submodels of the $(A,D)$ series 
\begin{itemize}
    \item Spin $0$: $(1,1)\,,\,(\e,\e)\,,\,(\s_1,\s_1)\,,\,(\s_2,\s_2)\,,$ $\,(X,X)\,,\,(Y,Y)\,,$ $(Z_1,Z_1)\,,$ or $(Z_2,Z_2)$, 
    \item Spin $1$: $(\e,X)\,\text{or}\,\,(X,\e)$,
    \item Spin $3$: $(1,Y)\,\text{or}\,\,(Y,1)$.
\end{itemize}
With this information, we can now proceed as before. We start from the theory of the stress tensor alone and keep adding local fields sequentially. At each step, we impose closure of the OPE. We depict the set of $d=2$ CFTs with $m=5$, together with their indices, in Figure \ref{m5fig}.

\subsection{\texorpdfstring{$(A,A)$}{Lg} series for general  \texorpdfstring{$m\geq5$}{Lg}  \label{m6}}

It might seem that the process of classification gets increasingly complicated as we move to higher $m$ since the number of possible chiral representations grows accordingly. But it turns out the number of complete models and submodels do not grow with $m$ and we can achieve a general classification. As we have seen above, the modular invariants conform with the top nodes of the tree since they are complete. We fix this top node, and the bottom node, i.e. the stress tensor algebra, and then we seek for models in between using the corresponding OPE algebra. We are not treating fermionic models, that also exist for any $m$ \cite{hsieh2021fermionic}.

We start in this section by fixing the complete model to be the $(A,A)$ diagonal series. These series contain only spin zero fields for all $m$. Their spectrum can be written schematically as follows
\be
\mathcal{S}_{(A_{m-1},A_{m})}=\frac{1}{2}\bigoplus^{m-1}_{r=1}\bigoplus^{m}_{s=1} (r,s)\otimes (r,s)\,.\label{AA}
\ee
Note the factor $1/2$ is necessary because we are summing over all possible values of $r$ and $s$ and therefore each field is counted for twice. We will use the standard notation for the diagonal models defined by a given $m$ as $(A_{m-1},A_m)$. We write the diagonal spin-zero fields appearing in the spectrum as
\be
\varphi^{0}_{(r,s)}= (r,s)\otimes (r,s)\,. \label{spin0}
\ee
The upper-index valued zero means these are spinless fields. The OPE algebra for these spin zero fields coincides with (\ref{fusion}). Specifically, we can write the OPE \footnote{These coincide with the OPE for diagonal fields that we used in section \ref{m345} for $m=3,4,5$.} involving $\varphi^{0}_{(r,s)}$ as 
\be 
\varphi^{0}_{(r,s)}\p \varphi^{0}_{(r',s')} = \sum^{r_{\text{max}}}_{\substack{r''=1-|r-r'|\\\text{Mod 2}}}\,\,\,\,\sum^{s_{\text{max}}}_{\substack{s''=1-|s-s'|\\ \text{Mod 2}}} \varphi^{0}_{(r'',s'')}\,.
\label{fusionAA} 
\ee  

The task now is to find all subsets of fields of the form (\ref{spin0}) contained in the spectrum (\ref{AA}) that are closed under the fusion rules (\ref{fusionAA}). For each $m\geq5$ we can find eight different theories, with only one being complete. Each of these submodels can be easily checked to be algebraically closed. A computer search shows there are no additional ones.  The possible subcategories for these models have already been classified in \cite{Kawahigashi:2003gi}  and we find exactly one submodel for each possible subcategory.\footnote{Ref. \cite{Kawahigashi:2003gi} arrives at such classification by chracterizing $Q$-systems of DHR endomorphismss. In turn, these $Q$-systems instruct us of the possible extensions of the models \cite{Longo:1994xe}. The techniques used are very different from ours. In particular, they use so-called $\alpha$-induction techniques in the context of inclusion of algebras \cite{kawahigashi2001multi,Kawahigashi:2002px,Kawahigashi:2003gi}. We use standard OPE techniques.} Using the formulas of the previous section, we can compute the global index as a function of $m$ for each model using (\ref{esteindice}), and check that it coincides with the one of the adequate tensor category.  Below we use some nomenclature from tensor categories but it turns out to be very simple. All of them come from the standard affine $su(2)_k$ fusion category, together with its even part and associated Jones indices. For notational convenience, we now define the Jones index (of total quantum dimension) of the category $su(2)_{m-2}$ to be
\be\label{lieaf}
    \hat{\mu}(m)\equiv\frac{m^2}{4}\sin^{-4}\left(\frac{\pi}{m}\right)\,.
    \ee
We describe the $su(2)_k$ fusion category and the derivation of this formula in appendix \ref{su2}. 
 
\begin{figure}[H]
\centering
    \includegraphics[width=0.9\textwidth]{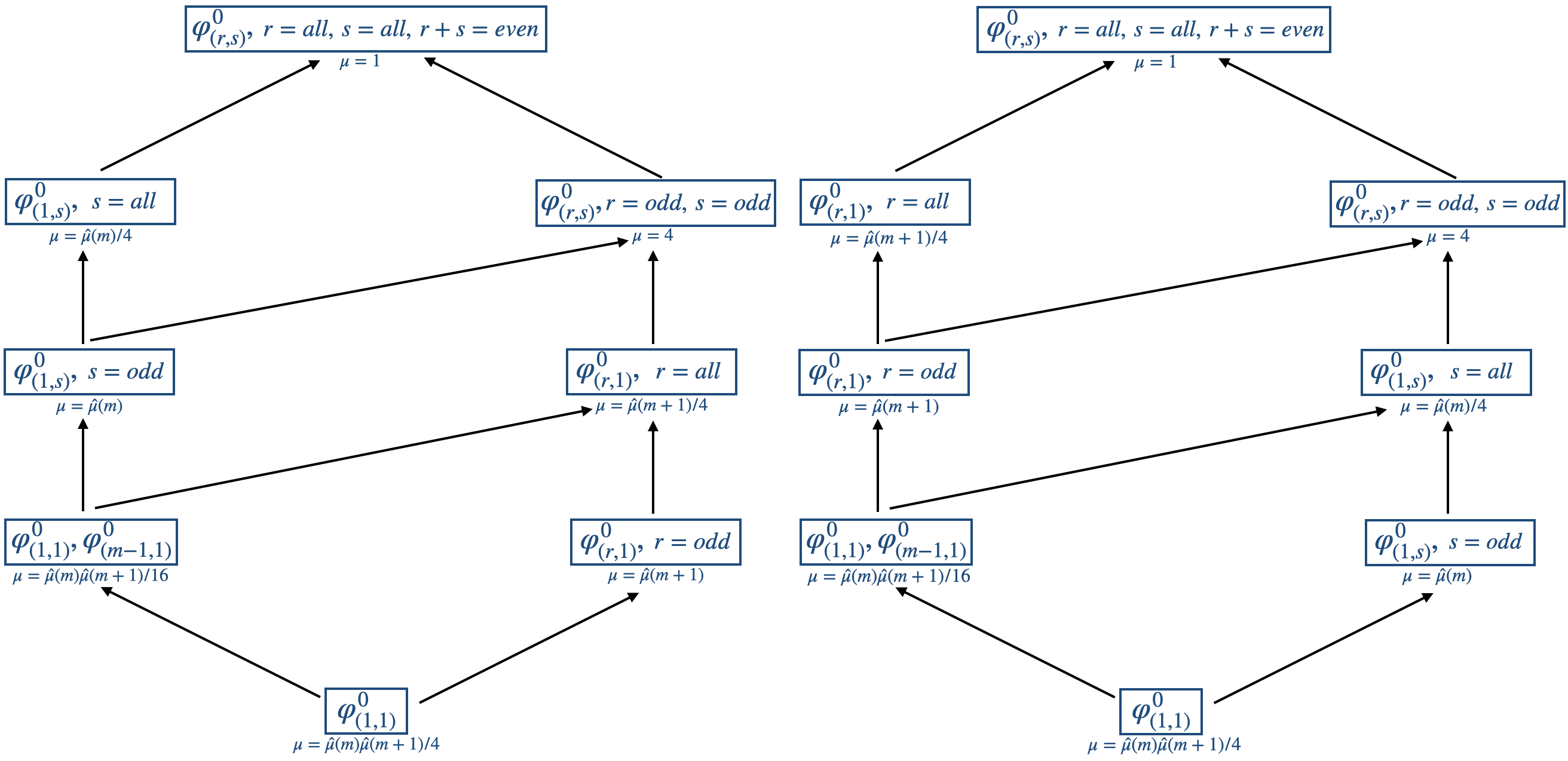}
    \caption{All possible $d=2$ CFTs corresponding to the $(A,A)$ series and their sualgebras for odd $m\geq5$ (left) and for even $m\geq6$ (right). We include all the corresponding indexes in terms of $\hat{\mu}(m)=m^2\sin^{-4}\left(\pi/{m}\right)/4$. The associated categories of superselection sectors can be found in the text.}
    \label{modelsdiag}
\end{figure}
\noindent The eight diagonal models we find for any $m$ are:
\begin{itemize}
    \item  The complete model including all spin zero fields allowed by the spectrum (\ref{AA}) for a given $m$. This is always a modular invariant model and has $\mu=1$. It is indeed the $(A_{m-1},A_m)$ model.
    
    \item A model including fields of the form $\varphi^{0}_{(r,s)}$ for all $r$ odd and $s$ odd. This can be obtained from the fixed point (uncharged) algebra under the $\mathbb{Z}_2$ symmetry that transforms the fields as $\varphi^{0}_{(r,s)}\to (-1)^{1+r} \varphi^{0}_{(r,s)}$ for the choice of $r+s$ even in the Kac table. This model has $\mu=4$ for any value of $m$, as corresponds to the action of a $\mathbb{Z}_2$ group on the complete model in $d=2$. The superselection sectors appearing after the orbifold can be described by the tensor category of the abelian group $\mathbb{Z}_2$. 
    
    \item A model that only includes the fields of the form $\varphi^{0}_{(1,s)}$ for all possible values of $s$. This model has an index $ \mu=\hat{\mu}(m)/4$, where $\hat{\mu}(m)$ has been defined above (\ref{lieaf}). In this case, the superselection sectors of the model can indeed be described by the tensor category $su(2)_{m-2}^{\text{even}}$, see appendix \ref{su2} for the notion of this category.

    \item A smaller model similar to the previous one in which we only include the fields of the form $\varphi^{0}_{(1,s)}$, for $s$ odd. This model can be computed to have an index $\mu=\hat{\mu}(m)$ and has a category of supeselection sectors $su(2)_{m-2}$.

    \item A model composed by fields of the form $\varphi^{0}_{(r,1)}$, for all allowed values of $r$. This model has an index $\mu=\hat{\mu}(m+1)/4$. Its superselection structure is controlled by the category $su(2)_{m-1}^{\text{even}}$.

    \item A model including the fields of the form $\varphi^{0}_{(r,1)}$, for $r$ odd. This has index $ \mu=\hat{\mu}(m+1)$. The superselection sectors are given by the tensor category $su(2)_{m-1}$.
    
    \item A model composed only by the stress tensor $\varphi^{0}_{(1,1)}$ and the field $\varphi^{0}_{(m-1,1)}$. This model has $\mu=\hat{\mu}(m)\hat{\mu}(m+1)/16$ and we can associate it to $su(2)_{m-2}^{\text{even}}\times su(2)_{m-1}^{\text{even}}$ in KL table.
    
    \item The conformal net of a given $m$ spanned by the stress tensor $\varphi^{0}_{(1,1)}$. This always has   $\mu=\hat{\mu}(m)^2\hat{\mu}(m+1)^2/4$.
\end{itemize}

The method to arrive at this classification mixed computer and induction techniques, mostly based on the OPE and the closure of the algebra. But once the classification is given, it is simple to convince oneself all these are well-defined closed  $d=2$ CFTs algebras. We sum up the results in the following table. Tensor categories follow by matching the indices and fusions with KL Table II \cite{Kawahigashi:2003gi}.\footnote{We note the category described in row $5$ of Table \ref{tabla1} was overlooked in writing Table II of \cite{Kawahigashi:2003gi}. We thank the authors for this clarification.}  We obtain
\begin{table}[H]
    \centering
    \renewcommand{\arraystretch}{1.4}
\begin{tabular}{cccc}
\cline{1-4}
\multicolumn{1}{|c|}{}              & \multicolumn{1}{c|}{Fields included}         & \multicolumn{1}{c|}{Tensor Category}   & \multicolumn{1}{c|}{Global Index ($\mu$)}               \\ \cline{1-4}
\multicolumn{1}{|c|}{$1$}           & \multicolumn{1}{c|}{$\varphi^{0}_{(r,s)}$ for all $r$ and $s$} & \multicolumn{1}{c|}{Id}     & \multicolumn{1}{c|}{$1$}               \\ \cline{1-4}
\multicolumn{1}{|c|}{$2$} & \multicolumn{1}{c|}{$\varphi^{0}_{(r,s)}$ for $r$ odd and $s$ odd} & \multicolumn{1}{c|}{$\mathbb{Z}_2$}  & \multicolumn{1}{c|}{$4$}      \\ \cline{1-4}
\multicolumn{1}{|c|}{$3$}      & \multicolumn{1}{c|}{$\varphi^{0}_{(1,s)}$ for all $s$} & \multicolumn{1}{c|}{ $su(2)_{m-2}^{\text{even}}$}  & \multicolumn{1}{c|}{$\frac{m^2}{16} \sin^{-4}\big(\frac{\pi}{m}\big)$}    \\ \cline{1-4}
\multicolumn{1}{|c|}{$4$}      & \multicolumn{1}{c|}{$\varphi^{0}_{(1,s)}$ for $s$ odd}  & \multicolumn{1}{c|}{ $su(2)_{m-2}$}  & \multicolumn{1}{c|}{$\frac{m^2}{4} \sin^{-4}\big(\frac{\pi}{m}\big)$}    \\ \cline{1-4}
\multicolumn{1}{|c|}{$5$}      & \multicolumn{1}{c|}{$\varphi^{0}_{(r,1)}$ for all $r$} & \multicolumn{1}{c|}{$su(2)_{m-1}^{\text{even}}$}  & \multicolumn{1}{c|}{$\frac{(m+1)^2}{16} \sin^{-4}\big(\frac{\pi}{m+1}\big)$}    \\ \cline{1-4}
\multicolumn{1}{|c|}{$6$}      & \multicolumn{1}{c|}{$\varphi^{0}_{(r,1)}$ for $r$ odd} & \multicolumn{1}{c|}{$su(2)_{m-1}$} & \multicolumn{1}{c|}{$\frac{(m+1)^2}{4} \sin^{-4}\big(\frac{\pi}{m+1}\big)$}     \\ \cline{1-4}
\multicolumn{1}{|c|}{$7$}      & \multicolumn{1}{c|}{$\varphi^{0}_{(1,1)}$ and $\varphi^{0}_{(m-1,1)}$} & \multicolumn{1}{c|}{$su(2)_{m-2}^{\text{even}}\times su(2)_{m-1}^{\text{even}}$} & \multicolumn{1}{c|}{$\frac{m^2(m+1)^2}{256} \sin^{-4}\big(\frac{\pi}{m}\big)\sin^{-4}\big(\frac{\pi}{m+1}\big)$}     \\ \cline{1-4}
\multicolumn{1}{|c|}{$8$}      & \multicolumn{1}{c|}{$\varphi^{0}_{(1,1)}$} & \multicolumn{1}{c|}{$(A_{m-1},A_m)$}   & \multicolumn{1}{c|}{$\frac{m^2(m+1)^2}{64} \sin^{-4}\big(\frac{\pi}{m}\big)\sin^{-4}\big(\frac{\pi}{m+1}\big)$}   \\ \cline{1-4}
\end{tabular}
\caption{All (sub)models of the $(A_{m-1},A_m)$ modular invariants with their tensor categories and global index.}
\label{tabla1}
\end{table}

An important remark follows. Although we can find these eight models for any given $m$, the inclusion of these algebras inside each other changes with the parity of $m$. We describe these inclusions graphically for $m$ even or odd in  Fig. \ref{modelsdiag}.

We also note that the same structure is applicable to the bosonic subalgebras of the triclritical Ising for $m=4$ and the Ising model for $m=3$. The only difference is that for such particular case there are coincidences between the nodes. For $m=4$, we have that the algebras 2 and 3 coincide, as well as 6 and 7. For $m=3$ the algebras 2, 3, 6, and 7  coincide, as well as 4 and 8, and also 1 and 5.

\subsection{\texorpdfstring{$(D,A)$}{Lg} and \texorpdfstring{$(A,D)$}{Lg} series for general  \texorpdfstring{$m\geq5$}{Lg} 
\label{m7}}

The natural next step is to fix the top node in the tree to be one of the $(D,A)$ or $(A,D)$ series, and find all the models in between those and the stress tensor. The modular invariant $(D,A)$ and $(A,D)$ series can be characterized by an integer parameter $n$ and the following spectrums:

\begin{itemize}
    \item For models with $m=4n+2$:
\be
\mathcal{S}_{(D_{2n+2},A_{4n+2})}=\frac{1}{2}\Big[\bigoplus^{4n+1}_{\substack{r=1\\ \text{Mod 2}}}\bigoplus^{4n+2}_{s=1} (r,s)\otimes (r,s)\Big] \oplus \frac{1}{2}\Big[\bigoplus^{4n+1}_{\substack{r=1\\ \text{Mod 2}}}\bigoplus^{4n+2}_{s=1} (r,s)\otimes (4n+2-r,s)\Big]\,.  \label{SAD1}
\ee
\item For models with $m=4n+1$:
\be
\mathcal{S}_{(A_{4n},D_{2n+2})}=\frac{1}{2}\Big[\bigoplus^{4n}_{r=1}\bigoplus^{4n+1}_{\substack{s=1\\ \text{Mod 2}}} (r,s)\otimes (r,s)\Big] \oplus \frac{1}{2}\Big[\bigoplus^{4n}_{r=1}\bigoplus^{4n+1}_{\substack{s=1\\ \text{Mod 2}}} (r,s)\otimes (r,4n+2-s)\Big]\,. \label{SAD2}
\ee
\item For models with $m=4n$:
\be
\mathcal{S}_{(D_{2n+1},A_{4n})}=\frac{1}{2}\Big[\bigoplus^{4n-1}_{\substack{r=1\\ \text{Mod 2}}}\bigoplus^{4n}_{s=1} (r,s)\otimes (r,s)\Big] \oplus \frac{1}{2}\Big[\bigoplus^{4n-2}_{\substack{r=2\\ \text{Mod 2}}}\bigoplus^{4n}_{s=1} (r,s)\otimes (4n-r,s)\Big]\,.  \label{SAD3}
\ee
\item For models with $m=4n-1$:
\be
\mathcal{S}_{(A_{4n-2},D_{2n+1})}=\frac{1}{2}\Big[\bigoplus^{4n-2}_{r=1}\bigoplus^{4n-1}_{\substack{s=1\\ \text{Mod 2}}} (r,s)\otimes (r,s)\Big] \oplus \frac{1}{2}\Big[\bigoplus^{4n-2}_{r=1}\bigoplus^{4n-2}_{\substack{s=2\\ \text{Mod 2}}} (r,s)\otimes (r,4n-s)\Big]\,. \label{SAD4}
\ee
\end{itemize}

For $n=1$ we recover the bosonic part of the cases discussed in section \ref{m345}. In particular, the complete spectrum $\mathcal{S}_{(A_{2},D_{3})}=\mathcal{S}_{(A_{2},A_{3})}$ corresponds to the Ising model and $\mathcal{S}_{(D_{3},A_{4})}=\mathcal{S}_{(A_{3},A_{4})}$ represents the Tricritical Ising model.  Also, the full spectrum $\mathcal{S}_{(A_{4},D_{4})}$ represents the Three States Potts model described by the partition function (\ref{Zpotts}). The remaining $\mathcal{S}_{(D_{4},A_{4})}$, still within the case $n=1$, we have not discussed yet.

To proceed, we  extend the notation for the fields given in (\ref{spin0}) to $\varphi^{\epsilon}_{(r,s)}$, where $(r,s)$ take the usual values and $\epsilon=0,1$. The value of $\epsilon$ naturally denotes if the field corresponds to the diagonal or non-diagonal part of the spectrum respectively. For example for the $(D_{2n+1},A_{4n})$ series 
\be 
\varphi^{0}_{(r,s)}= (r,s)\otimes (r,s)\,, \quad \varphi^{1}_{(r,s)}= (r,s)\otimes (4n-r,s)\;. \label{notf}
\ee
Following then Ref. \cite{Ribault:2016sla}, the OPE algebra rules that preserve diagonality can be written  as 
\be 
\varphi^{\epsilon}_{(r,s)}\p \varphi^{\epsilon'}_{(r',s')} = \sum^{r_{\text{max}}}_{\substack{r''=1-|r-r'|\\\text{Mod 2}}}\,\,\,\,\sum^{s_{\text{max}}}_{\substack{s''=1-|s-s'|\\ \text{Mod 2}}} \varphi^{\textrm{Mod}\,[\epsilon+\epsilon',2]}_{(r'',s'')}\,,
\label{fusionnondiag} 
\ee 
where $\textrm{Mod}\,[x,2]$ is $0$ for even $x$ and $1$ for odd $x$.  This is a natural extension of $(\ref{fusionAA})$. Note that for $(A_{4},D_{4})$ these OPE are compatible with the OPEs of section \ref{m5} derived using the $\mathbb{Z}_3$ symmetry of the Three States Potts model. Indeed, the cases $(D_{2n+2},A_{4n},)$ and $(A_{4n},D_{2n+2})$ involve two copies of the same field as we saw for $m=5$ in section \ref{m5}. These are accounted since they appear in both parts of the spectrums (\ref{SAD1}) and (\ref{SAD2}). More specifically, they are given by  $\varphi^{\epsilon}_{(2n+1,s)}$ for $m=4n+2$ or by $\varphi^{\epsilon}_{(r,2n+1)}$ for $m=4n+1$. In both cases, they obey the fusion rules (\ref{fusionnondiag}).

Again, the task is to find closed sets of fields for each of the spectrums  (\ref{SAD1}), (\ref{SAD2}), (\ref{SAD3}), and (\ref{SAD4}) under the OPE (\ref{fusionnondiag}) for different values of $n$. In what follows, we do this in the different cases. We include, for the purpose of clarity, the submodels of the $(A,A)$ series and their inclusions inside the models of the $(D,A)$ or $(A,D)$ series respectively.

\subsubsection{\texorpdfstring{$(A_{4n+1},A_{4n+2})$}{Lg} and \texorpdfstring{$(D_{2n+2},A_{4n+2})$}{Lg} series for the \texorpdfstring{$m=4n+2$}{Lg} case \label{m8}}

\begin{figure}[H]
    \centering
    \includegraphics[width=1\textwidth]{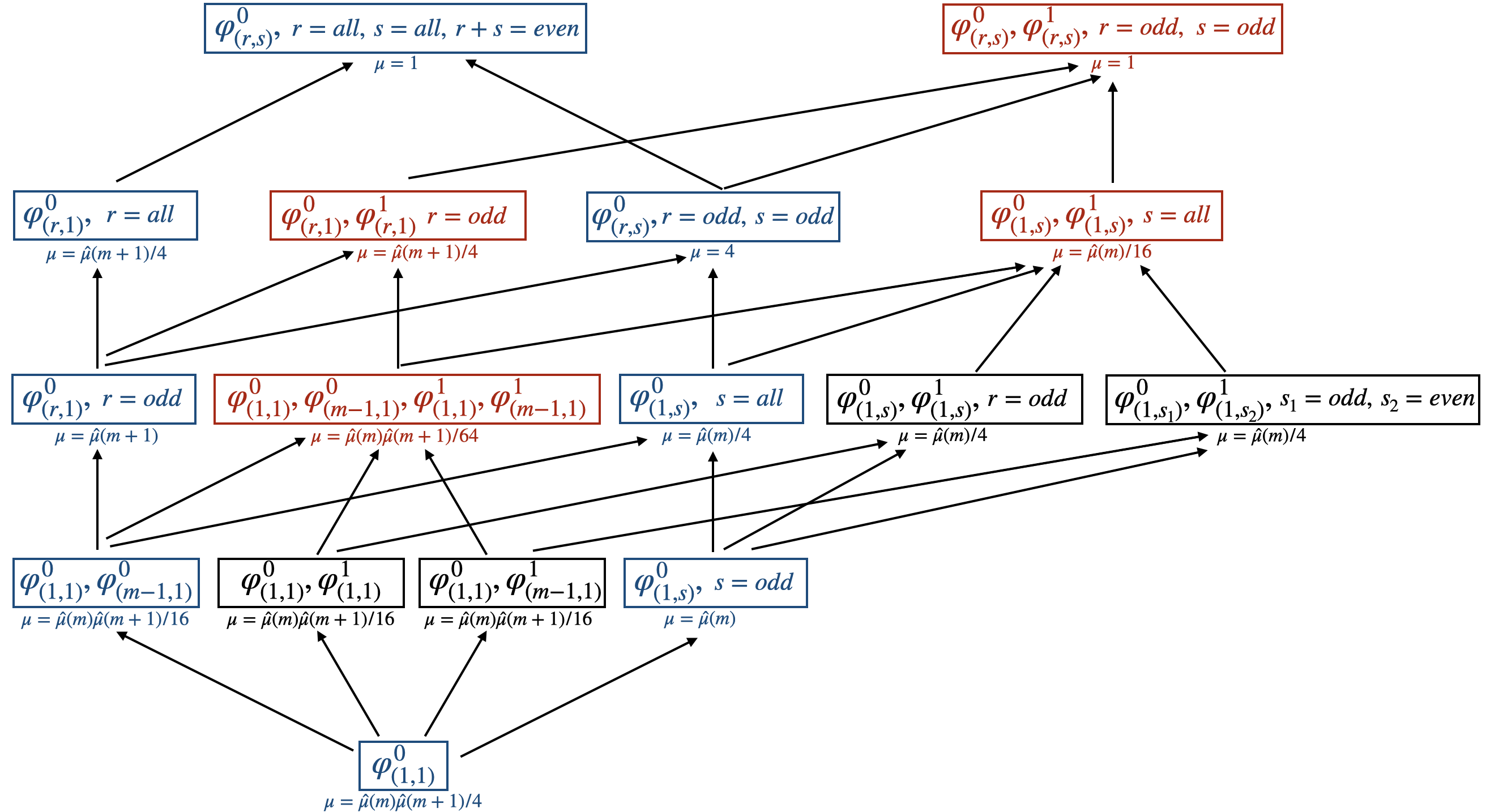}
    \caption{All possible $d=2$ bosonic CFTs corresponding to (sub)algebras of the $(A_{4n+1},A_{4n+2})$ and $(D_{2n+2},A_{4n+2})$ modular invariants for $m=4n+2$, with all the global Jones indexes in terms of $\hat{\mu}(m)=m^2\sin^{-4}\left(\pi/{m}\right)/4$. The blue boxes represent algebras composed only of spin zero fields. The red boxes describe algebras that have bosonic fields of spin $\geq 1$  that respect parity symmetry, while the black boxes are the ones that do not.}\label{4n2}
\end{figure}

We begin considering the case $m=4n+2$. These are models constructed from parts of the spectrums (\ref{AA}) and (\ref{SAD1}). In total, we find sixteen models. We have the eight diagonal models already discussed in section \ref{m6} and eight non-diagonal models. The non-diagonal models include four that respect parity symmetry and the rest do not. There are, of course, two complete models that correspond to the modular invariants $(A_{4n+1},A_{4n+2})$ and $(D_{2n+2},A_{4n+2})$. We sum up the inclusions and Jones indexes in Figure \ref{4n2}.
Furthermore, in the table below we describe the tensor categories of the superselection sectors of the non-diagonal models with parity symmetry, matching KL classification  \cite{Kawahigashi:2003gi}. These correspond to the red boxes in Figure \ref{4n2}. Note there are non parity symmetric models (black boxes in Fig. \ref{4n2}) that are not included in that classification. The categories $D_{2n}^{\textrm{even}}$ are described in the appendix \ref{su2}.

\begin{table}[H]
    \centering
    \renewcommand{\arraystretch}{1.4}
\begin{tabular}{cccc}
\cline{1-4}
\multicolumn{1}{|c|}{}              & \multicolumn{1}{c|}{Fields included}         & \multicolumn{1}{c|}{Tensor Category}   & \multicolumn{1}{c|}{Global Index ($\mu$)}               \\ \cline{1-4}
\multicolumn{1}{|c|}{$1$}           & \multicolumn{1}{c|}{$\varphi^{0}_{(r,s)}$ and $\varphi^{1}_{(r,s)}$ for $r$ and $s$ odd} & \multicolumn{1}{c|}{Id}     & \multicolumn{1}{c|}{$1$}                 \\ \cline{1-4}
\multicolumn{1}{|c|}{$2$}      & \multicolumn{1}{c|}{$\varphi^{0}_{(1,s)}$ and $\varphi^{1}_{(1,s)}$ for all $s$ }  & \multicolumn{1}{c|}{ $D_{2n+2}^{\text{even}}$}  & \multicolumn{1}{c|}{$\frac{m^2}{64} \sin^{-4}\big(\frac{\pi}{m}\big)$}    \\ \cline{1-4}
\multicolumn{1}{|c|}{$3$}      & \multicolumn{1}{c|}{$\varphi^{0}_{(r,1)}$ and $\varphi^{1}_{(r,1)}$ for $r$ odd} & \multicolumn{1}{c|}{$su(2)_{4n+1}^{\text{even}}$}  & \multicolumn{1}{c|}{$\frac{(m+1)^2}{16} \sin^{-4}\big(\frac{\pi}{m}\big)$}      \\ \cline{1-4}
\multicolumn{1}{|c|}{$4$}      & \multicolumn{1}{c|}{$\varphi^{0}_{(1,1)}$, $\varphi^{0}_{(m-1,1)}$, $\varphi^{1}_{(1,1)}$, and  $\varphi^{1}_{(m-1,1)}$} & \multicolumn{1}{c|}{$(D_{2n+2},A_{4n+2})$}   & \multicolumn{1}{c|}{$\frac{m^2(m+1)^2}{1024} \sin^{-4}\big(\frac{\pi}{m}\big)\sin^{-4}\big(\frac{\pi}{m+1}\big)$}   \\ \cline{1-4}
\end{tabular}
\caption{All (sub)models of the $(D_{2n+2},A_{4n+2})$ modular invariants that include non diagonal fields and respect parity symmetry with their tensor categories and global index.}
\label{tabla2}
\end{table}

\subsubsection{\texorpdfstring{$(A_{4n},A_{4n+1})$}{Lg} and \texorpdfstring{$(A_{4n},D_{2n+2})$}{Lg} series for the \texorpdfstring{$m=4n+1$}{Lg} case \label{m9}}

\begin{figure}[H]
    \centering
    \includegraphics[width=1\textwidth]{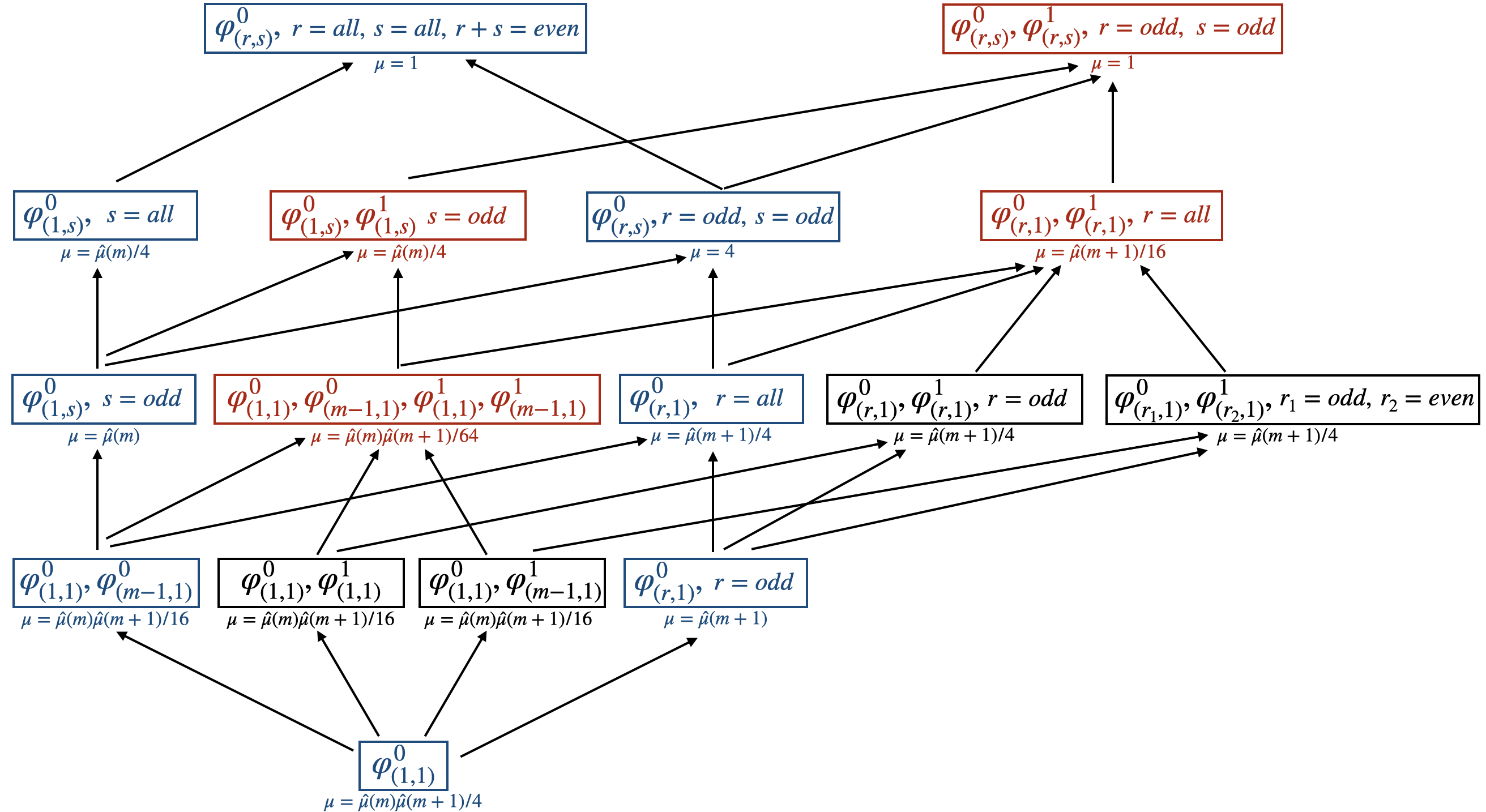}
    \caption{All possible $d=2$ CFTs corresponding to (sub)algebras of the $(A_{4n},A_{4n+1})$ and $(A_{4n},D_{2n+2})$ modular invariants for $m=4n+1$, with all the global Jones indexes in terms of $\hat{\mu}(m)=m^2\sin^{-4}\left(\pi/{m}\right)/4$. The blue boxes represent algebras composed only of spin zero fields. The red boxes describe algebras that have bosonic fields with spin that respect parity symmetry, while the black boxes are the ones that do not.} \label{4n1} 
\end{figure}

The next case is $m=4n+1$, involving fields of the spectrums (\ref{AA}) and (\ref{SAD2}). Again, we find sixteen models, including four new non-diagonal models that respect parity symmetry, and four that do not. We depict the possible $d=2$ CFTs, with the corresponding indexes and inclusion structure, in Figure \ref{4n1}. As usual, the top nodes correspond to the modular invariants $(A_{4n},A_{4n+1})$ and $(A_{4n},D_{2n+2})$.

For the sake of comparison with \cite{Kawahigashi:2003gi} the tensor categories of the superselection sectors of the non-diagonal models with parity symmetry are included in the following table
\begin{table}[H]
    \centering
    \renewcommand{\arraystretch}{1.4}
\begin{tabular}{cccc}
\cline{1-4}
\multicolumn{1}{|c|}{}              & \multicolumn{1}{c|}{Fields included}         & \multicolumn{1}{c|}{Tensor Category}   & \multicolumn{1}{c|}{Global Index ($\mu$)}               \\ \cline{1-4}
\multicolumn{1}{|c|}{$1$}           & \multicolumn{1}{c|}{$\varphi^{0}_{(r,s)}$ and $\varphi^{1}_{(r,s)}$ for $r$ and $s$ odd} & \multicolumn{1}{c|}{Id}     & \multicolumn{1}{c|}{$1$}                 \\ \cline{1-4}
\multicolumn{1}{|c|}{$2$}      & \multicolumn{1}{c|}{$\varphi^{0}_{(r,1)}$ and $\varphi^{1}_{(r,1)}$ for all $r$ }  & \multicolumn{1}{c|}{ $D_{2n+2}^{\text{even}}$}  & \multicolumn{1}{c|}{$\frac{(m+1)^2}{64} \sin^{-4}\big(\frac{\pi}{m+1}\big)$}    \\ \cline{1-4}
\multicolumn{1}{|c|}{$3$}      & \multicolumn{1}{c|}{$\varphi^{0}_{(1,s)}$ and $\varphi^{1}_{(1,s)}$ for $s$ odd} & \multicolumn{1}{c|}{$su(2)_{4n-1}^{\text{even}}$}  & \multicolumn{1}{c|}{$\frac{m^2}{16} \sin^{-4}\big(\frac{\pi}{m}\big)$}      \\ \cline{1-4}
\multicolumn{1}{|c|}{$4$}      & \multicolumn{1}{c|}{$\varphi^{0}_{(1,1)}$, $\varphi^{0}_{(m-1,1)}$, $\varphi^{1}_{(1,1)}$, and  $\varphi^{1}_{(m-1,1)}$} & \multicolumn{1}{c|}{$(A_{4n},D_{n+2})$}   & \multicolumn{1}{c|}{$\frac{m^2(m+1)^2}{1024} \sin^{-4}\big(\frac{\pi}{m+1}\big)\sin^{-4}\big(\frac{\pi}{m+1}\big)$}   \\ \cline{1-4}
\end{tabular}
\caption{All (sub)models of the $(A_{4n},D_{2n+1})$ modular invariants that include non diagonal fields and respect parity symmetry with their tensor categories and global index.}
\label{tabla3}
\end{table}

\subsubsection{\texorpdfstring{$(A_{4n-1},A_{4n})$}{Lg} and \texorpdfstring{$(D_{2n+1},A_{4n})$}{Lg} series for the \texorpdfstring{$m=4n$}{Lg} case \label{m10}}

\begin{figure}[H]
    \centering
    \includegraphics[width=1\textwidth]{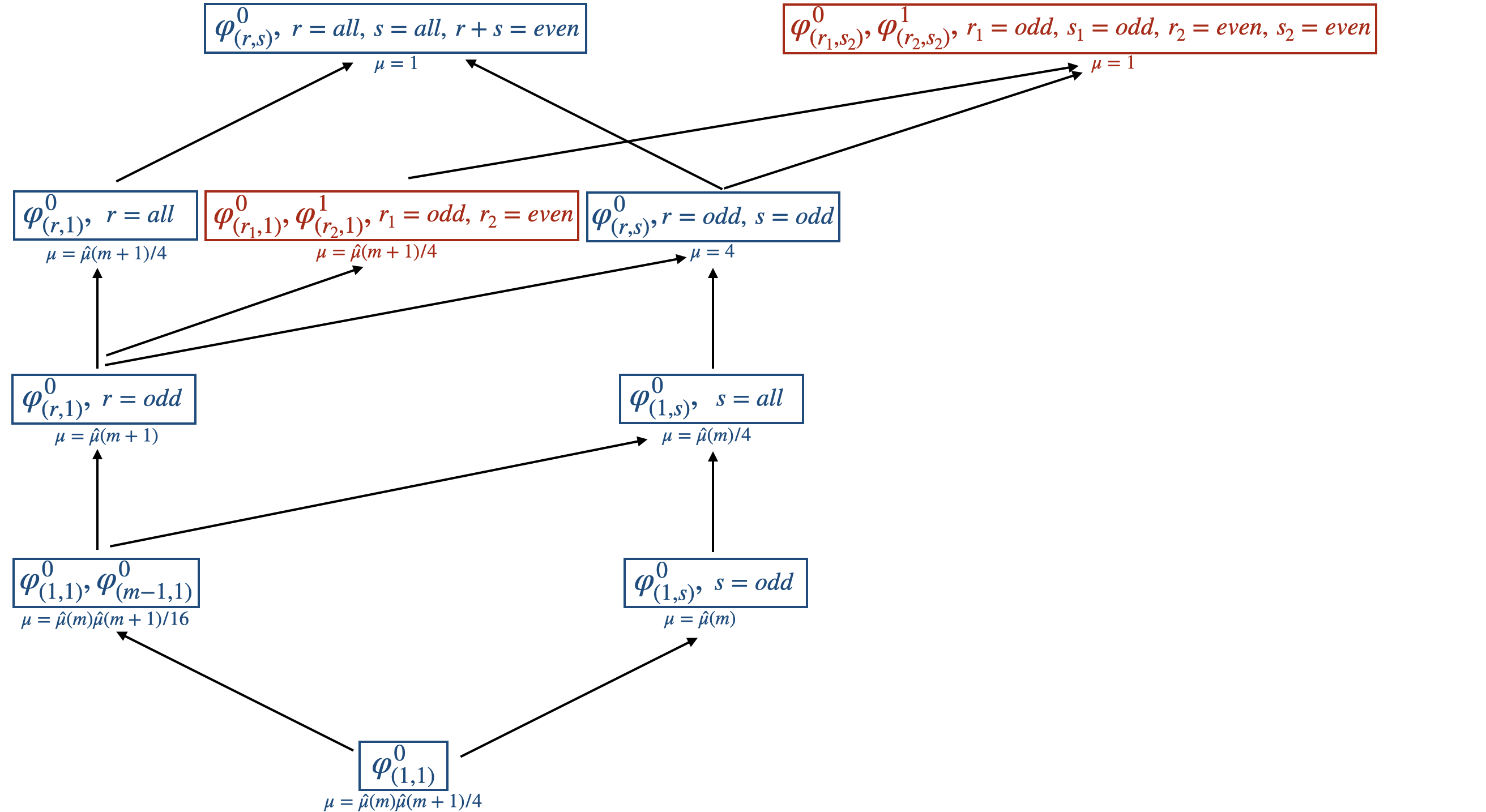}
     \caption{All possible $d=2$ CFTs corresponding to (sub)algebras of the $(A_{4n-1},A_{4n})$ and $(D_{2n+1},A_{4n})$ modular invariants for $m=4n$, with all the global Jones indexes in terms of $\hat{\mu}(m)=m^2\sin^{-4}\left(\pi/{m}\right)/4$. The blue boxes represent algebras composed only of spin zero fields. The red boxes describe algebras that have bosonic fields with spin.}\label{4n0}
\end{figure}

We return now to the $(D,A)$ series for $m=4n$. To this end, we consider the spectrums (\ref{AA}) and (\ref{SAD3}). We find ten models. This includes the eight diagonal models described before and two non-diagonal models with parity symmetry. There are two modular invariants models that correspond to  $(A_{4n-1},A_{4n})$ and $(D_{2n+2},A_{4n})$. Below we sum up the inclusion and indexes in Figure \ref{4n0}.
In the following table we characterize the new models in terms of their tensor categories:
\begin{table}[H]
    \centering
    \renewcommand{\arraystretch}{1.4}
\begin{tabular}{cccc}
\cline{1-4}
\multicolumn{1}{|c|}{}              & \multicolumn{1}{c|}{Fields included}         & \multicolumn{1}{c|}{Tensor Category}   & \multicolumn{1}{c|}{Global Index ($\mu$)}               \\ \cline{1-4}
\multicolumn{1}{|c|}{$1$}           & \multicolumn{1}{c|}{$\varphi^{0}_{(r_1,s_1)}$ for $r_1, s_1$ odd, and $\varphi^{1}_{(r_2,s_2)}$ for $r_2,s_2$ even} & \multicolumn{1}{c|}{Id}     & \multicolumn{1}{c|}{$1$}                 \\ \cline{1-4}
\multicolumn{1}{|c|}{$2$}      & \multicolumn{1}{c|}{$\varphi^{0}_{(r_1,1)}$  for $r_1$ odd and $\varphi^{1}_{(r_2,1)}$ for $r_2$ even}  & \multicolumn{1}{c|}{ $su(2)_{4n-1}^{\text{even}}$}  & \multicolumn{1}{c|}{$\frac{(m+1)^2}{16} \sin^{-4}\big(\frac{\pi}{m+1}\big)$}      \\ \cline{1-4}
\end{tabular}
\caption{All (sub)models of the $(D_{2n+1},A_{4n})$ modular invariants that include non diagonal fields and respect parity symmetry with their tensor categories and global index.}
\label{tabla4}
\end{table}
We remark that the tensor category $su(2)_{4n-1}^{\text{even}}$ of the new model coincides with the one of a submodel of the diagonal series. This is why it does not appear as a new possibility in KL classification. In other terms, there are two different models for this central charge that have the same category symmetry but different field composition (differing in spin and conformal dimensions for example).   

\subsubsection{\texorpdfstring{$(A_{4n-2},A_{4n-1})$}{Lg} and \texorpdfstring{$(A_{4n-2},D_{2n+1})$}{Lg} series for the \texorpdfstring{$m=4n-1$}{Lg} case \label{m11}}

\begin{figure}[H]
    \centering
    \includegraphics[width=1\textwidth]{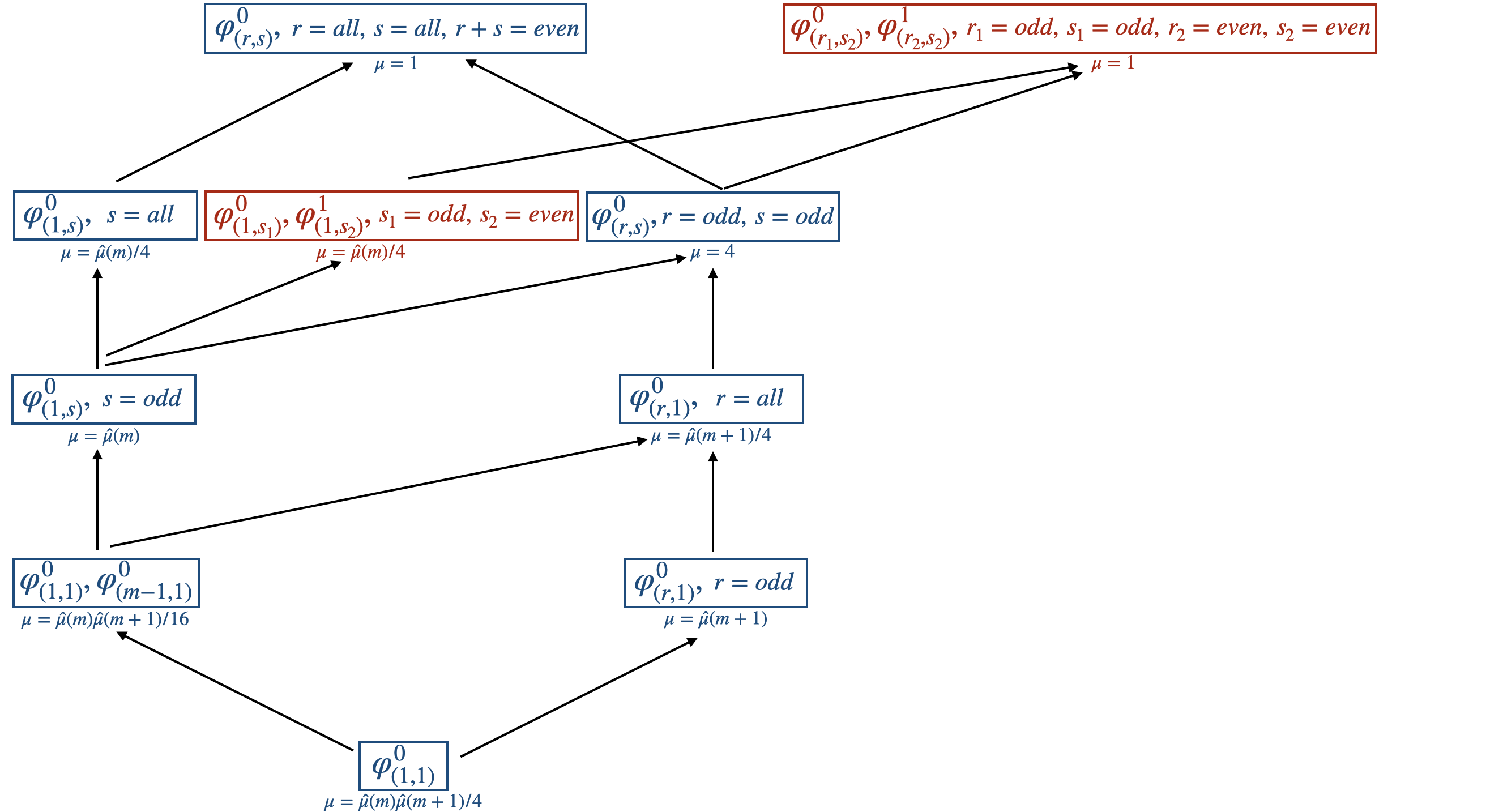}
    \caption{All possible $d=2$ CFTs corresponding to (sub)algebras of the $(A_{4n-2},A_{4n-1})$ and $(A_{4n-2},D_{2n+1})$ modular invariants for $m=4n+1$, with all the global Jones indexes in terms of $\hat{\mu}(m)=m^2\sin^{-4}\left(\pi/{m}\right)/4$. The blue boxes represent algebras composed only of spin zero fields. The red boxes describe algebras that have bosonic fields with spin.}\label{4n3}
\end{figure}

Finally, for the $(A,D)$ series associated with $m=4n-1$, we consider the spectrums (\ref{AA}) and (\ref{SAD4}) to look for building blocks. We again find ten models, with two new non-diagonal ones with parity symmetry. As always, there are two complete models that in this case correspond to  $(A_{4n-2},A_{4n-1})$ and $(A_{4n-2},D_{2n+1})$. We sum up the inclusions and indexes in Figure \ref{4n3}.
\vspace{-0.5cm}
The tensor categories associated with the new models are listed in the table below:
\begin{table}[H]
    \centering
    \renewcommand{\arraystretch}{1.4}
\begin{tabular}{cccc}
\cline{1-4}
\multicolumn{1}{|c|}{}              & \multicolumn{1}{c|}{Fields included}         & \multicolumn{1}{c|}{Tensor Category}   & \multicolumn{1}{c|}{Global Index ($\mu$)}               \\ \cline{1-4}
\multicolumn{1}{|c|}{$1$}           & \multicolumn{1}{c|}{$\varphi^{0}_{(r_1,s_1)}$ for $r_1,s_1$ odd, and $\varphi^{1}_{(r_2,s_2)}$ for $r_2,s_2$ even} & \multicolumn{1}{c|}{Id}     & \multicolumn{1}{c|}{$1$}                 \\ \cline{1-4}
\multicolumn{1}{|c|}{$2$}      & \multicolumn{1}{c|}{$\varphi^{0}_{(1,s_1)}$  for $s_1$ odd and $\varphi^{1}_{(1,s_2)}$ for $s_2$ even}  & \multicolumn{1}{c|}{ $su(2)_{4n-3}^{\text{even}}$}  & \multicolumn{1}{c|}{$\frac{m^2}{16} \sin^{-4}\big(\frac{\pi}{m}\big)$}      \\ \cline{1-4}
\end{tabular}
\caption{All (sub)models of the $(A_{4n-2},D_{2n+1})$ modular invariants that include non diagonal fields and respect parity symmetry with their tensor categories and global index.}
\label{tabla5}
\end{table}
\vspace{-0.5cm}
 Again, the category $su(2)_{4n-3}^{\text{even}}$ is shared by the new model found here and one of the diagonal series submodels. However, the associated CFTs differ in their spectrum of primary operators.

\vspace{-0.3cm}
\subsection{\texorpdfstring{$(A,E)$}{Lg} and \texorpdfstring{$(E,A)$}{Lg}  series: the \texorpdfstring{$E_6$}{Lg} example}\label{mE}
\vspace{-0.3cm}
Finally, we provide the first elements of the unitary $E$-series minimal models. These appear for $E_6$ and $m=11,12$. As above, we construct the possible submodels using the OPE. The fusion rules (OPE) of these models can be found in \cite{Nivesvivat:2025odb} with all other fusion rules of the $E$-series.\footnote{We thank S. Ribault and R. Nivesvivat for the communication of this information prior to the publication of their manuscript that includes the derivation of the E-series non-chiral fusion rules using a bootstrap approach. } We begin by analyzing the model that corresponds to $(E_6,A_{12})$ for $m=12$. The spectrum can be written as
\be
\mathcal{S}_{(E_6,A_{12})}=\bigoplus_{r=1,4,7}\bigoplus^{12}_{s=1} \Big[\phi^0_{(r,s)} \oplus \phi^1_{(r,s)}\Big]\,, \label{es12}
\ee
where we have used the notation for the fields
\begin{align}
&\phi^0_{(r,s)}=(r,s)\otimes (r,s)\,, &\quad &\phi^1_{(4,s)}=(4,s)\otimes (4,13-s)\,.  \\
&\phi^1_{(1,s)}=(1,s)\otimes (7,s)\,, &\quad &\phi^1_{(7,s)}=(7,s)\otimes (1,s)\,.
\end{align} 
\begin{figure}[H]
    \centering
    \includegraphics[width=1\textwidth]{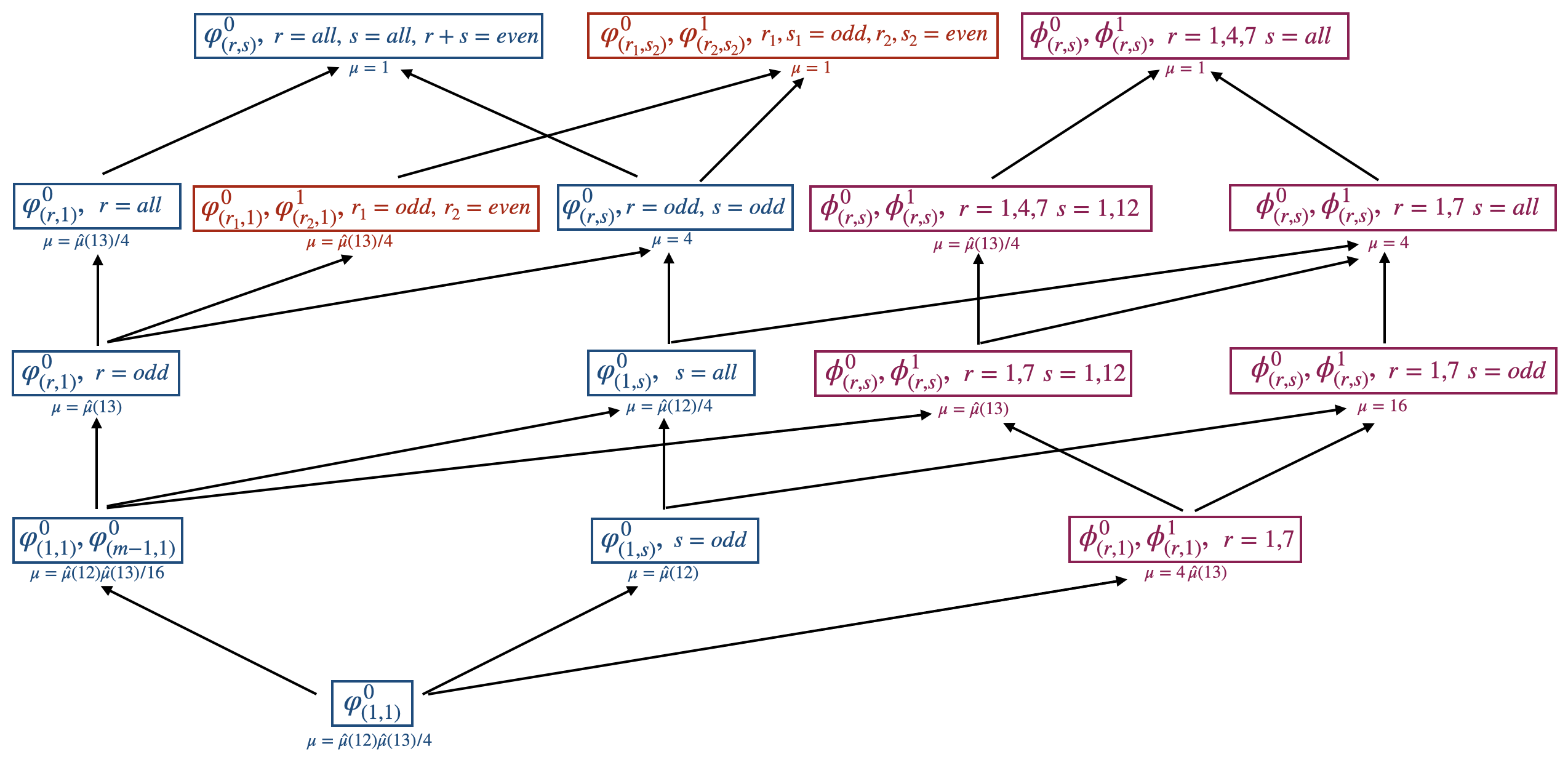}
    \caption{All models for $m=12$ with parity symmetry corresponding the (sub)algebras of the modular invariants $(A_{11},A_{12})$ (blue), $(D_{7},A_{12})$ (red) and $(E_6,A_{12})$ (purple).} \label{es1} 
\end{figure}
\noindent We have noted the fields as $\phi^0_{(r,s)}$ and $\phi^1_{(r,s)}$ instead of   $\varphi^0_{(r,s)}$ and $\varphi^1_{(r,s)}$. This is because the fusion rules of this model are different from (\ref{fusionAA}) or (\ref{fusionnondiag}). They only coincide for the fields $(1,s)\otimes (1,s)$, or more precisely $\phi^0_{(1,s)}=\varphi^0_{(1,s)}$.

From the spectrum (\ref{es12}) and the OPE we can build eighteen models. This includes four diagonal models (constructed from $\varphi^0_{(1,s)}$ entirely) already mentioned in the entries 3,4,7,8 of table \ref{tabla1}, as well as fourteen non diagonal models. Of these non diagonal models six are parity symmetric, and they can be described in terms of tensor categories as in table \ref{tabla6}:
\begin{table}[H]
    \centering
    \renewcommand{\arraystretch}{1.4}
\begin{tabular}{cccc}
\cline{1-4}
\multicolumn{1}{|c|}{}              & \multicolumn{1}{c|}{Fields included}         & \multicolumn{1}{c|}{Tensor Category}   & \multicolumn{1}{c|}{Global Index ($\mu$)}               \\ \cline{1-4}
\multicolumn{1}{|c|}{$1$}           & \multicolumn{1}{c|}{$\phi^0_{(r,s)}$ and $\phi^1_{(r,s)}$ for $r=1,4,7$ and all $s$} & \multicolumn{1}{c|}{Id}     & \multicolumn{1}{c|}{$1$}                 \\ \cline{1-4}
\multicolumn{1}{|c|}{$2$}      & \multicolumn{1}{c|}{$\phi^0_{(r,s)}$ and $\phi^1_{(r,s)}$ for $r=1,7$ and all $s$}  & \multicolumn{1}{c|}{ $\mathbb{Z}_2$}  & \multicolumn{1}{c|}{$4$} \\ \cline{1-4}
\multicolumn{1}{|c|}{$3$}      & \multicolumn{1}{c|}{$\phi^0_{(r,s)}$ and $\phi^1_{(r,s)}$ for $r=1,7$ and $s$ odd}  & \multicolumn{1}{c|}{ $su(2)_2$}  & \multicolumn{1}{c|}{$16$}        \\ \cline{1-4}
\multicolumn{1}{|c|}{$4$}      & \multicolumn{1}{c|}{$\phi^0_{(r,s)}$ and $\phi^1_{(r,s)}$ for $r=1,4,7$ and $s=1,12$}  & \multicolumn{1}{c|}{ $su(2)^{\text{even}}_{11}$}  & \multicolumn{1}{c|}{$\frac{169}{16}\sin^{-4}\left( \frac{\pi}{13}\right)$}
\\ 
\cline{1-4}
\multicolumn{1}{|c|}{$5$}      & \multicolumn{1}{c|}{$\phi^0_{(r,s)}$ and $\phi^1_{(r,s)}$ for $r=1,7$ and $s=1,12$}  & \multicolumn{1}{c|}{ $\mathbb{Z}_2\times su(2)^{\text{even}}_{11}$}  & \multicolumn{1}{c|}{$\frac{169}{4}\sin^{-4}\left( \frac{\pi}{13}\right)$}       \\ 
\cline{1-4}
\multicolumn{1}{|c|}{$6$}      & \multicolumn{1}{c|}{$\phi^0_{(r,1)}$ and $\phi^1_{(r,1)}$ for $r=1,7$}  & \multicolumn{1}{c|}{ $(E_6,A_{12})$}  & \multicolumn{1}{c|}{${169}\sin^{-4}\left( \frac{\pi}{13}\right)$}       \\ 
\cline{1-4}
\end{tabular}
\caption{All (sub)models of the $(E_6,A_{12})$ modular invariant that include non diagonal fields and respect parity symmetry with their tensor categories and global index.}
\label{tabla6}
\end{table}
This again coincides with \cite{Kawahigashi:2003gi} and gives a non trivial cross check for the fusion rules in \cite{Nivesvivat:2025odb}.  There are eight non diagonal models without parity symmetry. For completeness, we list all of them below:
\begin{itemize}
    \item $\{\phi^0_{(1,s)},\phi^1_{(1,s)},\,s=1,2,3,\dots,11,12\}$ and $\{\phi^0_{(1,s)},\phi^1_{(7,s)},\,ss=1,2,3,\dots,11,12\}$, 
    \item $\{\phi^0_{(1,s)},\phi^1_{(1,s)},\,s=1,3,5,7,9,11\}$ and $\{\phi^0_{(1,s)},\phi^1_{(7,s)},\,s=1,3,5,7,9,11\}$, 
    \item $\{\phi^0_{(1,s)},\phi^1_{(1,s)},\,s=1,12\}$ and $\{\phi^0_{(1,s)},\phi^1_{(7,s)},\,s=1,12\}$, 
    \item $\{\phi^0_{(1,1)},\phi^1_{(1,1)}\}$ and $\{\phi^0_{(1,1)},\phi^1_{(7,1)}\}$. 
\end{itemize}
The spectrum of the model $(A_{10},E_6)$ for $m=11$ is given by
\be
\mathcal{S}_{(A_{10},E_6)}=\bigoplus^{10}_{r=1}\bigoplus_{s=1,4,7}\Big[\phi^0_{(r,s)}\oplus \phi^1_{(r,s)}\Big] \,,
\ee
where now the fields are defined as
\be 
\phi^0_{(r,s)}=(r,s)\otimes (r,s)\,, \quad \phi^1_{(r,4)}=(r,4)\otimes (10-r,4)\,,
\ee
\be 
\phi^1_{(r,1)}=(r,1)\otimes (r,7)\,, \quad \phi^1_{(r,7)}=(r,7)\otimes (r,1)\,.
\ee
where we have that $\phi^0_{(r,1)}=\varphi^0_{(r,1)}$.

\begin{figure}[H]
    \centering
    \includegraphics[width=1\textwidth]{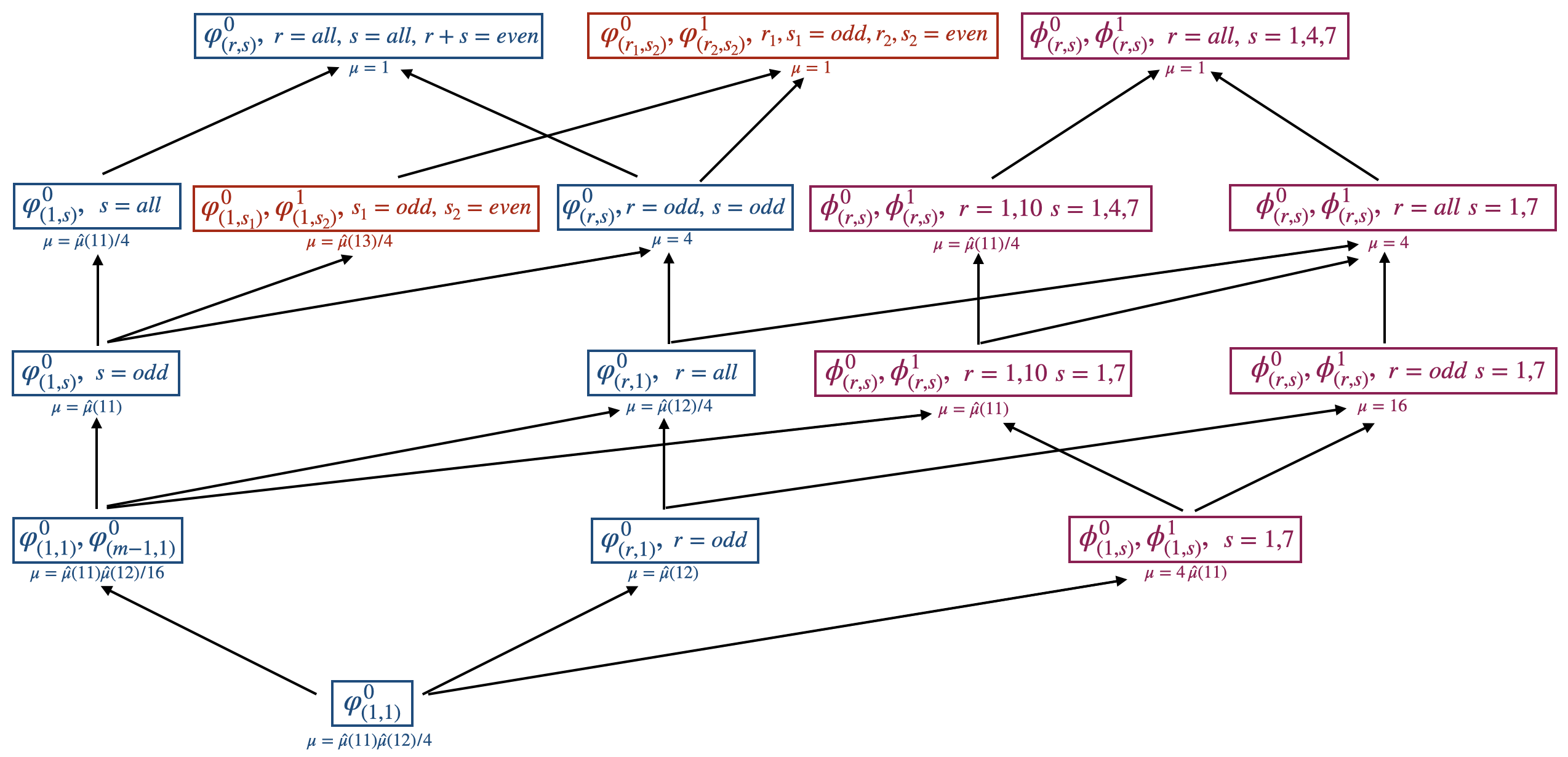}
    \caption{All models for $m=11$ with parity symmetry corresponding the (sub)algebras of the modular invariants $(A_{10},A_{11})$ (blue), $(A_{10},D_{7})$ (red) and $(A_{10},E_6)$ (purple).} \label{es2} 
\end{figure}

Analogously, from this spectrum, we can build eighteen models. These include the four diagonal models corresponding to entries 3,4,7,8 of table \ref{tabla1},  and fourteen non diagonal models. We describe the  tensor categories and indices of the six non diagonal models that respect parity symmetry in \ref{tabla7}:

\begin{table}[H]
    \centering
    \renewcommand{\arraystretch}{1.4}
\begin{tabular}{cccc}
\cline{1-4}
\multicolumn{1}{|c|}{}              & \multicolumn{1}{c|}{Fields included}         & \multicolumn{1}{c|}{Tensor Category}   & \multicolumn{1}{c|}{Global Index ($\mu$)}               \\ \cline{1-4}
\multicolumn{1}{|c|}{$1$}           & \multicolumn{1}{c|}{$\phi^0_{(r,s)}$ and $\phi^1_{(r,s)}$ for all $r$ and  $s=1,4,7$} & \multicolumn{1}{c|}{Id}     & \multicolumn{1}{c|}{$1$}                 \\ \cline{1-4}
\multicolumn{1}{|c|}{$2$}      & \multicolumn{1}{c|}{$\phi^0_{(r,s)}$ and $\phi^1_{(r,s)}$ for all $r$  and $s=1,7$}  & \multicolumn{1}{c|}{ $\mathbb{Z}_2$}  & \multicolumn{1}{c|}{$4$} \\ \cline{1-4}
\multicolumn{1}{|c|}{$3$}      & \multicolumn{1}{c|}{$\phi^0_{(r,s)}$ and $\phi^1_{(r,s)}$ for $r$ odd and $s=1,7$}  & \multicolumn{1}{c|}{ $su(2)_2$}  & \multicolumn{1}{c|}{$16$}        \\ \cline{1-4}
\multicolumn{1}{|c|}{$4$}      & \multicolumn{1}{c|}{$\phi^0_{(r,s)}$ and $\phi^1_{(r,s)}$ for $r=1,10$ and $s=1,4,7$}  & \multicolumn{1}{c|}{ $su(2)^{\text{even}}_{9}$}  & \multicolumn{1}{c|}{$\frac{121}{16}\sin^{-4}\left( \frac{\pi}{11}\right)$}
\\ 
\cline{1-4}
\multicolumn{1}{|c|}{$5$}      & \multicolumn{1}{c|}{$\phi^0_{(r,s)}$ and $\phi^1_{(r,s)}$ for $r=1,10$ and $s=1,7$}  & \multicolumn{1}{c|}{ $\mathbb{Z}_2\times su(2)^{\text{even}}_{9}$}  & \multicolumn{1}{c|}{$\frac{121}{4}\sin^{-4}\left( \frac{\pi}{11}\right)$}       \\ 
\cline{1-4}
\multicolumn{1}{|c|}{$6$}      & \multicolumn{1}{c|}{$\phi^0_{(1,s)}$ and $\phi^1_{(1,s)}$ for $s=1,7$}  & \multicolumn{1}{c|}{ $(A_{10},E_6)$}  & \multicolumn{1}{c|}{${121}\sin^{-4}\left( \frac{\pi}{11}\right)$}       \\ 
\cline{1-4}
\end{tabular}
\caption{All (sub)models of the $(A_{10},E_6)$ modular invariant that include non diagonal fields and respect parity symmetry with their tensor categories and global index.}
\label{tabla7}
\end{table}
Again, there are also eight models without parity symmetry that we list below
\begin{itemize}
    \item $\{\phi^0_{(r,1)},\phi^1_{(r,1)},\,r=1,2,3,4,5,6,7,9,10\}$ and $\{\phi^0_{(r,1)},\phi^1_{(r,7)},\,r=1,2,3,4,5,6,7,9,10\}$ 
    \item $\{\phi^0_{(r,1)},\phi^1_{(r,1)},\,r=1,3,5,7,9\}$ and $\{\phi^0_{(r,1)},\phi^1_{(r,7)},\,r=1,3,5,7,9\}$ 
    \item $\{\phi^0_{(r,1)},\phi^1_{(r,1)},\,s=1,10\}$ and $\{\phi^0_{(r,1)},\phi^1_{(r,7)},\,s=1,10\}$ 
    \item $\{\phi^0_{(1,1)},\phi^1_{(1,1)}\}$ and $\{\phi^0_{(1,1)},\phi^1_{(1,7)}\}$ 
\end{itemize}
The inclusion relations for the different models of $m=11,12$ is shown in figures \ref{es1} and \ref{es2} (non parity symmetric models not shown).

\section{Selection rules for RG flows \label{mrgflows}}

Having classified minimal models in this way, we can now derive selection rules for RG flows between them. 
The starting observation is the following. If we consider a flow that was triggered by a certain relevant scalar perturbation $\varphi$, then we can think the theory where the flow is taking place is the one generated by $\varphi$ and the stress tensor. Let us call the ``neutral'' algebra ${\cal N}$ to this model.\footnote{ The logic of this nomenclature parallels that of a QFT with a standard symmetry group $G$, perturbed by an operator which is invariant under the symmetry. In such case, the flow can be studied in the invariant or neutral algebra. In our case, this is by definition the algebra of stress tensor plus the perturbation, and the symmetry is by definition the set of topological operators leaving invariant this algebra.}
 This can be a non-modular invariant theory, i.e. one node in the trees described above that is not at the top. Then, even if correlation functions are altered, the whole algebraic structure above such a node acts just as a spectator. It cannot change with the scale.\footnote{ Again, as in the previous footnote, this parallels the situation of a QFT with a standard symmetry group $G$, perturbed by an operator which is invariant under the symmetry. In such case, the algebraic structure of charged operators under the symmetry cannot change with the scale.} This is understood as the idea of transportability of Haag duality violations, i.e., the fact that non local ``charged'' operators can be deformed and scaled by the action of the neutral algebra and remain non local operators in larger regions. This is just a generalization of a more familiar phenomenon, that an (exact) symmetry will remain so under changes of scale.     

 However, we note that even if these structures are preserved when changing the scale, it is possible they effectively disappear at the IR fix point. This is the case if some (or all) charged sectors become massive and are not represented any more at the IR fix point. These possible massive scenarios at the IR will give place to topological sectors.\footnote{We will comment more about this possibility in the discussion section below.} However, in this paper we will not deal with massive cases. We will assume the UV CFT is mapped completely to the IR CFT. That is, with more precision, no correlation function of an interpolating field corresponding to an UV field has exponentially decreasing correlators at large distances. So the full UV is ``visible'' from the IR. Under these conditions, there will be a preservation of sectors between the UV and IR fix points. In this case, we expect:

\begin{itemize}
   \item The global and relative Jones indices of the CFT trees described in the previous section that are at or above the perturbed node are preserved by the RG flow.
    \item   The structure of completions and their inclusions at or above the perturbed node is preserved by the RG flow. In particular, this also implies the identification of the charged classes of operators from the $UV$ to the $IR$.
    \item   The categories of DHR superselection sectors at or above the perturbed node are preserved by the RG flow.
\end{itemize}

However, the internal structure of the neutral algebra ${\cal N}$ will in general change between fix points. The internal structure or symmetries in ${\cal N}_{UV}$ is explicitly broken by the perturbation. New structures inside ${\cal N}_{IR}$ can be interpreted as emergent symmetries.

The strategy for analyzing selection rules is the following. For a given $m$ and considering any of the possible submodels given by the boxes in Figs. \ref{4n2}, \ref{4n1}, \ref{4n0}, \ref{4n3}, \ref{es1} or \ref{es2}, we look for relevant scalar fields that belong exclusively to the chosen box that we call ${\cal N}_{UV}$. This means the scalar belongs to ${\cal N}_{UV}$ but not to any of the smaller included models. Then, turning on the perturbation for this scalar field we know that this model has an intrinsic RG flow to another one ${\cal N}_{IR}$ at the IR. Because of the c-theorem this IR model has smaller $m$. We search for a model with smaller $m$ having the same category and index. The categories and possible (partial) completions of the model should also match. This means that if there are more than one completions each complete theory (and submodels above ${\cal N}$) will run to a corresponding completion.       

We consider the following facts for generic $m$, that is, not considering the exceptional cases.\footnote{We have not consider fermionic completions, but this analysis would not add to the present investigation. The perturbation is a scalar and the algebra generated cannot contain fermion fields. The possible RG will then be contemplated in the cases analyzed below. } To understand the models at the UV we have to count the exclusive relevant scalars for each of the submodels. We first look at the submodels of the diagonal completion. We name the submodels of the diagonal series with the row number in the table \ref{tabla1} of section \ref{m6}.  We have:

\vspace{-0.4cm}
\begin{itemize}

\item \underline{Model 1:}  (the complete $(A_{m-1},A_m)$ model) contains $m-3$ exclusive relevant scalars.

\item  \underline{Model 2:} ($\mathbb{Z}_2$ invariant part) contains $m-4$ exclusive relevant scalars. 

\item \underline{Model 3:}  (algebra of $(1,s)$ for all $s$) contains just two relevant scalars, $(1,3)$ and $(1,2)$, and only the latter is exclusive. 

\item \underline{Model 4:}   (algebra of $(1,s)$ with odd $s$) contains a unique relevant scalar, the field $(1,3)$, that is exclusive of this model.

\item \underline{Model 5:}  (algebra $(r,1)$ for all $r$) has a unique relevant exclusive scalar, the field $(2,1)$.

\item \underline{Model 6:}  (algebra $(r,1)$ with odd $r$) does not contain relevant operators.

\item\underline{Model 7:} (the identity plus $(m-1,1)$) does not contain relevant scalars. 

\item\underline{Model 8:}  (the stress tensor) does not contain relevant scalars. 
\end{itemize}

In addition, for the non diagonal models with $m>5$ only the complete models contain exclusive relevant scalars.\footnote{These are:  $(D,A)$ $m=4n+2$: $(m/2,s) \times (m/2,s)$ for $s=m/2-1,m/2$, $(A,D)$ $m=4n+1$: $(r,(m+1)/2) \times (r,(m+1)/2)$ for  $r=(m+1)/2-1,(m+1)/2$, $(D,A)$ $m=4n$: $(m/2,s) \times (m/2,s)$ for $s=m/2-1,m/2$, $(A,D)$ $m=4n-1$: $(r,(m+1)/2) \times (r,(m+1)/2)$ for  $r=(m+1)/2+1,(m+1)/2$.} Therefore, with the exception of the complete non diagonal models,  models drawn in red or black in figures \ref{4n2}, \ref{4n1}, \ref{4n0} or \ref{4n3} do not contain exclusive relevant scalars. 

Therefore for the minimal ${\cal N}_{UV}$ of generic $m$ we restrict attention to complete models, or the submodels 2, 3, 4 and 5 of the table \ref{tabla1} of section \ref{m6}. Then, we match the category (and index) between ${\cal N}_{UV}$ and ${\cal N}_{IR}$  concluding the following.
\vspace{-0.3cm}
\begin{enumerate}
\item[\textbf{a.}] Perturbations with exclusive scalars of complete models may in principle trigger an RG to any complete model of the lower values of $m$. In this cases, the perturbation is charged with respect to all of the symmetries and the category of superselection sectors is trivial. The power of symmetries to constrain the RG flows is very limited in this cases. The same can be said for perturbations exclusive of model 2 with category $\mathbb{Z}_2$. This same category can be found for any smaller value of $m$ and the RG could end in model 2 of any of the smaller central charges.

\item[\textbf{b.}] Model 3 has category $su(2)_{m-2}^{\text{even}}$. This is only matched (for smaller $m$) by the model 5 of table \ref{tabla1} (diagonal), or the non diagonal models 3 of table \ref{tabla2} when $m_{UV}-1=4 n+2$, or 2 of table \ref{tabla4} when $m_{UV}-1=4 n$. All possibilities correspond to $m_{IR}=m_{UV}-1$, and the possible non diagonal IR models occur only for $m_{UV}$ odd. Then, there can be a flow from $m$ to $m-1$ started by the field $(1,2)$.

\item[\textbf{c.}] Model 4 has category $su(2)_{m-2}$ and this is only matched (for smaller $m$) by the model 6 of table \ref{tabla1} (diagonal), corresponding to $m_{IR}=m_{UV}-1$. Then, there can be a flow from $m$ to $m-1$ started by the field $(1,3)$. 

\item[\textbf{d.}] Model 5 has category $su(2)_{m-1}$ and this is not matched by any model for smaller $m$. Therefore, a flow started by the field $(2,1)$ must have some massive sector.
\end{enumerate}
\vspace{-0.3cm}
Regarding previous literature, as far as we know, the results of the items (b) and (d) above for general $m$ are new. On the other hand, the flow described by item (c) was previously studied in the literature \cite{Zamolodchikov1987,GaiottoDomain,NonDiagonalRavanini,NonDiagonalKlassen}, and here we highlight that is greatly constrained by symmetry reasons.  

Let us then consider the cases of items (b) and (c) in more detail. Both of them correspond to a change between $m_{UV}=m+1$ and $m_{IR}=m$, and a change of central charge 
\be
c_{UV}-c_{IR}=\frac{12}{m(m+1)(m+2)}\,.
\ee

In the case of item (c), which we will call $su(2)_{m-1}$ symmetric or Zamolodchikov's flow, for each value of $m$ we have an allowed RG flow preserving the category of superselection sectors
\be \label{1fz}
su(2)_{m_{UV}-2}=su(2)_{m_{IR}-1}=su(2)_{m-1}\,.
\ee
In this flow, the global index is preserved and takes the value
\be
\mu_{UV}=\mu_{IR}=\hat{\mu}(m+1)\,.
\ee
This  RG flow fully happens minimally within the algebras generated by
\be 
\left\{ \varphi^0_{(1,s)}\,/\,s\,\, odd\right\}_{UV} \,\, \to \,\, \left\{\varphi^0_{(r,1)}\,/\,r\,\, odd\right\}_{IR} \;. \label{rg2}
\ee
 Interestingly, the only relevant operator included in this $UV$ model (that is also the least relevant scalar for the given central charge) is $\varphi^0_{(1,3)}$ with 
\be 
h_{1,3}+\overline{h}_{1,3}= 2\left(\frac{m_{UV}-1}{m_{UV}+1}\right)= 2\left(\frac{m}{m+2}\right)\,.
\ee
As stated before, this result is consistent with the literature. Originally it was proven by Zamolodchikov \cite{Zamolodchikov1987} for large $m$ that the flow from $m+1$ to $m$ can be achieved by perturbing the theory with the field of Kac label $(1,3)$ in the ${UV}$. This was done via a calculation of the beta function and finding the proper $IR$ fixed point, using conformal perturbation theory. The anomalous dimension in the fixed point of the field corresponding to $(1,3)$ in the $UV$ is consistent with a flow to the $(3,1)$ in the $IR$. The selection rule is valid for any $m$ though. 

\begin{figure}[H]
    \centering
    \includegraphics[width=1\textwidth]{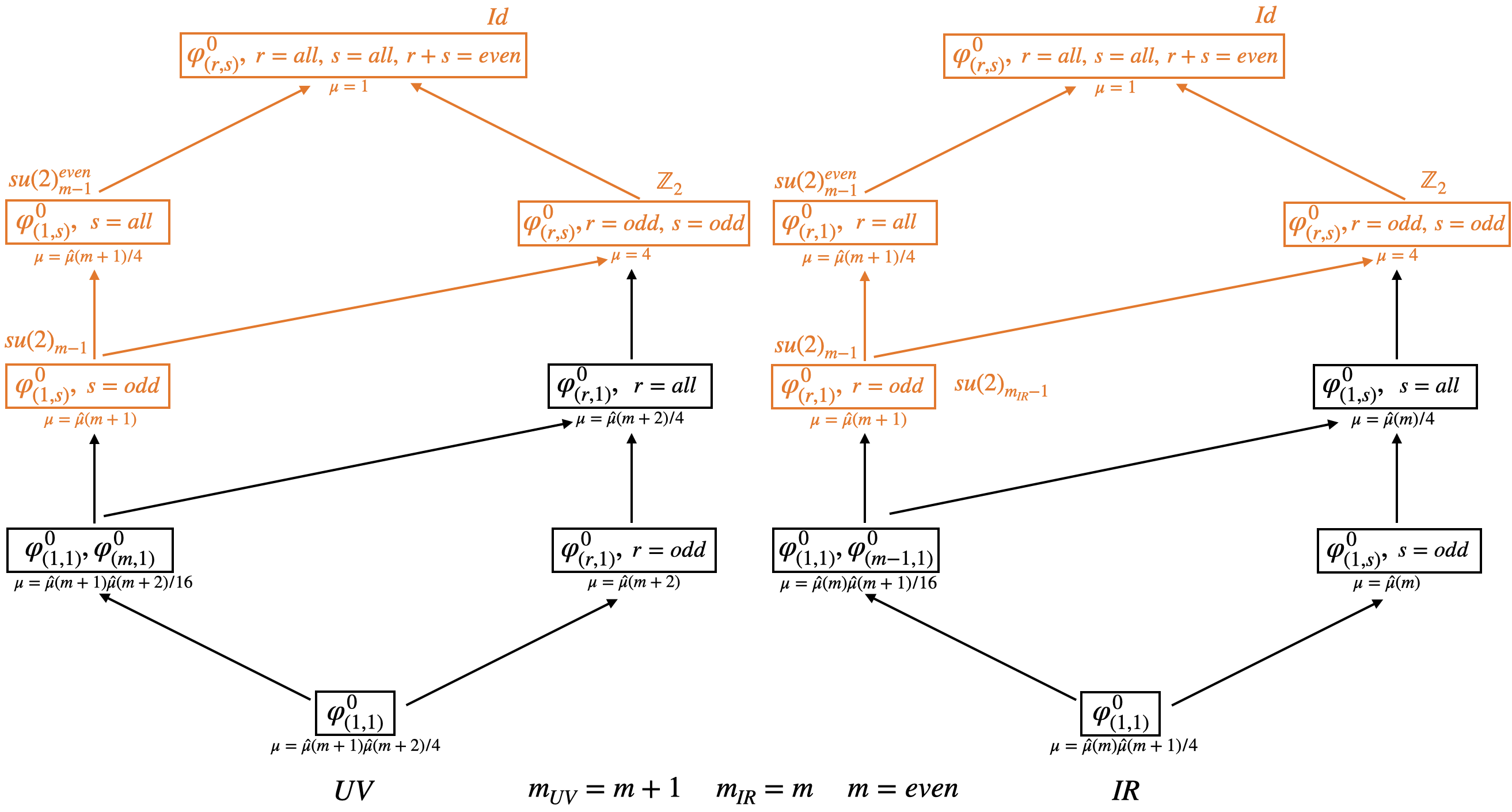}
    \caption{RG for $su(2)_{m-1}$ symmetric flows between $m_{UV}=m+1$ and $m_{IR}=m$ for $m$ even. The orange boxes highlight all possible diagonal extensions of the model containing the perturbation  with their corresponding superselection sectors tensor categories. This structure is preserved on the orange boxes corresponding to the IR diagram. }\label{frg1}
\end{figure}

\begin{figure}[H]
    \centering
    \includegraphics[width=1\textwidth]{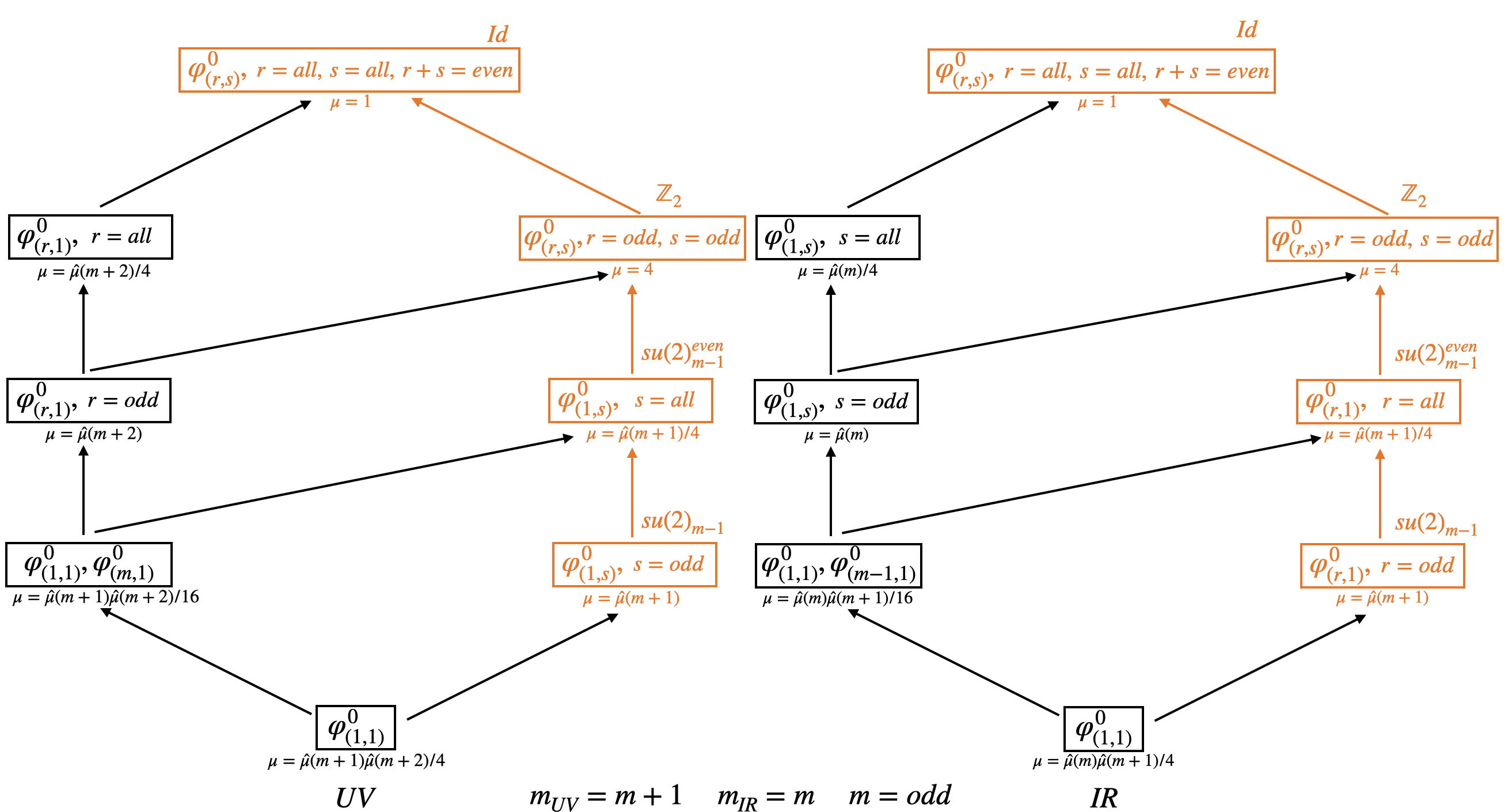}
    \caption{RG for $su(2)_{m-1}$ symmetric flows between $m_{UV}=m+1$ and $m_{IR}=m$ for $m$ odd. The orange boxes highlight all possible diagonal extensions of the model containing the perturbation  with their corresponding superselection sectors tensor categories. This structure is preserved on the orange boxes corresponding to the IR diagram.}\label{frg2}
\end{figure}

We depict these flows in Figure \ref{frg1} for $m$ even and Figure \ref{frg2} for $m$ odd inside the diagrams of the diagonal completions. Note how the whole structure of possible extension of the models with associated global indices and superselection sectors categories is preserved, even taking into account that there are important differences for different parity of $m$. 

The RG flow of item (b) is one preserving the category
\be 
su(2)_{m_{UV}-2}^{\text{even}}=su(2)_{m_{IR}-1}^{\text{even}}=su(2)_{m-1}^{\text{even}}\,. \label{cat1}
\ee
We can call this a $su(2)_{m-1}^{\text{even}}$ symmetric flow. The associated global index is
\be
\mu_{UV}=\mu_{IR}=\hat{\mu}(m+1)/4\,.
\ee
The mapping of the algebras, in the diagonal case, in this case is enlarged to include 
\be 
\left\{ \varphi^0_{(1,s)}\,/\,for\,\, all \,\,s \right\}_{UV} \,\, \to \,\, \left\{\varphi^0_{(r,1)}\,/\,for\,\, all \,\, r\right\}_{IR}  \;.\label{rg2}
\ee
For odd $m_{UV}$ it could also be the case that the IR algebra is a non diagonal subalgebra.  

Note that in this case, we have included a new relevant operator in the UV $\varphi^0_{(1,2)}$, that triggers the flow and has scaling dimension
\be 
h_{1,2}+\overline{h}_{1,2}= \frac{1}{2}\left(\frac{m_{UV}-2}{m_{UV}+1}\right)= \frac{1}{2}\left(\frac{m-1}{m+2}\right)\,.
\ee
This does not allow a perturbative evaluation for large $m$. This field is charged under a $\mathbb{Z}_2$ symmetry contained in the symmetry that preserved the algebra studied above. Consequently, it breaks that symmetry to a smaller one, and defines a new possible flow. 

Going back to the literature, a case studied in much detail is the flow between tricritical Ising and Ising model \cite{ZamolodchikovBethe1,ZamolodchikovBethe2,CardyTricritical,DelfinoTricritical,CalabreseTricritical}, using a large variety of methods. In particular Ref. \cite{CardyTricritical} established what happens when we perturb the tricritical Ising with each of the relevant operators included in the model. The perturbations with $\varepsilon$ or $\varepsilon'$, which correspond respectively to $(1,3)$ and $(1,2)$, with appropriate boundary conditions (and signs), trigger a flow to the Ising model. This coincides with our results in Figures \ref{43} and \ref{m4fig}. Note that the Ising model does not allow non diagonal completions, and therefore there is no ambiguity in our results for the tricritical Ising perturbed by $(1,2)$.  The perturbations with the $(2,1)$ field  $\sigma'$ give rise to a gapped phase. This is exhibited in our results considering that the subalgebras containing the $(2,1)$ field has $\mu =(5+\sqrt{5})^2/4$ which is not present in the Ising model. Also, this is in accordance with the general result of item (d). Equivalently, there is no allowed CFT to land on when we perturb with $\sigma'$.

\newpage
The analysis of Ref. \cite{Zamolodchikov1987} also proved that, in Zamolodchikov's flow,  to leading order in $m$, the fields in the UV of the form $(r,r\pm 1)$ are mixed with the IR fields $(r\pm 1,r)$ as
\be 
\left[\varphi^0_{(r-1,r)}\right]_{IR}=\sqrt{r^2-1}\, \left[\varphi^0_{(r,r+1)}\right]_{UV}- \left[\varphi^0_{(r,r-1)}\right]_{UV} \,, \label{map1}
\ee
\be 
\left[\varphi^0_{(r+1,r)}\right]_{IR}=\left[\varphi^0_{(r,r+1)}\right]_{UV}+ \sqrt{r^2-1}   \,\left[\varphi^0_{(r,r-1)}\right]_{UV}\,, \label{map2}
\ee
with the next orders given in \cite{GaiottoDomain,PoghossianNexto}. Analogously, there is a more complicated mixing (including derivatives) between the fields $(r,r\pm 2)$ and $(r,r)$ in the UV and $(r,r\pm 2)$ and $(r,r)$ in the IR. The question here is whether we can arrive at some understanding of these mappings by using generalized symmetry arguments, as developed above. 

Let's then take the standard case in which we perturb a diagonal model with the $\varphi^0_{(1,3)}$ field. The category of preserved superselection sectors is $su(2)_{m-1}$. This has generators $J_i$ for $i=1,2,\dots,m-1$, and divide the QFT into charged classes (multiplets of the symmetry) which do not change as we move from the $UV$ to the $IR$. These classes can be identified by comparing the fusion rules of $su(2)_{m-1}$ (see appendix \ref{su2}) and the fusion rules of the $m$ and $m+1$ minimal models. In the $UV$ this identification reads
\be 
J_{r-1}\equiv \Big\{\varphi^0_{(r,s)}\,:\, \forall s\, / \,r+s=even\Big\}\,,\quad r=1,2,\dots,m\;.
\ee
while in the $IR$  it can be understood in terms of fields as
\be 
J_{s-1}\equiv \Big\{\varphi^0_{(r,s)}\,:\, \forall r\, / \,r+s=even\Big\}\,,\quad s=1,2,\dots,m\;.
\ee
The non-trivial RG flow then implies there is a mapping of fields deduced from the mapping of sectors as
\be 
\Big\{\varphi^0_{(r,s)}\,:\, \forall s\, / \,s=odd\Big\}_{UV} \longleftrightarrow \Big\{\varphi^0_{(s,r)}\,:\, \forall s\, / \,s=even\Big\}_{IR} \;,\label{mapus1}
\ee
\be 
\Big\{\varphi^0_{(r,s)}\,:\, \forall s\, / \,s=odd\Big\}_{UV} \longleftrightarrow \Big\{\varphi^0_{(s,r)}\,:\, \forall s\, / \,s=odd\Big\}_{IR} \label{mapus2}\;.
\ee
If we only focus on the lowest scaling fields we recover 
\be 
\left\{\varphi^0_{(r,r+1)}\,,\varphi^0_{(r,r-1)}\right\}_{UV} \to\left\{\varphi^0_{(r+1,r)}\,,\varphi^0_{(r-1,r)}\right\}_{IR} \;,
\ee
together with
\be 
\left\{\varphi^0_{(r,r+2)}\,,\varphi^0_{(r,r)}\,,\varphi^0_{(r,r-2)}\right\}_{UV} \to\left\{\varphi^0_{(r+2,r)}\,,\varphi^0_{(r,r)}\,,\varphi^0_{(r-2,r)}\right\}_{IR}\;.
\ee
This is consistent with (\ref{map1}) and (\ref{map2}) and the other mappings in \cite{Zamolodchikov1987,GaiottoDomain,PoghossianNexto}. One can also recover similar mappings from the RG flow preserving the $su(2)_{m-1}^{\text{even}}$  or the $\mathbb{Z}_2$ categories. For instance, for the latter, we get that the two generators are represented as
\be 
J_{\text{neutral}}\equiv \Big\{\varphi^0_{(r,s)}\,:\, \forall r,s\, / \,r=odd \,\,\wedge s=odd \Big\}\;,
\ee
\be 
J_{\text{charged}}\equiv \Big\{\varphi^0_{(r,s)}\,:\, \forall r,s\, / \,r=even \,\,\wedge s=even \Big\}\;,
\ee
which is just saying that charged fields flow to charged fields. This statement is also included in (\ref{mapus1}) and (\ref{mapus2}).

\subsection{RG flows including \texorpdfstring{$(A,D)$}{Lg} and \texorpdfstring{$(D,A)$}{Lg} series models \label{rgad} }
Now, we proceed to describe the RG flow for the partial completions of the minimal flow algebra included in the allowed $(A,A)$, $(D,A)$, and $(A,D)$ series. More precisely, we re-analyze the flows between $m_{UV}=m+1$ and $m_{IR}=m$ dividing them into four possible scenarios. We remark that all these flows as just different partial completions of the same RG flow of the core $\mathcal{N}$ algebra. 

\begin{itemize}
\item \underline{$m_{UV}=4n+2$ and $m_{IR}=4n+1$:} This includes the flows of all possible minimal models that can be completed to $(A_{4n+1},A_{4n+2})$ and $(D_{2n+2},A_{4n+2})$ in the $UV$, to the ones that can be completed to $(A_{4n},A_{4n+1})$ and $(A_{4n},D_{2n+2})$ in the $IR$. This includes the diagonal ones mentioned in the previous section as well as e.g. an extra RG flow preserving the superselection sector tensor category $D^{\text{even}}_{2n+2}$ and the flow between both complete non-diagonal models.
\item \underline{$m_{UV}= 4n+1$ and $m_{IR}=4n$:} This includes the flows of all possible minimal models that can be completed to $(A_{4n},A_{4n+1})$ and $(A_{4n},D_{2n+2})$ in the $UV$, to the ones that can be completed to  $(A_{4n-1},A_{4n})$ and $(D_{2n+1},A_{4n})$ in the $IR$. In particular, this includes two extra  RG flows between non-diagonal minimal models, one preserving the category  $su(2)^{\text{even}}_{m-1}$ and one between complete models.
\item \underline{$m_{UV}=4n$ and $m_{IR}=4n-1$:} This includes the flows of all possible minimal models that can be completed to $(A_{4n-1},A_{4n})$ and $(D_{2n+1},A_{4n})$ in the $UV$, to the ones that can be completed to  $(A_{4n-2},A_{4n-1})$ and $(A_{4n-2},D_{2n+1})$ in the $IR$. In comparison with the previous section, the only extra information here is the RG flow between complete models.
\item \underline{$m_{UV}=4n-1$ and $m_{IR}=4n-2$:} This includes the flows of all possible minimal models that can be completed to $(A_{4n-2},A_{4n-1})$ and $(A_{4n-2},D_{2n+1})$  in the $UV$, to the ones that can be completed to  $(A_{4n-3},A_{4n-2})$ and $(D_{2n},A_{4n-2})$ in the $IR$. This e.g. again involves an extra non diagonal $su(2)^{\text{even}}_{m-1}$ and the RG flow between complete models.
\end{itemize}

\begin{figure}[H]
    \centering
    \includegraphics[width=1.1\textwidth]{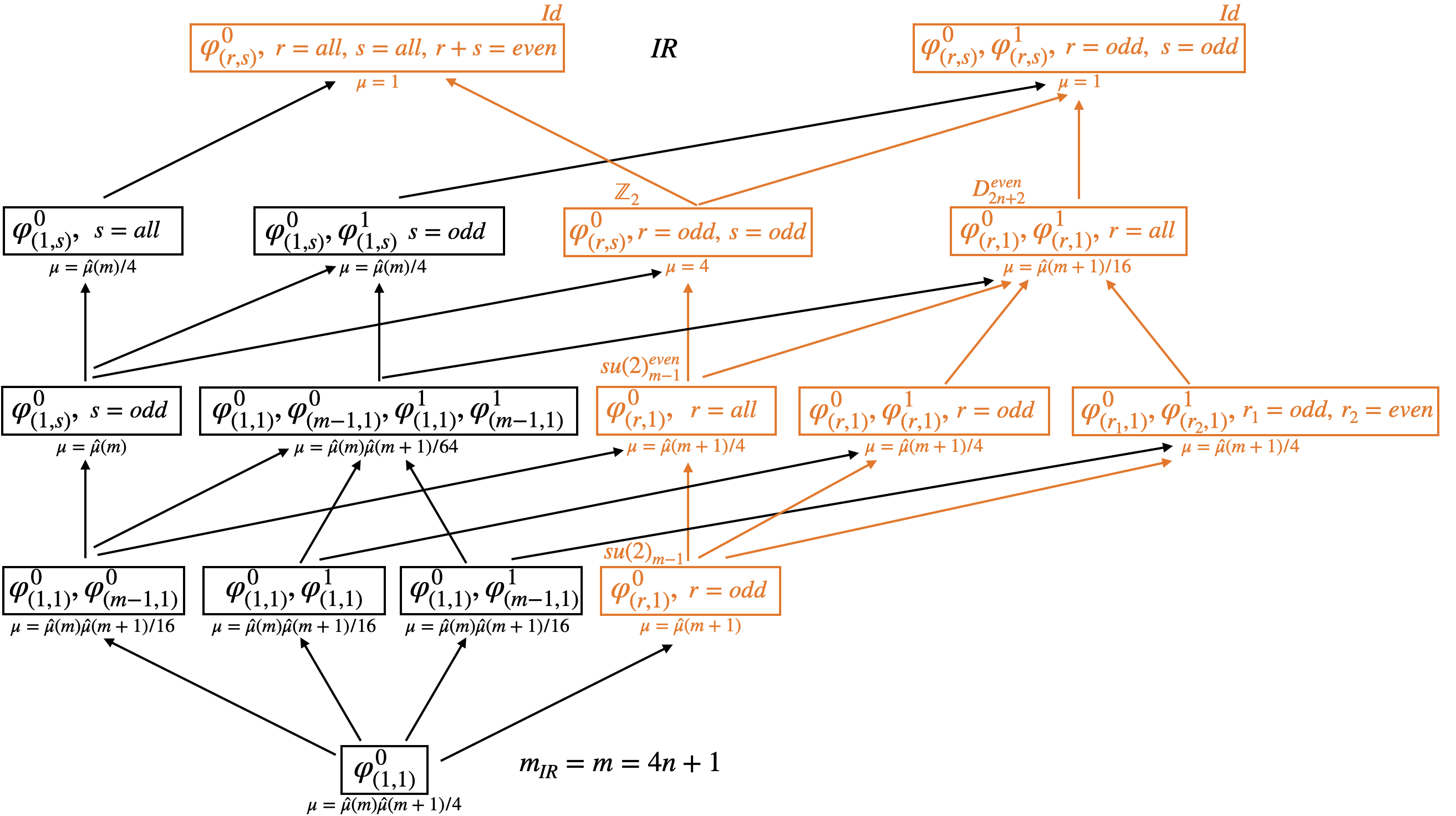}
    \caption{RG flow structure between $m_{UV}=4n+2$ (up) and $m_{IR}=4n+1$ (down). The orange boxes highlight all possible completions of the model of the highest global index involved in the RG flow with their corresponding superselection sectors tensor categories. }\label{rg34}
\end{figure}

\begin{figure}[H]
    \centering
    \includegraphics[width=1.1\textwidth]{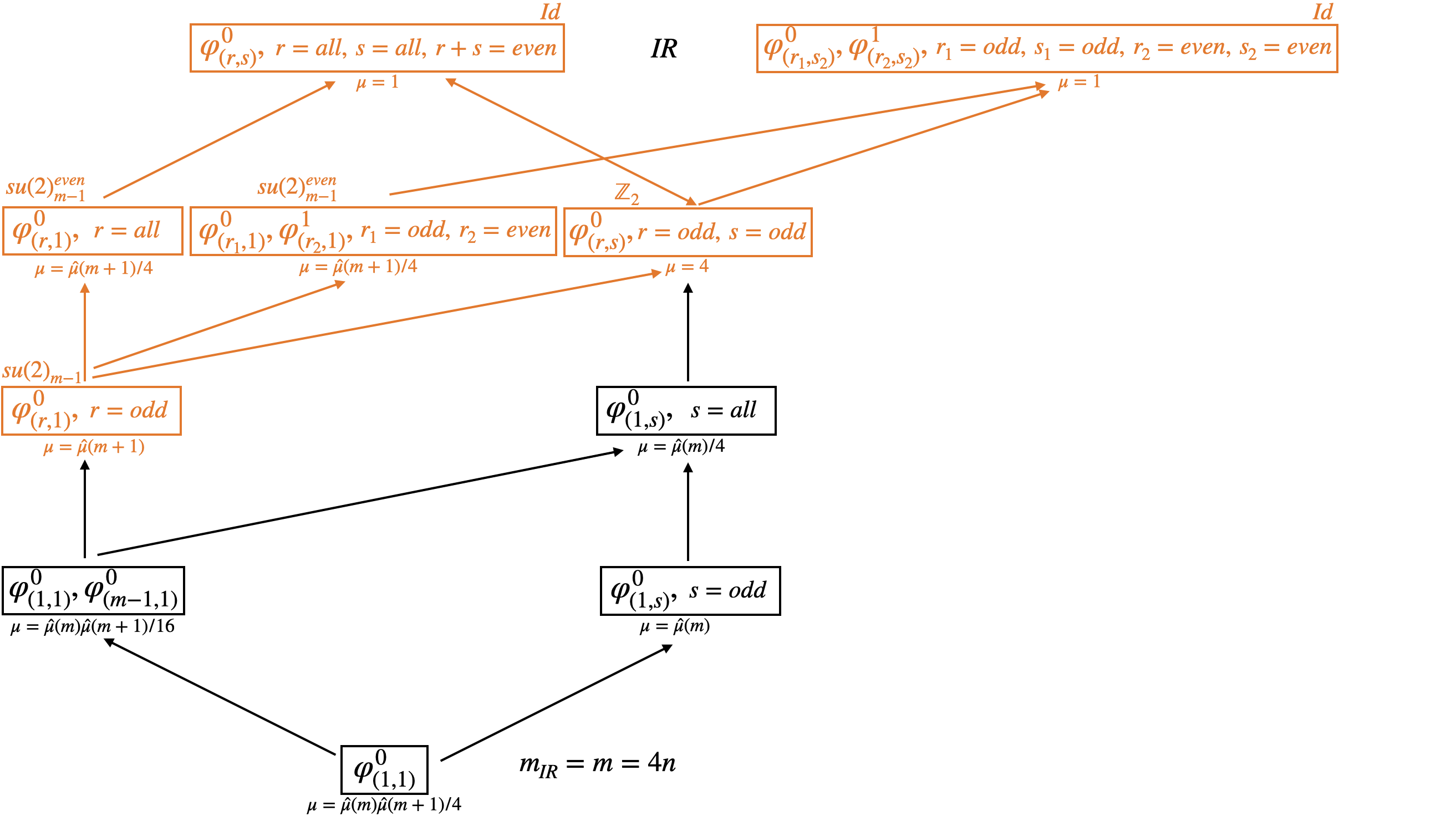}
    \caption{RG flow structure between $m_{UV}=4n+1$ (up) and $m_{IR}=4n$ (down). The orange boxes highlight all possible completions of the model of the highest global index involved in the RG flow with their corresponding superselection sectors tensor categories.} \label{rg56}
\end{figure}

\begin{figure}[H]
    \centering
    \includegraphics[width=1.1\textwidth]{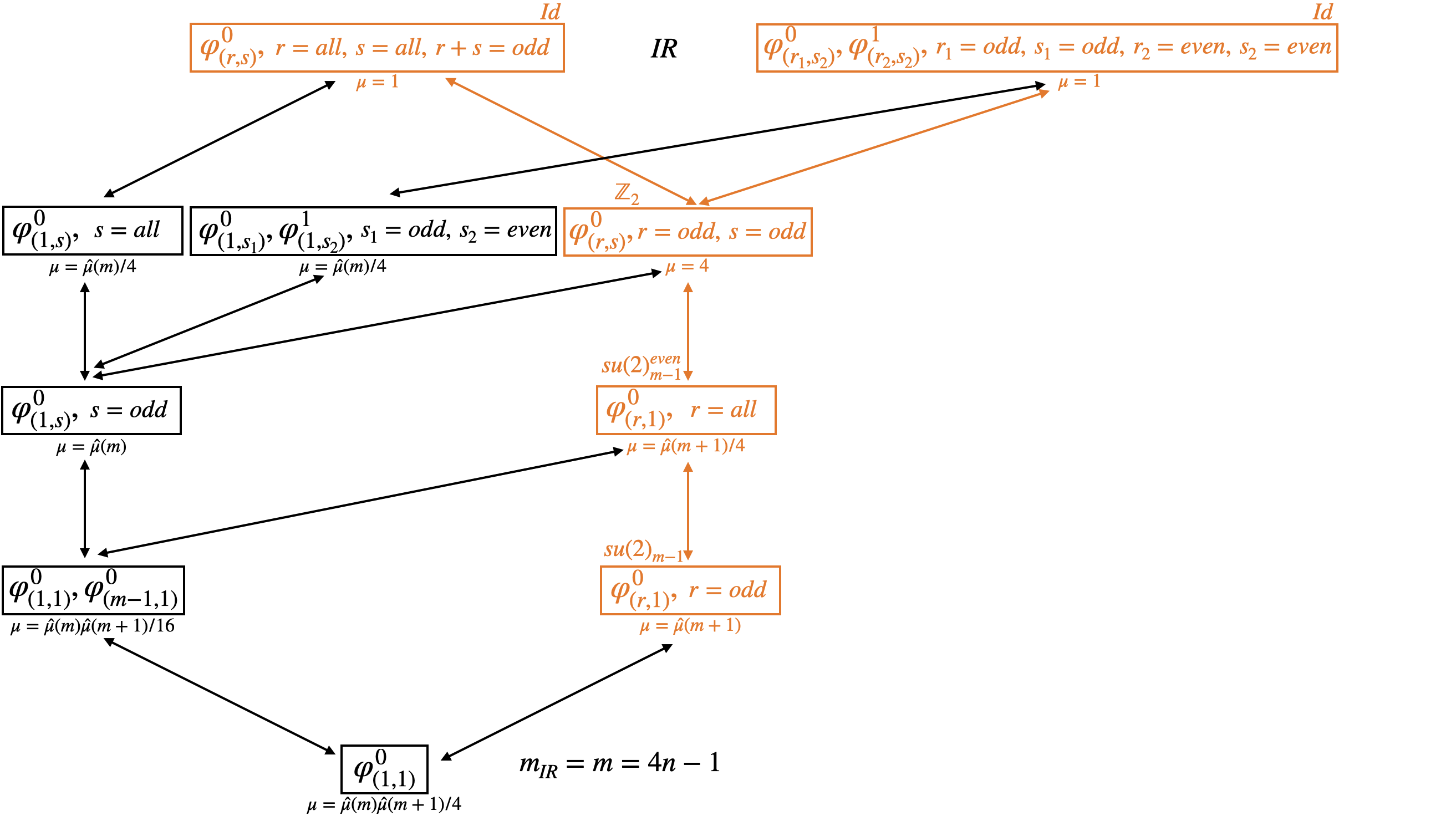}
    \caption{RG flow structure between $m_{UV}=4n$ (up) and $m_{IR}=4n-1$ (down). The orange boxes highlight all possible completions of the model of the highest global index involved in the RG flow with their corresponding superselection sectors tensor categories.}\label{rg78}
\end{figure}

\begin{figure}[H]
    \centering
    \includegraphics[width=1.1\textwidth]{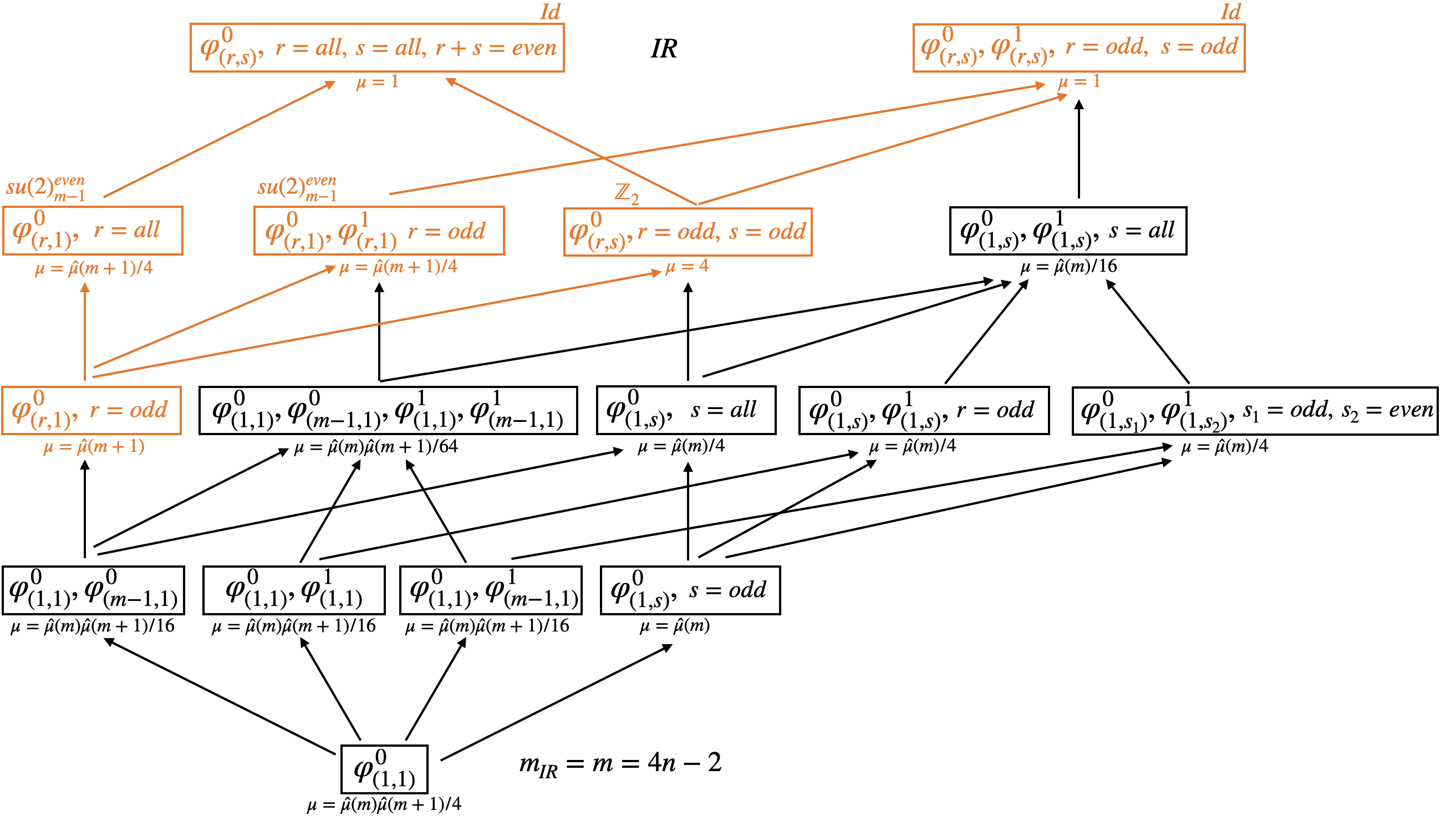}
    \caption{RG flow structure between $m_{UV}=4n-1$ (up) and $m_{IR}=4n-2$ (down). The orange boxes highlight all possible completions of the model of the highest global index involved in the RG flow with their corresponding superselection sectors tensor categories.}\label{rg910}
\end{figure}

Because the non diagonal submodels for $m>5$ do not include new exclusive relevant operators, we should think of these flows as extensions of the $su(2)_{m-1}$ and $su(2)_{m-1}^{\text{even}}$ flows triggered by the diagonal fields of kac label $(1,2)$ and $(1,3)$. Following this line, the Zamolodchikov flows are shown in figures \ref{rg34}, \ref{rg56}, \ref{rg78}, and \ref{rg910}. 
In all cases, we see that the structure of allowed completions and their categories is 
preserved. In particular, non diagonal models can only flow to non diagonal ones. This is consistent with the literature  \cite{NonDiagonalRavanini,NonDiagonalKlassen}.
The flows of symmetry $su(2)_{m-1}^{\text{even}}$ can be looked at from the same figures by going one step up in the tree in the UV diagram. Note that only for $m_{IR}$ even there are two $su(2)_{m-1}^{\text{even}}$ symmetric models, one of them is non diagonal. For both of these models the structure of the tree above only consists of the complete model.  So in this case the $su(2)_{m-1}^{\text{even}}$ symmetric flow could in principle change from a diagonal to a non diagonal model.

For $m=5$ we have to check the non diagonal subalgebras of the modular invariant $(A_4,D_4)$, the three states Potts model. These are depicted in Figure \ref{m5fig}. In this case, these seem to include a new exclusive relevant operator, because we are simply duplicating the field $(1,3)$. This is in $Z_1$ and $Z_2$. This starts a flow defined by the tensor category $su(2)_3^{\text{even}}$ that can have an $IR$ fixed point with the same category for $m=4$. This is one of the subalgebras of the tricritical Ising model with $\mu=(5+\sqrt{5}^2)/4$ depicted in Figure \ref{m4fig}. This overlaps with the usual diagonal  Zamolodchikov flow because for $m=4$ we have $(D_3,A_4)=(A_3,A_4)$. This result is expected from the literature \cite{NonDiagonalRavanini}.

 \begin{figure}[H]
    \centering
    \includegraphics[width=1.1\textwidth]{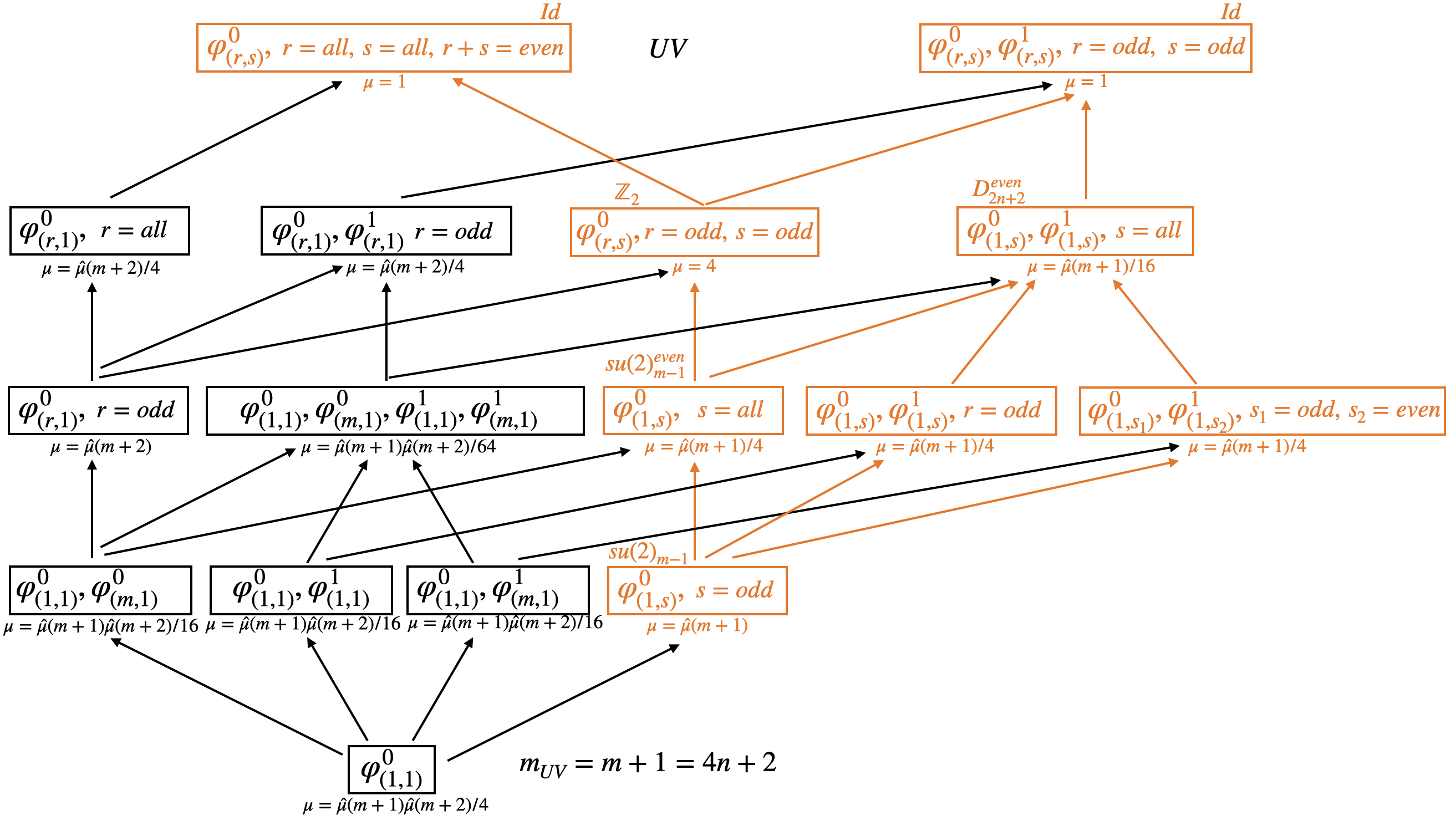}
\end{figure}

 \begin{figure}[H]
    \centering
    \includegraphics[width=1.1\textwidth]{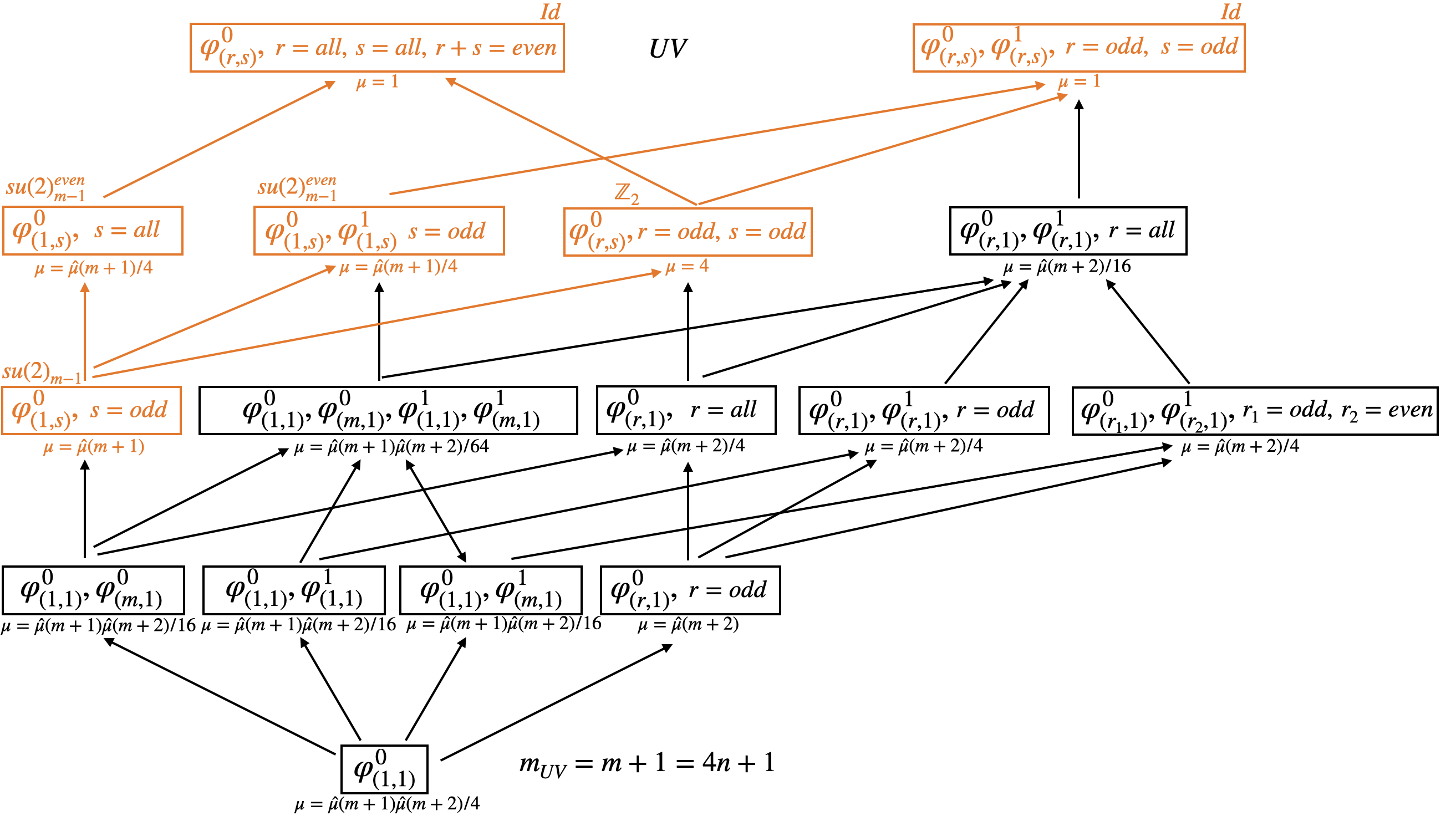}
\end{figure}

 \begin{figure}[H]
    \centering
    \includegraphics[width=1.1\textwidth]{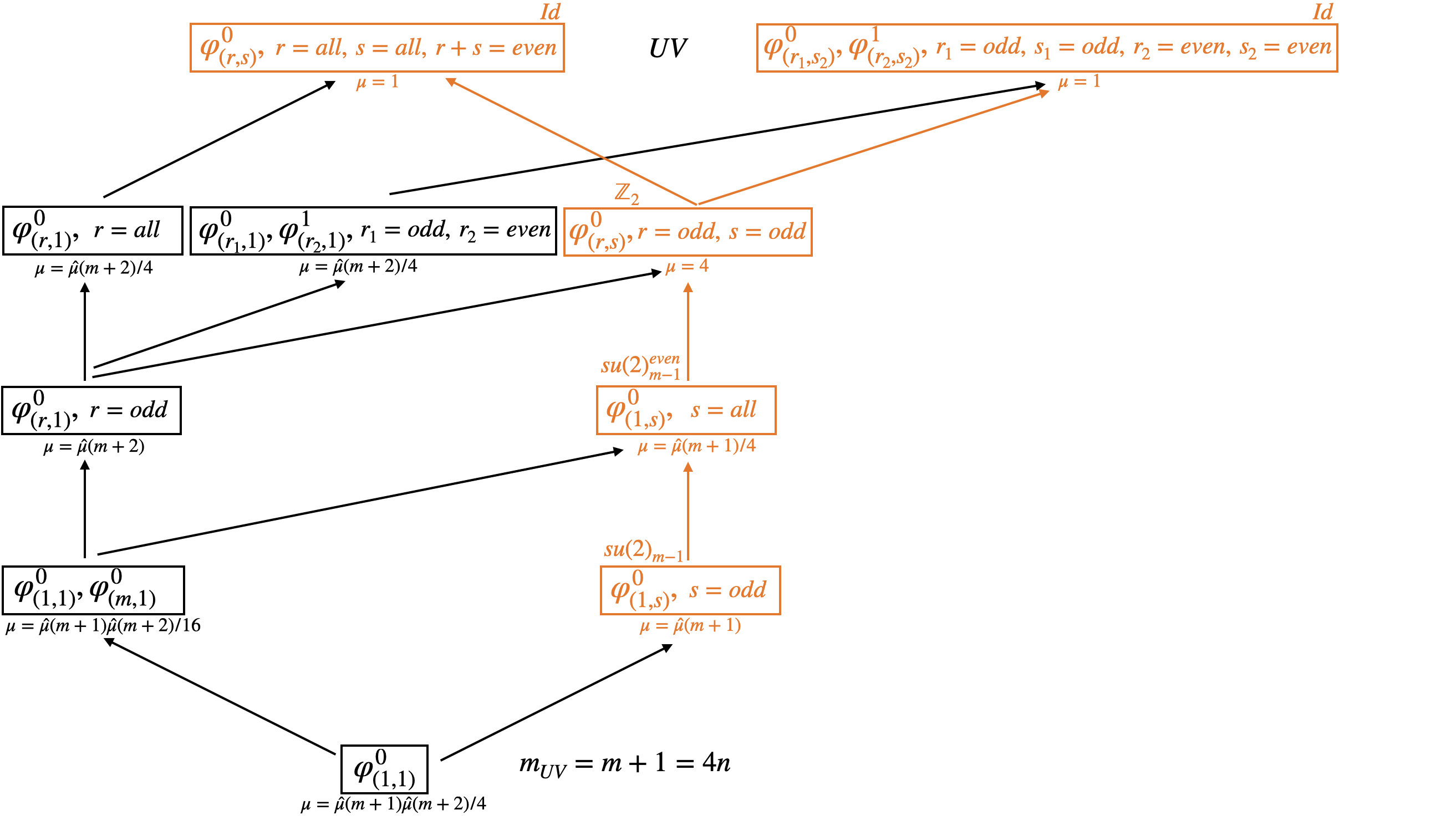}
\end{figure}

 \begin{figure}[H]
    \centering
    \includegraphics[width=1.1\textwidth]{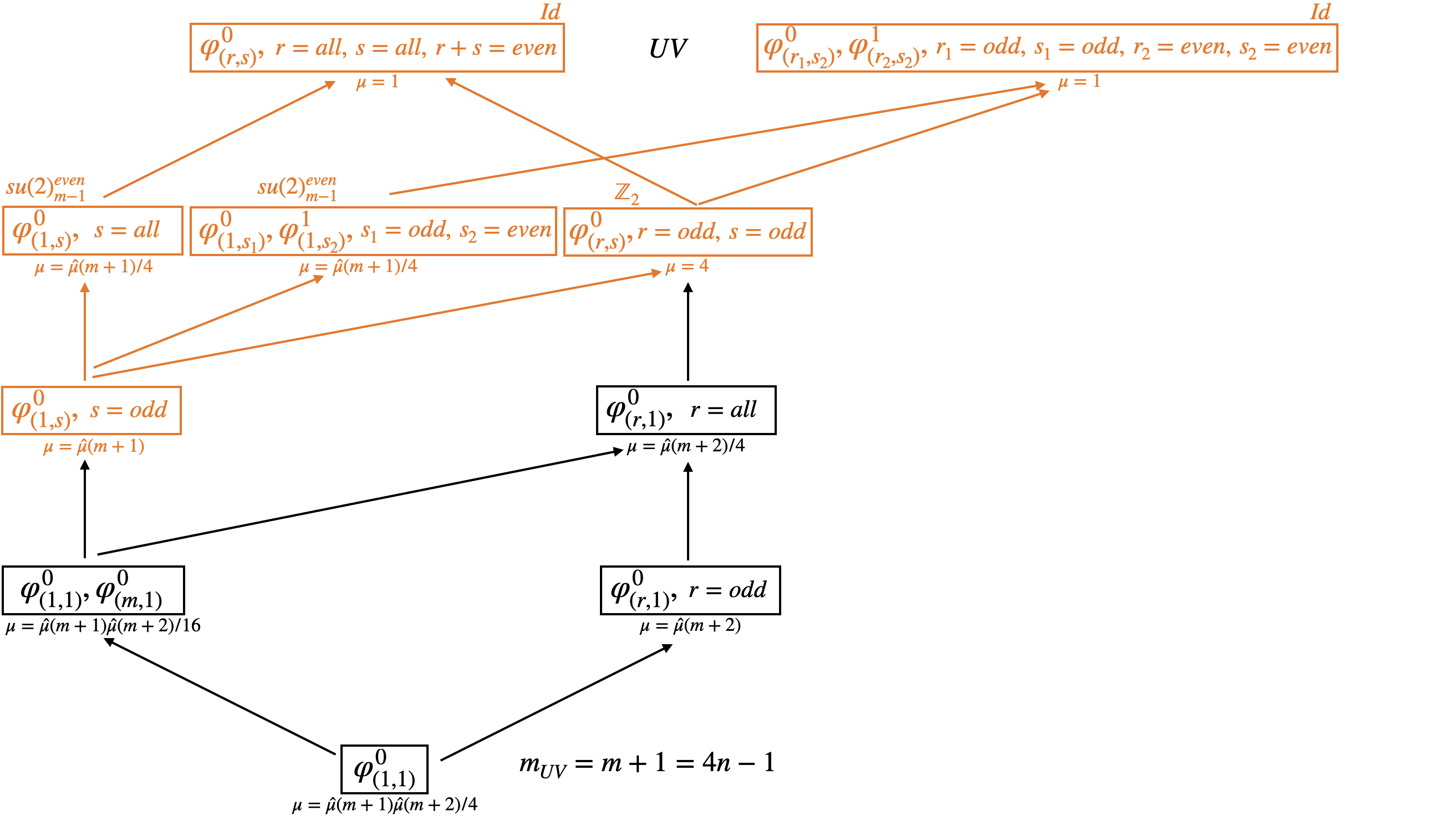}
\end{figure}

\newpage 
\subsection{RG flows including the \texorpdfstring{$(E_6,A_{12})$}{Lg} and \texorpdfstring{$(A_{10},E_6)$}{Lg} models}
Finally, we briefly explain the situation for the additional models of type $E$ for $m=11,12$. There are several new exclusive relevant scalars for the new complete and $\mathbb{Z}_2$ models. As above, for these models not much information about selection rules for the RG can be derived since these trivial categories appear for any $m$. In addition, there are new relevant exclusive scalars for the category $SU(2)_2$. These are the fields $(7,7)$ and $(7,9)$ for $m=12$, and $(5,7)$ and $(7,7)$ for $m=11$. This implies that there could be an RG flow preserving this category from $m=12$ to $m=11$. It is also possible that both $m=11,12$ could decay to the Ising model since the stress tensor for the Ising model has category $SU(2)_2$ and the same structure of completations. The same is true for the alegbra labeled by the category $SU(2)_2$ in the tricrital Ising model\footnote{See \cite{NakayamaGross} for a similar discusion in terms of topological defect lines.}

\begin{figure}[H]
    \centering
    \includegraphics[width=1.1\textwidth]{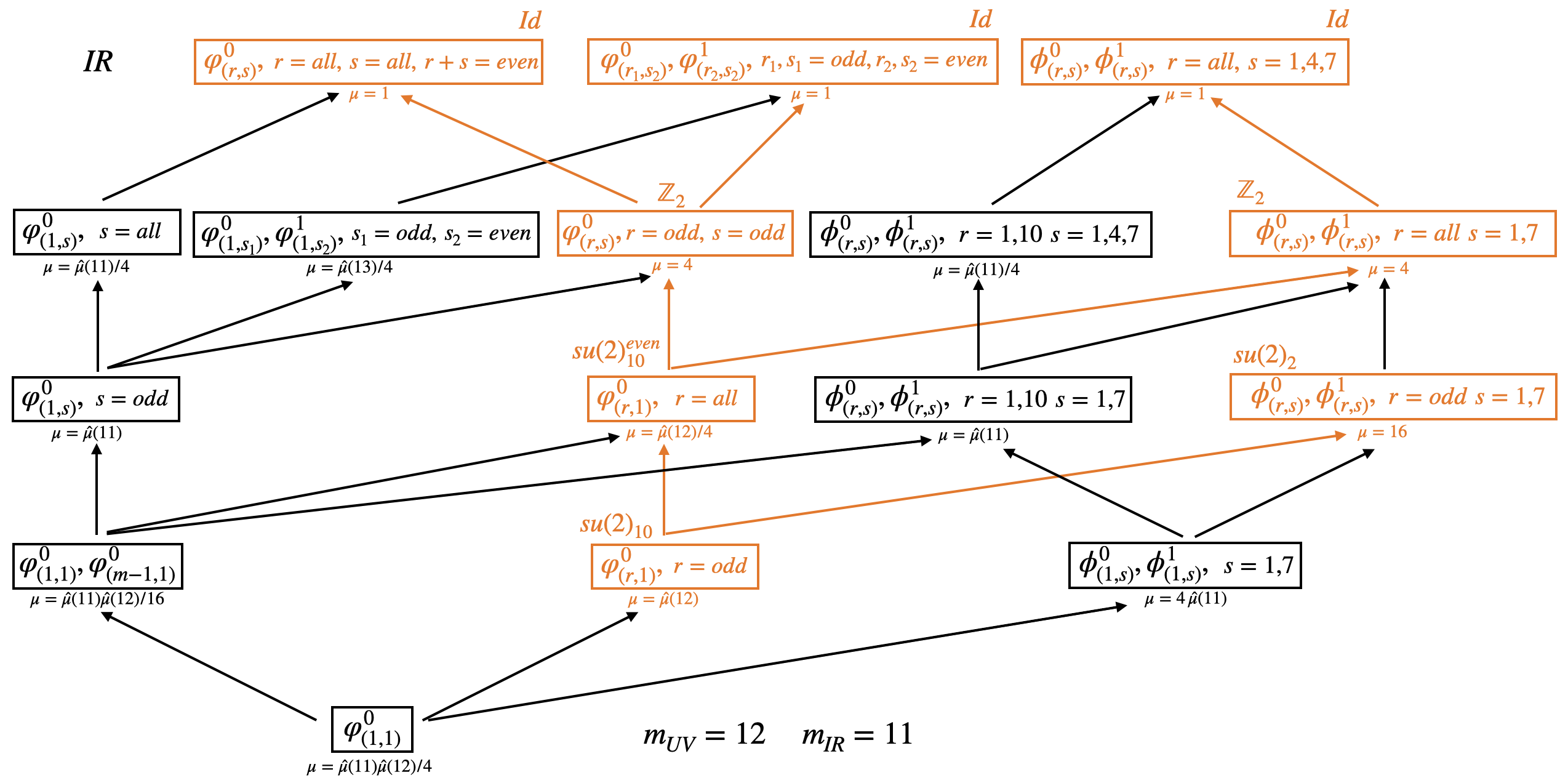}
    \caption{Zamolodchikov RG flow structure between $m_{UV}=12$ (up) and $m_{IR}=11$ (down). The orange boxes highlight all possible completions of the model of the highest global index involved in the RG flow with their corresponding superselection sectors tensor categories. In this case the (sub)models corresponding $E_6$ are included  }\label{rg1112}
\end{figure}

The compatibility of the Zamolodchikov and $SU(2)^{\textrm{even}}_{m-1}$ RG from $m=12$ to $m=11$  with the new model extensions can be seen in figure \ref{rg1112}. This is compatible with the known results between the $(E_6,A_{12})$ and $(A_{10},E_6)$ modular invariants \cite{NonDiagonalRavanini}. These new algebraic structures are also compatible with the flows from $m=13$ to $m=12$ and from $m=11$ to $m=10$ because the flows in this case happen in a branch that is isolated from the $E_6$ completions. 

There is a new flow allowed form $m_{UV}=13$ to $m_{IR}=12$ of the type  $SU(2)^{\textrm{even}}_{m-1}$ symmetric, on top of the two that are generally available for odd $m_{UV}$. This is because there is a new $E$-type model with this category for $m=12$. 

 \begin{figure}[H]
    \centering
    \includegraphics[width=1.1\textwidth]{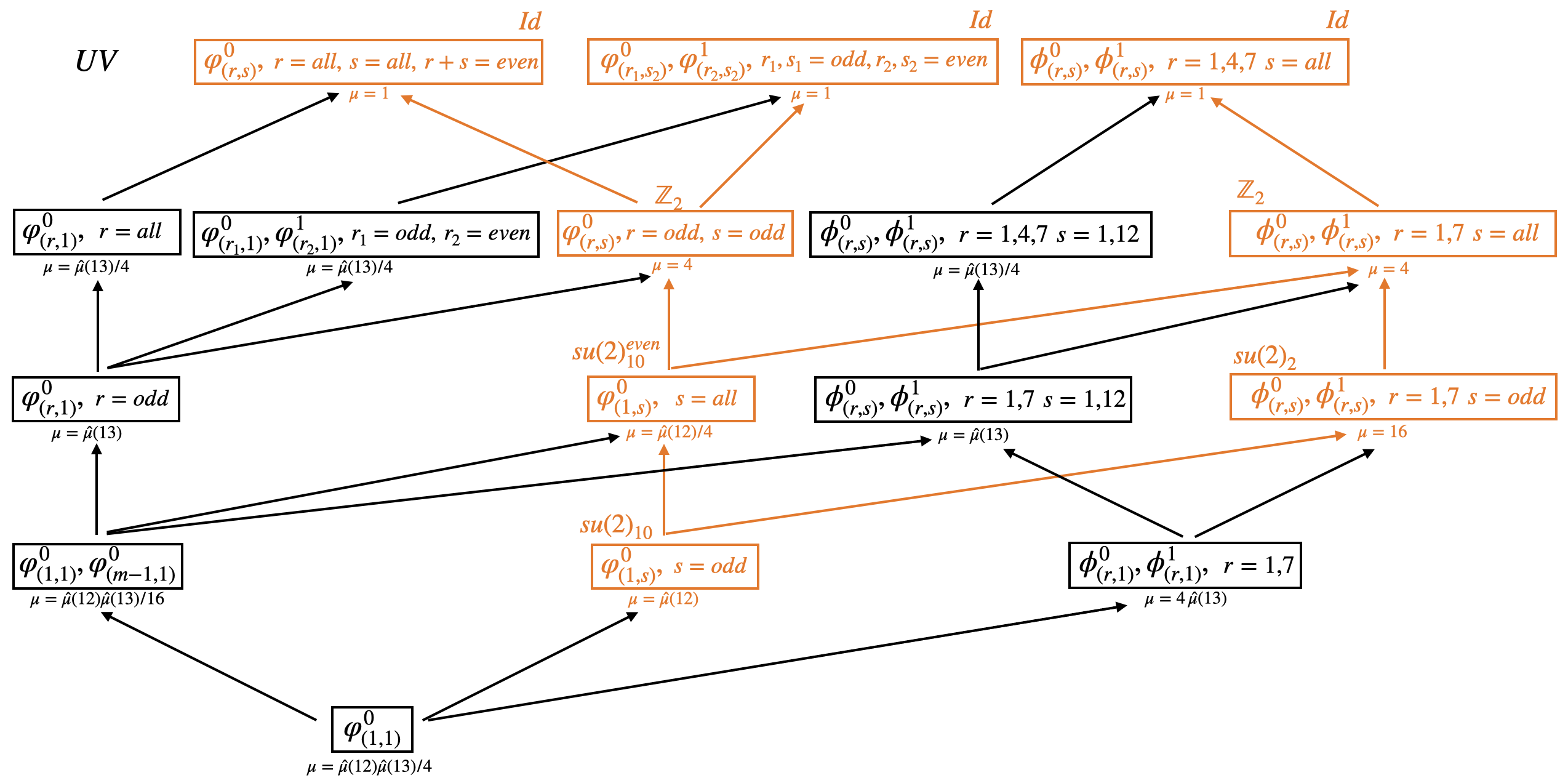}
\end{figure}

\section{Discussion}\label{discussionrg}

The recent results \cite{Benedetti:2024dku} equating completeness of the local operator algebras and modular invariance of the Euclidean theory on the torus suggested the existence of a finer classification of $d=2$ CFTs. The first objective of this article has been to provide such a classification for minimal models, i.e. $d=2$ CFTs with $c<1$. This classification has been achieved on a purely constructive approach. Starting with the appropriate chiral algebras we add primaries in all possible ways and enforce OPE closure and locality. Our results are consistent with the classification of allowed superselection sector categories obtained by Kawahigashi and Longo \cite{Kawahigashi:2002px,Kawahigashi:2003gi}. Albeit, very similar classifications at the end, we find it important to describe the conceptual and practical differences. While Ref. \cite{Kawahigashi:2002px,Kawahigashi:2003gi} classified possible subcategories of extensions of the Virasoro algebras with $c<1$, we classified $d=2$ CFTs  with $c<1$. Equivalently, we classified actual models as defined by their spectrum of primaries, not only the possible symmetries. In fact, we found different models for the same symmetry category, and also understood the precise inclusion structure of different models. Our classification naturally includes models without parity symmetry. Still, the observed fact that these two classifications are very similar expresses the power of symmetry for $d=2$ CFTs  with $c<1$, which almost determines the models themselves. Relatedly, our results can be seen as a rederivation of Ref. \cite{Kawahigashi:2002px,Kawahigashi:2003gi} classification (in a purely constructive manner we only see the appearance of the categories classified there). 

Given this finer understanding of minimal models, i.e. of fixed points of the RG flow with $c<1$, our second objective has been to provide a unified account on selection rules for RG flows connecting these theories. The fact that symmetry is so powerful in this zoo of theories, allowed to obtain strong selection rules for a large variety of RG flows. For example, given Zamolodchikov irreversibility theorem, some scalar perturbations fix the resulting central charge in the IR. Also, the fields can be organized into categorical multiplets, and non-trivial maps between UV and IR operators classes can be obtained. Another interesting aspects is the conservation of Jones indices and the structure of completions. Finally, some simple no-go results can be obtained, e.g. perturbing with such scalar primary cannot end in a non trivial IR local CFT.

\noindent There are several interesting avenues for future research:

\textbf{Massless algebras and a conjecture:}
In our analysis of RG selection rules, we have imposed that the UV CFT gives place to the IR CFT without massive sectors. In some cases this was shown not to be possible, implying a massive outcome of the RG flow. There is an interesting possibility for RG flows being only {\sl partially massless}. More precisely, this means that correlation functions of some interpolating field corresponding to UV primaries are massive (with exponentially decaying correlators at large distances) while others are massless (power law decaying correlators). When both happen at the same time an interesting situation occurs. When looked at large scales we have massless subalgebras that close into themselves. 
 This can be interpreted as that there is an IR symmetry, interpreted as the one keeping this massless algebra in itself.  This symmetry could then permeate to all scales of the full QFT including the UV. This is suggested by the stress tensor being neutral, i.e. massless in the above sense, because a massive stress tensor would give place to a gapped model. Algebraically, a violation of Haag duality for a certain region can be deformed to a Haag duality violation to deformed regions (in particular scaled regions) by the action of the stress tensor. Note that this argument applies as well to higher dimensional theories, where symmetries are simply given by internal group transformations. This idea seems to be supported by Feynmann diagrams of simple perturbative models.  Then the conjecture is that cases when both massless and massive fields happen simultaneously necessarily involve the existence of a global internal symmetry of the full theory. Massive fields are charged with respect to this symmetry while  massless ones are uncharged. This is in a certain way a precise version of the idea of naturalness.

\textbf{Comments on non-unitary models:} In this article we have not commented on non-unitary models. These models are not defined through local von Neumann algebras and Hilbert spaces and our basic starting points mentioned in the introduction are not valid. Still, it is interesting to note that the fusion rules (the OPEs) for non-unitary models are the same, the only things that change are the upper bounds for the sums of the indices $r$ and $s$ appearing in the chiral fields. More precisely, the bounds $r\leq m-1$ and $s\leq m$ do not apply anymore and we have $r\leq p-1$ and $s\leq q-1$, for some choice of $p$ and $q$. This means that the general structure of OPE subalgebras of these non-unitary models is similar to the one above. However, for a given $p$ and $q$ we will find a different zoo of exclusive relevant scalars. It would be interesting to understand if the associated selection rules work in the same way as above, i.e. if there is an effective notion of category of superselection sectors and Jones indices that clarify those selection rules. For example, it would be interesting to recover the results of \cite{Nakayama:2024msv} from this point of view.

\textbf{Extensions to rational theories with $c>1$:} Another direction for future research concerns extensions of these constructions to the case $c>1$. While non-rational theories might be out of reach in this vein, we expect rational theories to work in very much the same way as for minimal models. Of course, in such scenarios, we expect a wilder zoo of symmetry categories, associated with a larger space of affine chiral models, such as those appearing in coset of constructions. The classification of submodels of complete coset constructions, whether through OPE techniques as we used above, or whether using $Q$-system algebraic techniques as in \cite{Kawahigashi:2002px,Kawahigashi:2003gi}, is an important problem, as it would allow e.g. to study RG flows of the form ($g_l$ is an affine Lie algebra)
\be  
UV\rightarrow\frac{g_l\times g_{l+m}}{g_l}\;,\,\,\,\,\,\,\,\,\,  
IR\rightarrow\frac{g_l\times g_{m-l}}{g_m}\;,
\ee
and many others, again in a unified fashion. It would be interesting to verify whether one can arrive at similar results as in \cite{GaiottoDomain,Delmastro:2021otj} using these concepts.

\textbf{Gapped flows and Topological Quantum Field Theories:} Related to the first item in this list, it is interesting to ask for the fate of symmetries whose associated charged operators are massive. In some cases, one expects that the symmetry gets codified into a low energy Topological Field Theory, see e.g. \cite{Reshetikhin:1989qg,Smirnov:1991uw,OBrien:2017wmx,Aasen:2016dop,Aasen:2020jwb,ShaoMinimal,CordovaMinimal} and references therein. In principle, a natural guess is that the category defining the TQFT is the dual to the category of superselection sectors of the non-complete node associated with the perturbation. It would be interesting to see if the previous classification of incomplete models helps in the understanding of the classification of possible TQFTs arising in the low energy limit of $d=2$ QFTs.

\section*{Acknowledgements} 
We are grateful for insightful discussions with Yichul Choi, Paul Fendley, Davide Gaiotto, Yasuyuki Kawahigashi, Roberto Longo, Javier Molina-Villaplana, Brandon  Rayhaun, Sylvain Ribault, Gonzalo Torroba and Yunqin Zheng. We are especially grateful to Rongvoram Nivesvivat and Sylvain Ribault for sharing their upcoming work on E-series minimal models fusion rules. The work of H.C and J.M is partially supported by CONICET, CNEA and Universidad Nacional de Cuyo, Argentina. H.C and J.M wish to thank the hospitality from Perimeter Institute where part of this work was developed. V.B acknowledges the support of a RFA Fellowship from the Abdus Salam International Centre for Theoretical Physics (ICTP), Trieste, Italy. The work of V.B and J.M  was performed in part at the Aspen Center for Physics during the workshop ``The Microscopic Origin of Black Hole Entropy'', supported by a grant from the Simons Foundation (1161654, Troyer). 

\appendix

\section{Algebraic inclusions, Haag duality, and entropic order parameters}\label{HaagDuality}

In this appendix we briefly review the notion of Jones index associated with algebraic inclusions \cite{J,KOSAKI1986123,L11,Longo:1994xe}, and its application to Haag duality in QFT \cite{Casini:2020rgj,Review,Casini:2019kex,kawahigashi2001multi,Benedetti:2024dku}. This way we introduce in more detail some of the concepts and quantities computed above. We start by describing general aspects of algebraic inclusions, following mostly Ref. \cite{Longo:1994xe}. We then describe the application of these mathematical tools in the context of $d=2$ CFTs, following \cite{Benedetti:2024dku}.

\subsection{Algebraic inclusions and the Jones index}

A constant theme in this article has been the situation in which we have an inclusion of type III Von Neumann algebras
\be
\mathcal{N}\subset\mathcal{M}\;.
\ee
In this situation there might be a space of conditional expectations $\varepsilon : \mathcal{M}\rightarrow \mathcal{N}$. These are linear maps from $\mathcal{M}$ to $\mathcal{N}$ forced to satisfy
\be
\hspace{-1mm} \varepsilon\left(n_{1}\,m\,n_{2}\right)=n_{1}\varepsilon\left(m\right)n_{2}\,,\hspace{3mm} \forall m\in\mathcal{M},\,\forall n_{1},n_{2}\in\mathcal{N}.\label{ce_def_prop}
\ee
For the classification of these spaces of conditional expectations for general algebras see \cite{L11,Magan:2020ake} and references therein.

Key further information appears through the subfactor machinery described in \cite{Longo:1994xe}. We have the Jones ladder associated to the original inclusion
\be
\cdots\supset\mathcal{M}_1\supset\mathcal{M}\supset\mathcal{N}\supset\mathcal{N}_1\supset\cdots\;,
\ee
and its commutant counterpart
\be
\cdots\subset\mathcal{M}_1'\subset\mathcal{M}'\subset\mathcal{N}'\subset\mathcal{N}_1'\subset\cdots\;.
\ee
These ladders are built by adjoining subsequent Jones projections to the given algebras, e.g. $\mathcal{M}_1 =\mathcal{M}\vee e_\mathcal{N}$, where $e_\mathcal{N}\in \mathcal{N}'$ is the Jones projection associated with the inclusion $\mathcal{N}\subset\mathcal{M}$. This Jones projection is the operator that projects into the Hilbert space generated by $\mathcal{N}$ from the vacuum.

Now, given $\mathcal{N}\subset \mathcal{M}$ and a cyclic and separating vector for both algebras $\vert \Omega\rangle$, Ref. \cite{L11} introduces the ``canonical endomorphism''  as
\be  
\gamma (\mathcal{M})\equiv j_{\mathcal{N}}j_{\mathcal{M}} (\mathcal{M}) \subset \mathcal{M}\;,
\ee
where $j_{\mathcal{M}}(m)\equiv J_{\mathcal{M}} \,m \, J_{\mathcal{M}}$ and $j_{\mathcal{N}}(n)\equiv J_{\mathcal{N}} \,n \,J_{\mathcal{N}}$, are the modular conjugations associated with each algebra and the vector $|\Omega\rangle$. This is an endomorphism for $\mathcal{M}$ for which $\gamma (\mathcal{M})=\mathcal{N}_1$. It naturally restricts to the canonical endomorphism $\rho$ associated with $\mathcal{N}\supset\mathcal{N}_1$ as 
$\rho (\mathcal{N})\equiv \gamma\vert_{\mathcal{N}}\subset \mathcal{N}$. We also have $\rho (\mathcal{N})= \mathcal{N}_2$, and so forth. Then the canonical endomorphism jumps two steps on the Jones ladder.

For strict inclusions $\mathcal{N}\subset\mathcal{M}$, the canonical endomorphism is not an irreducible endomorphism, in the precise sense that
\be 
\gamma(\mathcal{M})'\cap \mathcal{M}\neq \mathds{1}\;.
\ee
If $\mathcal{N}\subset \mathcal{M}$ has  finite Jones index $\lambda$, the canonical endomorphism can be expressed as a finite direct sum of irreducible endomorphisms $\rho_r$ of $\mathcal{N}$ in the standard ``group theory'' manner
\be \label{candec}
\rho \simeq \oplus_r N_r\, \rho_{r} \;,
\ee
Here $N_r$ is the multiplicity with which representation $\rho_{r}$ appears in $\rho$. For irreducible subfactors $\mathcal{N}' \cap\mathcal{M}= \mathds{C}$, the identity endomorphism appears only once.

More generally, there is a natural composition of endomorphisms of a von Neumann algebra $\mathcal{N}$. We have
\be 
\rho_r\circ\rho_{r'}(n)\equiv \rho_r (\rho_{r'}(n))\:,\quad\,\,\,n\in \mathcal{N}\;.
\ee
This is well defined since $\rho_r(\mathcal{N})\subseteq \mathcal{N}$. As for any endomorphism, this can be decomposed as a direct sum of irreducible sectors
\be 
\rho_r\circ\rho_{r'}\simeq \oplus_{r''} \textbf{N}_{rr'}^{r''}\, \rho_{r''}\;.
\ee
For general endomorphisms of von Neumann algebras, this composition might not be commutative, i.e. $\textbf{N}_{rr'}^{r''}\neq \textbf{N}_{r'r}^{r''}$. For DHR endomorphisms we have a composition that is commutative due to causality so that in such case
\be 
\textbf{N}_{rr'}^{r''}= \textbf{N}_{r'r}^{r''}\;,\quad\,\,\,\,\rho_r\in \textrm{DHR}\;.
\ee
Equivalently, $\textrm{DHR}$ endomorphisms form a fusion category, see \cite{haag2012local} for a full account.

A natural question in this context is if the irreducible endomorphisms appearing in the decomposition of the canonical endomorphism (\ref{candec}) close on themselves with respect to this fusion. In the case of $d>2$, the reconstruction theorem \cite{Doplicher:1990pn} ensures the category of DHR endomorphisms is the dual of a compact group $G$, i.e. the irreducible endomorphisms are labeled by irreps $r$ of $G$ and the canonical endomorphisms is
\be
\rho \simeq \oplus_r d_r\, \rho_{r}\;,
\ee
where $d_r$ is the dimension of the irrep $r$ of $G$. In this case, since the fusion of endomorphisms is the same as the fusion of the irreps in the group, the set of irreducible endomorphisms appearing in the canonical endomorphism closes under fusion.

But in $d=2$, and in general subfactors as well, this is not the case, and we typically see scenarios in which the $\rho_r$ appearing in the canonical $\rho$ generate further irreducible endomorphisms under fusion which were not present in $\rho$. We will see very explicit examples below in the application to $d=2$ CFTs. Still, for scenarios of finite index, this fusion will stop at some point, generating a finite (close in itself) fusion subcategory of the full category of endomorphisms of the algebra. It is said in this case that the fusion has finite depth.

Let us come back to the canonical endomorphism (\ref{candec}). Such decomposition implies we can find a finite set of partial isometries $\omega_r^i\in \mathcal{N}$, $i=1,\cdots, N_r$, satisfying
\be 
\omega_r^{i\dagger}\omega_{j}^{j}=\delta_{ij}\delta_{rs}\,, \quad\sum_{r,i}\, \omega_r^i \,\omega_r^{i\dagger}=\mathds{1} \,,\quad \omega_r^i\, \rho_{r}(n) =\rho(n)\,\omega_r^i\,,\quad n\in \mathcal{N}\;.
\ee
In terms of these isometries, the canonical endomorphism explicitly reads
\be 
\rho(n)=\sum_{r,i}\, \omega_{r}^i\,\rho_r(n)\,\omega_{r}^{i\dagger}\;.
\ee 
Notice the isometry $\omega_{\mathds{1}}\equiv\omega$ that intertwines the identity representation with the canonical endomorphism is unique for irreducible subfactors. The conditional expectation is also unique in this scenario. It reads $\varepsilon (m)= \omega^{\dagger} \gamma (m) \omega$. Ref. \cite{Longo:1994xe} then shows that we can reconstruct $\mathcal{M}$ from $\mathcal{N}$ and a further isometry $v\in \mathcal{M}$, which intertwines the identity representation and the canonical endomorphism $\gamma$ of $\mathcal{M}$. The range projection of this partial isometry is the Jones projection $vv^{\dagger}=e_{\mathcal{M}'}\in \mathcal{M}$ for the dual inclusion $\mathcal{M}'\subset \mathcal{N}'$.

The reconstruction in terms of the isometry $v$ makes also explicit the existence of charged operators $\psi_r^i$ intertwining the identity representation and the irreducible sectors $\rho_r$. Explicitly we can define\footnote{We can make them isometries by renormalizing the operators as $\psi_r^i\rightarrow\sqrt{\frac{\lambda}{d_r}}\psi_r^i$.}
\be 
\psi_r^i\equiv \omega_{r}^{i\dagger}\,v\;.
\ee
Since $v$ intertwines the identity with $\gamma$ in $\mathcal{M}$, it also intertwines the identity with $\rho$ in $\mathcal{N}$, and then we conclude
\be 
\psi_r^i\,n=\rho_r(n)\,\psi_r^i\;,\quad\,\,\,\, n\in \mathcal{N}\;.
\ee
The fact that the irreducible endomorphisms $\rho_r$ appearing in $\rho$ might not close under the fusion of endomorphisms naively suggests that the charge operators $\psi_r^i$ also generate irreducible sectors $r$ under the OPE that were nor present in the canonical endomorphism. But this cannot be the case since such operators will be outside $\mathcal{M}$, which is not possible since $\mathcal{M}$ is an algebra and $\psi_r^i \in \mathcal{M}$. To see this more explicitly, Ref. \cite{Longo:1994xe} shows that any element $m\in \mathcal{M}$ can be written as $n\,v$ for some $n\in \mathcal{N}$. The precise formula uses the conditional expectation and reads
\be 
m=\lambda\,\varepsilon (mv^*)\,v\;.
\ee
Since $\psi_r^i \in \mathcal{M}$, we have that $\psi_r^i\,\psi_{r'}^j \in \mathcal{M}$. Applying the previous formula one can arrive at \cite{Longo:1994xe}
\be 
\psi_r^i\,\psi_{r'}^j =\lambda\,\sum\limits_{r'',k,l}\,(C_l)_k^{ij}\,T_e\,\psi_{r''}^k\;,
\ee
where $T_e$ is a basis of intertwiners for $\rho_{r''}\rightarrow\rho_r\rho_{r'}$, only charges $r''$ contribute which are contained in $\rho$, and we have defined the generalized ``Clebsch-Gordan coefficients'' $(C_l)_k^{ij}\in \mathcal{N}$. The charged operators $\psi_r^i$ creating the $\rho_r$ that appear in the canonical endomorphism do close an algebra under the OPE. This closure does not furnish a proper fusion category but it is what we use in the text above to construct local $d=2$ CFTs.

Finally, a further important aspect of the Jones ladder is that inclusions are anti-isomorphic in a zig-zag manner. This is due to the modular conjugations, which serve to map inclusions to one another. For example $\mathcal{N}_1\subset\mathcal{N}$ is anti-isomorphic to $\mathcal{M}'\subset\mathcal{N}'$ since $j_{\mathcal{N}}(\mathcal{N})=\mathcal{M}$ and $j_{\mathcal{N}}(\mathcal{N}_1)=\mathcal{M}'$. Therefore the canonical endomorphism of $\mathcal{M}'\subset\mathcal{N}'$ decomposes similarly to \ref{candec} as
\be \label{candec}
\tilde{\rho} \simeq \oplus_r N_r\, \tilde{\rho}_{r} \;,
\ee
where $\tilde{\rho}_{r}$ are the irreducible endomorphisms of $\mathcal{N}'$. Both $N_r$ and the dimensions $d_r=\tilde{d}_r$ are equal for both inclusions.\footnote{See \cite{Longo:1994xe} for the proper definition of dimension of an endomorphism.} Then we have an algebra of dual partial isometries $\tilde{\omega}_r^{i}\in \mathcal{N}'$ intertwining $\tilde{\rho}_{r}$ in $\tilde{\rho}$. This can be used to build projectors into the different representations
\be 
P_r^i\equiv \tilde{\omega}_r^{i} \tilde{\omega}_r^{i\dagger}\in  \mathcal{N}'\cap \tilde{\rho}(\mathcal{N}')'\;.
\ee
These projectors belong to the relative commutant. Ref. \cite{Longo:1994xe} shows that
\be \label{procon}
\varepsilon (e_\mathcal{N})=\varepsilon (vv^{\dagger})=\frac{1}{\lambda}\mathds{1}\,,\,\,\,\,\,\,\,\,\,\,\,\,\,\,\,\,\,\varepsilon' (P_r^i)=\frac{d_r}{\lambda}\;,
\ee
where $\varepsilon'$ is the dual conditional expectation, i.e. the one associated to the inclusion $\mathcal{M}'\subset \mathcal{N}'$, and where we have introduced the Jones index $\lambda$ associated with the conditional expectation $\varepsilon$. In these scenarios, this index is proven to be the dimension of the canonical endomorphism
\be \label{lrho}
\lambda=\sum_r N_r d_r\;.
\ee
Jones celebrated result \cite{Jones1983} is that indices lying in between $1$ and $4$ belong to a discrete series $4 \cos(\pi/n)^2, n=3,4,\cdots$, that gives $1,2,(3+\sqrt{5})/2,3,\cdots$. Any value is possible for indices greater than $4$.

There are a couple of interesting examples where the formula (\ref{lrho}) give familiar results. The first is the case of a symmetry group $G$, i.e in the inclusion $\mathcal{M}\supset\mathcal{N}$, the small algebra $\mathcal{N}$ is the invariant part under the action of the symmetry group $G$. In this case, we have $N_r=d_r$ and then $\lambda=\sum_r d_r^2=G$, the order of the group. Another is the inclusion of a chiral algebra into a diagonal modular invariant completion. In this case $N_r=1$, but $d_r=(d_{\textrm{chiral}})^2_r$. Then $\lambda=\sum_r (d_{\textrm{chiral}})^2_r$ is the total quantum dimension of the model.

\subsection{Application to $d=2$ CFTs}

 In $d=2$ CFTs, a general region $R$ is a multi-interval region, and we should study the violation of Haag duality and its structure for such regions. For chiral theories, these inclusions were studied in \cite{kawahigashi2001multi}, where they were related to the category of DHR superselection sectors \cite{Doplicher:1971wk,Doplicher:1973at}, equivalently the modular tensor category associated with the chiral algebra.\footnote{In $d>2$, the category of DHR sectors has permutation exchange symmetry. It is said to be symmetric. This is the starting point for the DHR reconstruction theorem \cite{Doplicher:1990pn,Longo:1994xe}. For $d=2$, the category of DHR sectors does not admit generically a permutation symmetry. Instead, it is a braided category \cite{FRS,cmp/1104179464}, with a representation of the braiding group.} A more recent analysis is Ref. \cite{Benedetti:2024dku}, which we now follow.

 Consider a $d=2$ CFT, and a two interval region $R=R_1\cup R_2$ defined by the segments $(a_1,a_2)$ and $(b_1,b_2)$ with conformal ratio $x\equiv
 {(b_1-a_1)(b_2-a_2)}/{(a_2-a_1)(b_2-b_1)}\in (0,1)$. We can extract the Jones index of the region $R$ from the bootstrap data in the following manner. The Renyi entropy is defined as
 \be 
S_n=(1-n)^{-1}\log \textrm{tr} \rho^n\;.
 \ee
 Then we can define the following mutual information for $R$
 \be 
I_n\equiv S_n(R_1)+S_n(R_1)-S_n(R_1\cup R_2)\equiv -\frac{(n+1)c}{6n}\log (1-x)\,+U_n(x)\;,
 \ee
 where we have written it in terms of the conformal ratio since it is a finite universal quantity. This expression defines the function $U_n(x)$. If Haag duality is satisfied for two intervals then this function must satisfy the crossing symmetry
 \be 
U_n(x)=U_n(1-x)\Longleftrightarrow \textrm{Haag duality holds for} R\;.
 \ee
 We now specify to $n=2$. In this case, the manifold computing the Reny entropy is a genus one surface. It can be conformally mapped to a torus of radius $1$ and height $l$ \cite{Headrick:2010zt}. The relation between the conformal ratio $x$ and the height $l$ is
 \be 
x=\left(\frac{\theta_2(il)}{\theta_3(il)}\right)^4\;.
 \ee
Defining the partition function in the usual manner
\be 
Z(\tau)\equiv tr\,q^{L_0-c/24}\,\bar{q}^{\bar{L}_0-c/24}\;,\,\,q\equiv e^{i2\pi\tau}\;,
\ee
after some algebra one arrives at the following expression for the crossing assymetry
\be 
U_2(x)-U_2(1-x)=\log\,Z(il)-\log\,Z(i/l)\;.
\ee
Therefore, a violation of Haag duality implies a violation of S duality and viceversa. Equivalently, the local meaning of modular invariance is completeness of the QFT, in the sense mentioned above \cite{Benedetti:2024dku}.

The crossing assymetry is in general a complicated function. But in the limit of touching intervals, i.e $x\rightarrow 1$ we can obtain a universal expression. Write the partition function in terms of the coupling matrix $M_{rs}$ that combines the chiral representation $r$ with the anti-chiral representation $s$ so that
\be 
Z=\sum\,M_{rs}\chi_r(\tau)\bar{\chi}_s(\overline{\tau})\;,
\ee
where $\chi_r(\tau)\equiv tr_r\,e^{2\pi i\tau(L_0)-c/24}$. The chiral characters transform under modular transformations in the usual way
\be 
\chi_r(-1/\tau)=\sum\limits_{s}\,S_{rs}\,\chi_s(\tau)\;,\,\,\,\,\,\,\chi_r(\tau+1)=\sum\limits_{s}\,T_{rs}\,\chi_s(\tau)\,.
\ee
Using the definition of the quantum dimension and modular $S$ matrices
\be 
\lim\limits_{\tau\rightarrow 0}\frac{\chi_r(\tau)}{\chi_1(\tau)}=\frac{S_{0r}}{S_{00}}=d_r\,\,\,\,\quad\,\,\,\,\lim\limits_{\tau\rightarrow 0}\frac{\chi_1(\tau)}{\chi_1(-1/\tau)}=S_{00}\;,
\ee
we obtain
\be 
\lim\limits_{x\rightarrow 1}U_2(x)-U_2(1-x)=\lim\limits_{l\rightarrow 0}\log\left(\frac{Z(\tau)}{Z(-1/\tau)}\right)=\log\left(\frac{\sum_{ij}M_{ij}d_i d_j}{\sum_i\,d_i^2}\right)\,.
\ee
It turns out that this number has a precise physical and mathematical meaning in the theory. A technical derivation shows that
\be 
\mu^{-1/2}=\lim\limits_{l\rightarrow 0}\frac{Z(\tau)}{Z(-1/\tau)}=\frac{\sum_{ij}M_{ij}d_i d_j}{\sum_i\,d_i^2}\;,
\ee
where $\mu$ is the Jones index associated with the inclusion of algebras for the two interval region $R$, namely
\be 
\mu=\textrm{Jones index of}  \,\,\,{\cal A}(R_1\cup R_2)\subseteq  \hat{{\cal A}}(R_1\cup R_2)\;.
\ee
This is the global index $\mu$ we have been computing for the classification of local minimal models above. It is intrinsic to the theory. In can only take the values classified by Jones and mentioned above. For $\mu=1$ the two algebras are the same and Haag duality is satisfied. The model is complete. Such coupling matrices are modular invariant. For $\mu>1$ the model is incomplete and $S$-duality is violated.\footnote{For a region $R$ composed of $n$ intervals one has an associated Jones index $\mu^{n-1}$.}

Also, an insightful relation was obtained in Ref. \cite{kawahigashi2001multi} for inclusion of theories $\mathcal{T}\in\mathcal{C}$ of the sort we have been discussing in the article. For such inclusion, we have three natural Jones indices, the global indices of each theory, i.e. $\mu_{\mathcal{T}}$ and $\mu_{\mathcal{C}}$, and the index $\lambda$ of the inclusion of theories. The latter measures the relative size of both theories. These indices are related as
\be\label{mugi}
\mu_T=\mu_C \, \lambda^2\,,
\ee
 a relation that we have also used above.

Finally, it is worth noticing that $d=2$ CFTs furnish an example of the subtle discrepancy between the fusion of DHR endomorphisms appearing in the canonical endomorphism and the OPE of charged operators creating those endomorphisms from the vacuum. Let's take the Ising model as an explicit example. All other models work the same way. The Ising model contains three chiral sectors $1,\varepsilon, \sigma$. These are also the chiral DHR sectors of the chiral algebra. This model also contains three primary local fields $(1,1),(\varepsilon,\varepsilon),(\sigma,\sigma)$, by combining the same chiral and antichiral sectors. It also contain associated diagonal DHR sectors $\rho_{(1,1)},\rho_{(\varepsilon,\varepsilon)},\rho_{(\sigma,\sigma)}$. The canonical endomorphism is
\be \label{canis}
\rho_{\textrm{Ising}}\sim \rho_{(1,1)}\oplus \rho_{(\varepsilon,\varepsilon)}\oplus \rho_{(\sigma,\sigma)}\;.
\ee
The OPE of the local fields closes in itself
\be
(\sigma,\sigma)(\sigma,\sigma)\sim (1,1) + (\varepsilon,\varepsilon)\;.
\ee
There are several ways to see this. In particular, one can use the OPE that follows from the solution of the theory. However, as DHR endomorphisms, the fusion does not close in itself. Indeed we have
\be
\rho_{(\sigma,\sigma)}\circ\rho_{(\sigma,\sigma)}\sim \rho_{(1,1)}\oplus \rho_{(1,\varepsilon)}\oplus \rho_{(\varepsilon,1)}\oplus \rho_{(\varepsilon,\varepsilon)}\;.
\ee
The new DHR sectors,  $\rho_{(1,\varepsilon)}$ and $\rho_{(\varepsilon,1)}$, that are not part of the the canonical endomorphisms (\ref{canis}), are created by charged fields (fermionic in this case), $(1,\varepsilon)$ and $(\varepsilon,1)$, that are not created in the OPE of any of the fields of the theory $(1,1),(\varepsilon,\varepsilon),(\sigma,\sigma)$. In fact this is clearly impossible since the resulting theory would be non-local.

\section{Fusion, quantum dimensions and Jones indexes for \texorpdfstring{$su(2)_n$}{Lg} and \texorpdfstring{$D_{2n}^{\textrm{even}}$}{Lg}\label{su2}}
The affine lie algebra $su(2)_n$ has $n+1$ sectors or charges $\alpha_i$ for $i=0,1,2,\dots,n$ . The fusion rules are 
\be 
\alpha_i \otimes \alpha_j =\bigoplus\limits_{k=\vert i-j\vert,i+j+k \,\textrm{even}}^{\textrm{min}(i+j,2n-i-j)}\alpha_{k}\;,
\ee
The modular S matrix of an affine lie algebra $su(2)_{n}$ can be computed from the modular transformations of the characters to be 
\be 
S_{ij}=\sqrt{\frac{2}{(n+2)}}\sin\left(\frac{\pi(i+1)(j+1)}{n+2}\right)\,,\quad 0\leq i,j \leq n\;.
\ee
The quantum dimensions follow from this formula in the usual way. They read
\be 
d_{i}= \frac{S_{0i}}{S_{00}} = \frac{\sin\left({\pi(i+1)}/({n+2})\right)}{\sin\left({\pi}/{(n+2)}\right)}\,,\quad 0\leq i \leq n\;.
\ee
As mentioned above, a local field is a combination of a chiral and antichiral representation of the category of superselection sectors. A field with chiral sectors $(i,j)$ has then dimension $d_{ij}=d_i\,d_j$. If the partition function defining the model is
\be 
Z=\sum\,M_{ij}\chi_i(\tau)\bar{\chi}_{j}(\overline{\tau})\;,
\ee
then the canonical endomorphism reads
\be
\rho\simeq \oplus  M_{ij}\, \rho_{i}\bar{\rho}_{j}\;.
\ee
As reviewed in the previous section, the dimension of the canonical endomorphism is then
\be 
d_\rho=\sum M_{ij} d_i d_j\;.
\ee
For a diagonal complete model this would be
\be 
d_\rho=\sum\limits_{i}  d_i^2\;.
\ee
Using equation (\ref{mugi}), the global index is
\be
{\mu}\big[su(2)_{n}\big]=\Big(\sum_{i=0}^n d_{i}^2\Big)^2= \frac{(n+2)^2}{4}\sin^{-4}\left(\frac{\pi}{n+2}\right)\;.
\ee
Finally, the category $su(2)_n^{even}$ is the one generated by all the even sectors of $su(2)_n$ namely $J_i$ for $i=0,2,4,\dots,2[n/2]$. For this we have
\be
{\mu}\big[su(2)_{n}^{\text{even}}\big]=\Big(\sum_{\substack{i=0\\\text{Mod 2}}}^{2[n/2]} d_{i}^2\Big)^2= \frac{(n+2)^2}{16}\sin^{-4}\left(\frac{\pi}{n+2}\right)\;.
\ee
We can verify that the index of the inclusion of the even part into the complete one is 
\be 
\lambda=\sqrt{\frac{{\mu}\big[su(2)_{n}\big]}{{\mu}\big[su(2)_{n}^{\text{even}}\big]}}=2\;,
\ee
and associated $\mu=4$, as expected.

Moving now to the $D^{\text{even}}_{2n}$ case. To our knowledge this was first discussed in \cite{BE3,BE2,BE4}.\footnote{In such references this category is called $D_{2\varrho+2}$.} We start with the simplest $n=1$ case, i.e. $D_4$. This is the category that appears in the conformal embedding of chiral algebras $su(2)_{4}\subset su(3)_1$. If we call $\chi_i$ with $i=0,\cdots,4$ the chiral characters of the affine $su(2)_{4}$ algebra, the corresponding  modular invariant is
\be 
Z_{D_4}=\vert \chi_0+\chi_4\vert^2+2\vert\chi_2\vert^2\;.
\ee
We then have three extended blocks, one formed by $\xi_0$ and $\chi_4$, and the $\chi_2$ is duplicated. The $D_4$ category has, then, three elements $\chi^D_0$, $\chi^D_{2,1}$, $\chi^D_{2,2}$. Using $\alpha$ induction techniques, Ref. \cite{BE3,BE2,BE4} showed the fusion category associated to these objects is that of $\mathds{Z}_3$, as expected for the chiral algebra $su(3)_1$. The global index is then $\mu_{D_4}=9$.

Part of the previous structure goes over to the general scenario. For convenience, we consider without loss of generality $n\rightarrow n+1$. The category $D^{\text{even}}_{2n+2}$ is given in terms of the $su(2)_{4n}$. In the general case, the category does not come from a conformal inclusion as in the $D_4$ case, but the structure is similar. Again label the chiral representations of $su(2)_{4n}$ as $\chi_i$, with $i=0,\cdots,4n$. Then the corresponding modular invariant is
\be
Z_{D^{\text{even}}_{2n+2}}=\frac{1}{2}\,\sum\limits_{\substack{4n\geq i\geq 0\\ i\in 2\mathds{Z}}}\vert\chi_i+\chi_{4n-i}\vert^2\;.
\ee
From the perspective of $su(2)_{4n}$, what this is doing is
\begin{itemize}
    \item Do not include odd representations. This is the origin of $\textrm{even}$ in $D^{\text{even}}_{2n+2}$.
    \item Construct combined blocks from $\chi_i$ and $\chi_{4n-i}$, with $i$ even.
    \item The chiral representation $\chi_{2n}$ get duplicated.
\end{itemize}
The fusion category $D^{\text{even}}_{2n+2}$ has then $n+2$ elements. We can think of these elements in the following way, which then helps with the fusion rules. Take $2n+2$ objects: $\alpha_i$, with $i=0,1,\cdots, 2n-1 $, $\alpha_{2n}^{(1)}$ and $\alpha_{2n}^{(2)}$. Then $D^{\text{even}}_{2n+2}$ is composed of the $\alpha_i$ with $i$ even plus $\alpha_{2n}^{(1)}$ and $\alpha_{2n}^{(2)}$. The fusion rules are as in the $su(2)_{4n}$ case, namely
\be 
\alpha_{i}\otimes \alpha_{j}=\bigoplus\limits_{k=\vert i-j\vert,i+j+k \,\textrm{even}}^{\textrm{min}(i+j,4n-i-j)}\alpha_{k}\;,
\ee
where we identify $\alpha_j$ with $\alpha_{4n-j}$ and $\alpha_{2n}=\alpha_{2n}^{(1)}\oplus \alpha_{2n}^{(2)}$. The fusion rules involving the duplicated $\alpha_{2n}^{(i)}$ are more involved and can be found in Ref. \cite{BE2}.

The resulting index can be recovered from the fusion rules. To be precise, the fusion rules fix the dimensions of $\alpha_{i}$ and $\alpha_{2n}^{(1)}$, and $\alpha_{2n}^{(2)}$. In this case, these dimensions are simple in terms of the dimensions of $su(2)_{4n}$. To be specific, the objects $\alpha_i$, with $i=0,1,\cdots, 2n-1$, have the same dimensions $d_i$ as they would have in the  $su(2)_{4n}$ category. In addition, the $\alpha_{2n}^{(i)}$ have half of the dimension $d_{2n}$. This can be thought of as a consequence of $\alpha_{2n}=\alpha_{2n}^{(1)}\oplus \alpha_{2n}^{(2)}$. In this manner,  way we obtain for the category $D^{\text{even}}_{2n+2}$ the following global index 
\be
{\mu}\big[D^{\text{even}}_{2n+2}\big]=\Big(2\left(\frac{d_{2n}^2}{2}\right)^2+\sum_{\substack{i=0\\\text{Mod 2}}}^{2n-2} d_{i}^2\Big)^2= \frac{(4n+2)^2}{64}\sin^{-4}\left(\frac{\pi}{4n+2}\right)\;.
\ee

\newpage
\bibliographystyle{utphys}
\bibliography{EE}

\end{document}